\documentclass[prd,aps,showpacs,floats,floatfix,superscriptaddress,twocolumn,nofootinbib]{revtex4}

\usepackage{graphicx}
\usepackage{color}
\usepackage{amssymb}
\usepackage{amsmath}

\usepackage{longtable}

\def\laq{\raise 0.4ex\hbox{$<$}\kern -0.8em\lower 0.62ex\hbox{$\sim$}}
\def\gaq{\raise 0.4ex\hbox{$>$}\kern -0.7em\lower 0.62ex\hbox{$\sim$}}

\newcommand{\beq}{\begin{equation}}
\newcommand{\eeq}{\end{equation}}
\newcommand{\bea}{\begin{eqnarray}} 
\newcommand{\eea}{\end{eqnarray}}
\newcommand{\bse}{\begin{subequations}} 
\newcommand{\ese}{\end{subequations}}
\newcommand{\ba}{\begin{array}}
\newcommand{\ea}{\end{array}}

\newcommand{\mytextrm}[1]{{}}

\newlength{\sizeonefig}
\newlength{\sizetwofig}
\newlength{\sizeonefigb}
\newlength{\sizetwofigb}
\begin{document}

\title{Anatomy of the binary black hole recoil: A multipolar analysis}

\author{Jeremy D. Schnittman} 

\affiliation{Maryland Center for Fundamental Physics,
Department of Physics, University of Maryland, College
  Park, Maryland 20742 }

\author{Alessandra Buonanno} 

\affiliation{Maryland Center for Fundamental Physics,
Department of Physics, University of Maryland, College
  Park, Maryland 20742 } 

\author{James R. van Meter}

\affiliation{Gravitational Astrophysics Laboratory, NASA 
Goddard Space Flight Center, 8800 Greenbelt Rd., Greenbelt, MD 20771}

\affiliation{Center for Space Science \& Technology, 
University of Maryland Baltimore
County, Physics Department, 1000 Hilltop Circle, Baltimore, MD 21250}

\author{John G. Baker} 

\affiliation{Gravitational Astrophysics Laboratory, NASA 
Goddard Space Flight Center, 8800 Greenbelt Rd., Greenbelt, MD 20771}

\author{William D. Boggs}

\affiliation{Department of Physics, University of Maryland, College
  Park, Maryland 20742 }

\author{Joan Centrella} 

\affiliation{Gravitational Astrophysics Laboratory, NASA 
Goddard Space Flight Center, 8800 Greenbelt Rd., Greenbelt, MD 20771}

\author{Bernard J. Kelly} 

\affiliation{Gravitational Astrophysics Laboratory, NASA 
Goddard Space Flight Center, 8800 Greenbelt Rd., Greenbelt, MD 20771}

\author{Sean T. McWilliams}

\affiliation{Department of Physics, University of Maryland, College
  Park, Maryland 20742 }

\begin{abstract}
We present a multipolar analysis of the gravitational recoil computed 
in recent numerical simulations of binary black hole (BH) coalescence, for both unequal
masses and non-zero, non-precessing spins. We show that multipole 
moments up to and including $\ell=4$ are sufficient to accurately
reproduce the final
recoil velocity (within $\simeq 2\%$) and that only a few dominant modes contribute 
significantly to it (within $\simeq 5\%$). We describe how the relative 
amplitudes, and more importantly, the relative phases, of these few modes 
control the way in which the recoil builds up 
throughout the inspiral, merger, and ringdown phases.
We also find that the numerical results can 
be reproduced by an ``effective Newtonian'' formula for the multipole moments 
obtained by replacing the radial separation in the Newtonian formulae
with an effective radius computed from the numerical data.
Beyond the merger, the numerical results are reproduced by a 
superposition of three Kerr quasi-normal modes (QNMs). Analytic formulae, 
obtained by expressing the multipole moments in terms of the
fundamental QNMs of a Kerr BH, are able to explain the onset and
amount of ``anti-kick'' for each of the simulations. Lastly, we apply
this multipolar analysis to help explain the remarkable difference
between the amplitudes of planar and non-planar kicks for equal-mass
spinning black holes.
\end{abstract}


\date{\today}

\maketitle

\section{Introduction}
\label{intro}

After the recent breakthrough in numerical relativity
(NR)~\cite{FP,CLMZ,Bakeretal1}, a number of different
groups are now able to evolve binary black holes (BHs)
through merger~\cite{sperhake,gonzalez,szilagyi}. Recently, a great deal of effort has been directed
towards the computation of the recoil velocity of the final
BH~\cite{HSL,recoil,recoilJena,recoilPSU,recoilAEI,recoilGoddard,Bigrecoil,recoilFAU,recoilRI}.
The fundamental cause of this recoil is a net linear
momentum flux in the gravitational radiation, due to some asymmetry in
the system~\cite{Bonnor,Peres,Bekenstein,Cooperstock,Fitchett}, typically unequal
masses or spins in the case of BH binaries. The recoil has
great astrophysical importance because it
can affect the growth of supermassive black holes (SMBHs) in the early
universe~\cite{HM,Merrittetal,Volonteri07,Schnittman07b}. In those scenarios dark-matter haloes
grow through hierarchical mergers. The SMBHs at the centers
of such haloes are expected to merge unless they have been kicked out
of the gravitational potential well because the recoil velocity
gained in a prior merger is larger than the halo's escape
velocity.

Other astrophysical implications include the displacement of
the SMBH, along with its gaseous accretion disk, forming an
``off-center'' quasar~\cite{HDR}. These quasars might also have emission lines
highly red- or blue-shifted relative to the host galaxy due to the
Doppler shift of the recoil velocity~\cite{Bonning}. Additionally, these displaced
SMBHs could in turn displace a significant amount of stellar mass from
the galactic nucleus as they sink back to the center via dynamical
friction, forming a depleted core of missing mass on the order of
twice the SMBH mass~\cite{Merrittetal,Boylan_Kolchin,Lauer}. 

Numerical simulations have now been used to compute recoil velocities for
non-spinning unequal-mass BH binary
systems~\cite{HSL,recoil,recoilJena} in the range $m_2/m_1 = (1\cdots 4)$,
where $m_1$ and $m_2$ are the individual BH masses; 
for spinning, non-precessing binary
BHs~\cite{recoilAEI,recoilPSU,recoilGoddard}, and also for
precessing BHs with both
equal~\cite{Bigrecoil,recoilFAU} as well as unequal masses~\cite{recoilRI}.
Quite interestingly, there exist initial spin configurations for which 
the recoil velocity can be quite large, e.g., $\gaq \, 3000$ km/sec~
\cite{recoilRI,Bigrecoil,recoilFAU,recoilJena2}. However, it is not yet clear
whether those very large recoil velocities are astrophysically
likely~\cite{Schnittman04,Volonteri07,bogdanovic,loeb}. So far, due to limited
computational resources, the numerical simulations have explored a
rather small portion of the total parameter space. 

Analytic calculations, based on the post-Newtonian (PN) expansion of Einstein's 
field equations~\cite{LB} and PN-resummation techniques~\cite{DIS98,BD1,BD2,DJS,DJS2,BCD}, 
have made predictions for the recoil velocity~\cite{AW,LK,Favataetal,BQW,DG} before the
NR breakthrough. Since the majority of the linear momentum flux 
is emitted during the merger and ringdown (RD) phases, it is 
difficult to make definitive predictions for the recoil 
using {\it only} analytic methods. These methods need to be somehow 
calibrated to the NR results, so that they can be accurately extended 
during the transition from inspiral to RD. 
So far, in the non-spinning case, the PN model~\citep{BQW} 
has provided results consistent with NR all along the adiabatic 
inspiral; the effective-one-body (EOB) model~\cite{BD1,DJS,DIS98} 
can reproduce the total recoil, including the contribution 
from the RD phase, but with large uncertainties~\cite{DG}. 
In Ref.~\cite{Sopuerta}, perturbative calculations that make 
use of the so-called close-limit approximation~\cite{CLA} have been 
used to predict the recoil for unequal-mass binary BHs 
moving on circular and eccentric orbits. 
More recently, Ref.~\cite{SB} provided the first estimates of the
distribution of recoil velocities from spinning BH mergers using the 
EOB model, calibrated to the NR results. 

In this paper we present a diagnostic of the physics of the recoil, 
trying to understand how it accumulates during the inspiral, 
merger, and RD phases. The majority of the analysis is based on
several numerical simulations of
non-spinning, unequal-mass binary systems, as well as spinning,
non-precessing binary systems obtained by the Goddard numerical
relativity group. What we learn in this study will be used in a forthcoming paper to
improve the PN analytic models~\cite{BQW,DG,SB}, so that they can be used to interpolate
between NR results, efficiently and accurately covering the entire
parameter space.

We frame our understanding using the multipolar formalism originally
laid out by Thorne~\cite{KT,BD,BDS,BS,JS}. 
We work out which multipole moments contribute most significantly 
to the recoil. We employ analytic, but leading order, formulae
for the linear momentum flux during 
the inspiral phase, and express the multipole moments in terms of 
a linear superposition of quasi-normal modes (QNMs) during the RD
phase~\cite{RD}. These analysis tools help us understand why for some
binary mass and spin configurations 
the so-called ``anti-kick'' is larger than in other cases. By
anti-kick, we mean that the recoil velocity reaches a
maximum value before decreasing to a final, smaller velocity
asymptotically. As shown in Ref.~\cite{recoilGoddard}, even a
relatively small range of binary parameters can give rise to a large
variety of anti-kick magnitudes (and even the complete lack of an
anti-kick in some cases).

\begin{figure}
\includegraphics[width=0.48\textwidth,clip=true]{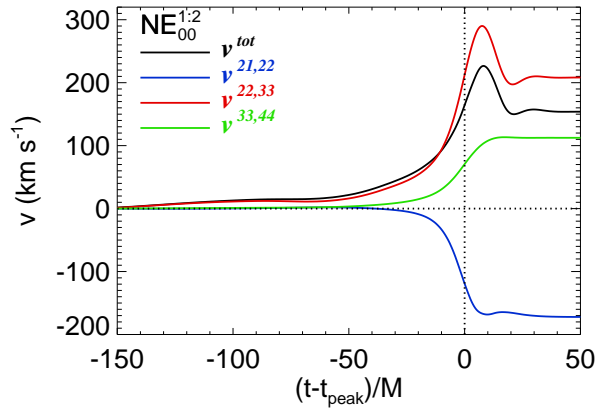}
\caption{\label{intro_fig} The recoil velocity as a function of time
  for a binary BH system with mass ratio 1:2 and no spins. The total
  recoil is plotted in black, along with the contributions from
  different mode pairs, described below in Sec.~\ref{multipoles}. We denote by $t_{\rm peak}$ the time at
  which the multipole $I^{22}$ reaches its maximum (see Section
  \ref{multipoles}).}
\end{figure}

An example of this multipole analysis is shown in Fig.~\ref{intro_fig}, 
which plots the recoil velocity as a
function of time (black curve), along with the separate contributions
from the mass-quadrupole--mass-octupole (red), mass-quadrupole--
current-quadrupole (blue), and mass-quadrupole--mass-hexadecapole (green)
moments. This plot corresponds to a non-spinning system with mass
ratio of 1:2. Note in particular how the modes add both
constructively and destructively to give the total recoil. For the
non-spinning, unequal-mass systems, the kick and anti-kick are
dominated by the mass-quadrupole--mass-octupole modes, but also receive
significant contributions from the other mode-pairs. For all of the
simulations presented in this paper, we scale the time axis around
$t_{\rm peak}$, the time at which the mass quadrupole mode reaches a
maximum, closely corresponding to the peak in gravitational wave power,
as well as the time that a single horizon is formed and the ringdown
phase begins. 

This paper is organized as follow. In Sec.~\ref{NS}, after introducing our 
definitions and notations, we review the binary parameters used 
in the numerical simulations and examine the main features of the 
numerical runs. In Sec.~\ref{multipoles} we discuss the multipolar expansion 
of the linear momentum, angular momentum and energy fluxes 
given in terms of the symmetric trace-free radiative mass 
and current moments, and show how to 
compute those fluxes from the multipole decomposition of the
Weyl scalar $\Psi_4$. 
In Sec.~\ref{multipolesNR}, we analyse the multipole content of 
the numerical waveforms during the inspiral and ringdown phases. 
In Sec.~\ref{quasi-Newtonian} we show that, by properly normalizing 
the binary radial separation, the multipole moments computed 
at leading order in an expansion in $1/c$ can approximate quite well 
the numerical results. Moreover, a superposition of three QNMs 
matches the RD phase. In Sec.~\ref{anatomy} we apply the tools 
developed in the previous sections to understand, using analytic 
expressions, how the kick builds up during the inspiral, 
merger, and ringdown phases. We also apply these methods to help explain
the large difference between planar and non-planar kicks from
equal-mass spinning BHs~\cite{recoilPSU,Bigrecoil,recoilRI}. Finally, Sec.~\ref{discussion} 
contains a brief discussion of our main results and future 
research directions. In Appendix ~\ref{app} we discuss recent results 
for mass ratio $1:4$. 

\section{Numerical simulations}
\label{NS}

In this section we introduce our definitions and notation, and review 
the main features of the numerical simulations. Throughout the paper,
we adopt geometrical units with $G=c=1$ (unless otherwise specified)
and metric signature $(-1,1,1,1)$. 

\subsection{Definitions and conventions}
Our complex null tetrad is defined using the time-like unit vector normal
to a given hypersurface $\hat{\tau}$, the radial unit vector $\hat{r}$, and 
ingoing ($\vec{\ell}$) and outgoing ($\vec{n}$) null vectors as 
\bse
\begin{eqnarray}
\label{eq:tetrad}
  \vec{\ell} &\equiv& \frac1{\sqrt{2}}(\hat{\tau} + \hat{r}), \\
  \vec{n} &\equiv& \frac1{\sqrt{2}}(\hat{\tau} - \hat{r})\,.
\end{eqnarray}
\ese
We define the complex null vectors $\vec{m}$ and $\vec{m}^*$ by
\bse
\bea
  \vec{m} &\equiv& \frac1{\sqrt{2}}(\hat\theta + i\hat\varphi), \\
  \vec{m}^* &\equiv& \frac1{\sqrt{2}}(\hat\theta - i\hat\varphi),
\eea
\ese
with the standard spherical metric at infinity $ds^2 =
-d\tau^2+dr^2+r^2(d\theta^2+\sin^2\theta
d\varphi^2)$. The orthogonality relations of this
tetrad are then 
\bse
\bea
\vec{\ell}\cdot\vec{\ell} &=& \vec{n}\cdot\vec{n} = \vec{m}\cdot\vec{m} = 
\vec{m}^*\cdot\vec{m}^* = 0 \, , \\
\vec{\ell}\cdot\vec{n} &=& -\vec{m}\cdot\vec{m}^* = -1 \, , \\
\vec{\ell}\cdot\vec{m} &=& \vec{\ell}\cdot\vec{m}^* =
\vec{n}\cdot\vec{m} = \vec{n}\cdot\vec{m}^* = 0 \, .
\eea
\ese

In terms of this tetrad, the complex Weyl scalar $\Psi_4$ is given by 
\begin{equation}
\label{eq:Psi4_Weyl}
  \Psi_4 \equiv C_{abcd}\,n^a(m^b)^*n^c(m^d)^*\,,
\end{equation}
where $C_{abcd}$ is the Weyl tensor and $*$ denotes complex conjugation.

To relate $\Psi_4$ to the gravitational waves (GWs), we note that in
the transverse-traceless (TT) gauge (see Chap.~35 in Ref.~\cite{MTW}), 
%
\begin{subequations}\label{eq:Riemann}
\begin{eqnarray}
\label{eq:Riemann1}
  \frac14(\ddot{h}^{TT}_{\hat\theta\hat\theta} 
      - \ddot{h}^{TT}_{\hat\varphi\hat\varphi}) &=&
      -R_{\hat{\tau}\hat\theta\hat{\tau}\hat\theta} =
      -R_{\hat{\tau}\hat\varphi\hat{r}\hat\varphi} =
      -R_{\hat{r}\hat\theta\hat{r}\hat\theta} \nonumber\\
      &=& R_{\hat{\tau}\hat\varphi\hat{\tau}\hat\varphi} =
      R_{\hat{\tau}\hat\theta\hat{r}\hat\theta} =
      R_{\hat{r}\hat\varphi\hat{r}\hat\varphi}, \\
\label{eq:Riemann2}
  \frac12\ddot{h}^{TT}_{\hat\theta\hat\varphi} =
      -R_{\hat{\tau}\hat\theta\hat{\tau}\hat\varphi} &=&
      -R_{\hat{r}\hat\theta\hat{r}\hat\varphi}=
      R_{\hat{\tau}\hat\theta\hat{r}\hat\varphi} =
      R_{\hat{r}\hat\theta\hat{\tau}\hat\varphi}\,.
\end{eqnarray}
\end{subequations}
%
Following usual convention, we take the $h_+$ and $h_\times$ polarizations
of the GW to be given by
\bse
\begin{eqnarray}
  \ddot{h}_+ &=& \frac12(\ddot{h}^{TT}_{\hat\theta\hat\theta} 
      - \ddot{h}^{TT}_{\hat\varphi\hat\varphi})\,, \\
\ddot{h}_\times &=& \ddot{h}^{TT}_{\hat\theta\hat\varphi}\,.
\end{eqnarray}
\ese
Since the Riemann and Weyl tensors coincide in vacuum regions of the
spacetime ($R_{abcd} = C_{abcd}$), we find by combining the above
equations:
\begin{equation}
\label{eq:Psi4_ddh_defn}
  \Psi_4 = -(\ddot{h}_+ - i\ddot{h}_\times)\,.
\end{equation}
Note that this expression for $\Psi_4$  is tetrad-dependent.  Here we assume
the tetrad given in Ref.~\cite{BCL}, Eqs.~(5.6).
It is also common for $\Psi_4$ to be scaled according
to an asymptotically Kinnersley tetrad (Ref.~\cite{BCL}, Eqs.~(5.9))
which introduces a factor of 2 as in Ref.~\cite{Baker:2006kr}.

It is most convenient to deal with $\Psi_4$ in terms of its harmonic
decomposition.  Given the definition of $\Psi_4$ in
Eq.~(\ref{eq:Psi4_Weyl}) and the fact that $\vec{m}^*$ carries a
spin-weight of $-1$, it is appropriate to decompose $\Psi_4$ in terms
of spin-weight $-2$ spherical harmonics $_{-2}Y_{\ell
m}(\theta,\varphi)$ \cite{goldberg}. There is some freedom in the definition of the
spin-weighted spherical harmonics.  Here, we define them 
as a linear combination of the scalar spherical 
harmonics $Y_{\ell m}$ and $Y_{(\ell -1)m}$, as in Ref.~\cite{wiaux}:
\begin{widetext}
\begin{equation}
  {}_{\pm2}Y_{\ell m}\left(\theta,\varphi\right)=\left[\frac{\left(\ell -2\right)!}{\left(\ell +2\right)!}\right]^{1/2}\left[\alpha_{(\ell m)}^{\pm}\left(\theta\right)Y_{\ell m}\left(\theta,\varphi\right)+\beta_{(\ell m)}^{\pm}\left(\theta\right)Y_{(\ell -1)m}\left(\theta,\varphi\right)\right]\,,
\end{equation}
for $\ell \geq2$ and $\vert m\vert\leq \ell $, and with the functional
coefficients 
\bse
\begin{eqnarray}
 \alpha_{(\ell m)}^{\pm}\left(\theta\right) & = & \frac{2m^{2}-\ell
  \left(\ell +1\right)}{\sin^{2}\theta}\mp2m\left(\ell
  -1\right)\frac{\cot\theta}{\sin\theta}+\ell \left(\ell
  -1\right)\cot^{2}\theta\,, \\ 
  \beta_{(\ell m)}^{\pm}\left(\theta\right) & = &
  2\left[\frac{2\ell +1}{2\ell -1}\left(\ell
  ^{2}-m^{2}\right)\right]^{1/2}\left(\pm\frac{m}{\sin^{2}\theta}+\frac{\cot\theta}{\sin\theta}\right)\,.
\end{eqnarray}
\ese
\end{widetext}
Finally, in the far field ($r \gg M$) we decompose the dimensionless
Weyl scalar $M r\Psi_4$ as 
\begin{equation}\label{eq:psi4Ylmdef}
  M r \Psi_4(t,\vec{r}) = 
\sum_{\ell m}{}_{-\!2}C_{\ell m}(t){}_{-\!2}Y_{\ell m}(\theta,\varphi)\,,
\end{equation}
where $M$ is the total mass of the binary 
system (see below for explanations), and $r$ is the radial distance to the
binary center of mass. In Eq. (\ref{eq:psi4Ylmdef}), and
throughout this paper, the notation $\sum_{\ell m}$ is shorthand for
$\sum_{\ell=2}^{\infty}\,\sum_{m=-\ell}^{\ell}$.

\subsection{Details of numerical simulations}
\label{details_NS}

We set up the simulations by placing 
the BHs on an initial Cauchy surface
using the Brandt-Br{\" u}gmann prescription \cite{Brandt97b};
the Hamiltonian constraint is solved using the second-order-accurate
multigrid solver {\tt AMRMG} \cite{Brown:2004ma}. We use the
Bowen-York \cite{Bowen80} framework to incorporate the BH 
spins and momenta, with the choice
of initial tangential momentum informed by the quasi-circular
PN approximation of Ref.~\cite{DJS2}, Eq.(5.3). These initial
conditions typically result in a small level of orbital eccentricity,
which is quickly damped by the radiation reaction losses. The
simulations described in Ref.~\cite{recoilGoddard} showed that the
final recoil varied by only a few percent over a wider range of
initial eccentricities. 

\begin{center}
\begin{table*}
 \caption{Parameters of the numerical simulations (see
   Sec.~\ref{details_NS} for explanations). All masses are normalized
   to an inital total mass of $M=1$.}
\label{table:idparams}
 \begin{tabular}{c| c c c c c c c c c c c c c }
  \hline  \hline
   Run & $m_{1}$ & $m_{2}$ & $\delta m$ & $q$ & $a_1/m_1$ & $a_2/m_2$
   & $\Delta^z$ & $\Delta^p$ & $\xi^z$ & $\Sigma_{33}^z$ & $M_{\rm f}$& 
$a_{\rm f}/M_{\rm f}$ & $v_{\rm f}$(km/s)\\
  \hline 
   ${\rm EQ}_{+-}$   & 0.503 & 0.503 & 0.0    & 1.0   & 0.198 & -0.198
   & -0.2 & 0.0 & 0.0 & 0.075 & 0.967 & 0.697 & 90\\
   ${\rm EQ}_{\rm planar}$   & 0.503 & 0.503 & 0.0    & 1.0   & 0.198 & -0.198
   & 0.0  & 0.2 & 0.0 & 0.0 & 0.967 & 0.697 & 690\\
   ${\rm NE}^{2:3}_{00}$ & 0.401 & 0.593 & -0.192    & 0.677 & 0.0   &
   0.0    & 0.0 & 0.0 & 0.0 & 0.0 & 0.960 & 0.675 & 100\\
   ${\rm NE}^{1:2}_{00}$ & 0.333 & 0.667 & -0.333 & 0.500 & 0.0   &
   0.0    & 0.0 & 0.0 & 0.0 & 0.0 & 0.966 & 0.633 & 140\\ 
   ${\rm NE}^{1:4}_{00}$ & 0.2 & 0.8 & -0.6 & 0.250 & 0.0   &
   0.0    & 0.0 & 0.0 & 0.0 & 0.0 & 0.980 & 0.478 & 150\\ 
   ${\rm NE}^{2:3}_{+-}$ & 0.399 & 0.610 & -0.210 & 0.655 & 0.201 &
   -0.194 & -0.2 & 0.0 & 0.002 & 0.072 & 0.971 & 0.640 & 190\\ 
   ${\rm NE}^{2:3}_{-+}$ & 0.399 & 0.610 & -0.212 & 0.653 & -0.201 &
   0.193 & 0.2 & 0.0 & -0.002 & -0.072 & 0.967 & 0.704 & 70\\ 
  \hline \hline
 \end{tabular}
\end{table*}
\end{center}

The parameters for the runs considered in this paper are shown in
Table~\ref{table:idparams}. We use the following notation: ${\rm EQ}$ 
and ${\rm NE}$ indicate equal-mass and unequal-mass runs, respectively.
The subscripts $0,+,-$ refer to zero spin, spin aligned, and spin
anti-aligned with the orbital angular momentum, respectively (the
EQ$_{\rm planar}$ run has spins in the orbital plane and anti-aligned
with each other). For the
unequal-mass cases we use a superscript to indicate the mass ratio
$m_1:m_2$. We denote by $m_1$ the BH horizon mass computed as 
\beq
m_1 = \sqrt{m_{{\rm irr},1}^2 + \frac{S_1^2}{4 m_{{\rm irr},1}^2}}\,,
\eeq
where $\mathbf{S}_1 = a_1 m_1 \mathbf{\hat{S}}_1 = S_1 \mathbf{\hat{S}}_1$ is
the spin angular momentum of BH 1,
$m_{{\rm irr},1} = \sqrt{A_1/16 \pi}$ is its irreducible mass~\cite{IRM}, and $A_1$
is its apparent horizon area. Similar definitions hold for BH 2.
The binary's total mass is $M= m_1+m_2$, $\delta m = m_1-m_2$, the mass 
ratio is $q = m_1/m_2 \le 1$, and the symmetric mass ratio is $\eta =
m_1 m_2/M^2$. Following Kidder~\cite{LK}, we further define the spin
vectors $\mathbf{S} = \mathbf{S}_1+\mathbf{S}_2$,
$\mathbf{\Delta} = M (\mathbf{S}_2/m_2-\mathbf{S}_1/m_1)$, and
$\mathbf{\xi}=\mathbf{S}+(\delta m/M)\mathbf{\Delta}$. The spin vector
$\Sigma^{z}_{33}$ is defined below in Sec.~\ref{diff_mom}.

The mass and spin parameters of the final BH are $M_{\rm f}$ and
$a_{\rm f}$. The values of $M_{\rm f}$ and $a_{\rm f}$ listed in
Table~\ref{table:idparams} are computed from the loss of energy and
angular momentum from the initial time to the end of the RD
phase. They are compatible with the values obtained by extracting the
fundamental QNMs (see below Sec.~\ref{RD_phase}). All spins are
orthogonal to the orbital plane, so
$\Delta^x=\Delta^y=0$ (the exception is a single run EQ$_{\rm planar}$
with planar spins discussed in Sec.~\ref{planarspins}. In
Table~\ref{table:idparams}, the spin components in the orbital plane
are represented by $\Delta^p \equiv
|\Delta^x+i\Delta^y|$.). Additionally, all runs have
$|a_1|/m_1=|a_2|/m_2$ with spins pointing in opposite directions, so
$\mathbf{\xi}\approx 0$ within the accuracy of the initial data.

The simulations were carried out using the moving puncture
method \cite{CLMZ,Bakeretal1} in the finite-differencing
code {\tt Hahndol} \cite{Imbiriba:2004tp}, which solves the Einstein equations
in a standard 3+1 BSSN conformal formulation.  Dissipation \cite{Huebner99}
terms (tapered to zero near the punctures) and constraint-damping \cite{Duez:2004uh} terms 
were added for robust
stability. We used the gauge condition recommended in Ref.~\cite{vanMeter:2006vi}
for moving punctures, fourth-order-accurate mesh-adapted
differencing \cite{Baker:2005xe}
for the spatial derivatives, and a fourth-order-accurate
Runge-Kutta algorithm for the time-integration. The adaptive mesh
refinement and most of the parallelization was handled by the
software package {\tt Paramesh} \cite{MacNeice00}, with fifth-order accurate
interpolation between mesh refinement regions.

The grid spacing in the finest refinement region around each 
BH is $h_f = 3M/160$.  We extract data for the radiation
at a radius $r_{\rm ext} = 45M$. The wave
extraction was performed by 4th order interpolation to a sphere followed
by angular integration with a Newton-Cotes formula.  We have found
satisfactory convergence of the results.  For example, for the 1:2
mass ratio run, for which a higher resolution of $h_f=1M/64$ was run in addition
to $h_f = 3M/160$, the rates of convergence of the Hamiltonian and momentum 
constraints are comparable to those found in our equal mass runs reported
in \cite{Baker:2006kr}, and the radiated momenta from the two resolutions agree to within $2\%$.
This was also true for a 2:3 mass ratio test case with aligned spins
(the NE++ run in Ref.~\cite{recoilGoddard}, which is representative
of the NE$_{+-}^{2:3}$ and NE$_{-+}^{2:3}$ runs here).

\section{Multipolar formalism}
\label{multipoles}

In this Section we review the most relevant results from
Thorne~\cite{KT}, showing how a multipole decomposition of the
gravitational radiation field can be used to calculate the energy,
angular momentum, and linear momentum fluxes from a BH binary
system. When restricting the analysis to leading order terms 
we shall often express the radiative multipole moments in terms of the
source multipole moments~\cite{BD,BDS,BS,JS}, so in much of the
discussion below we will use these two descriptions interchangebly.

\subsection{Linear momentum flux}
\label{linear_flux}

In the literature~\cite{HSL,recoil,recoilJena,recoilAEI,recoilRI,recoilPSU} 
it is common to compute the linear momentum flux, and then the recoil, 
using the following formula 
\begin{equation}
\label{dPdt_psi4}
\frac{dP_i}{dt} = \frac{r^2}{16\pi}\int d\Omega\, \frac{x_i}{r}\,
\left|\int_{-\infty}^t dt \Psi_4 \right|^2\,,
\end{equation}
where $r$ is the extraction radius and the antiderivative of
$\Psi_4$ is used because the linear momentum flux scales as the square 
of the first derivative of the wave strain, whereas $\Psi_4$ is 
proportional to the second derivative of the strain [see
Eq.~(\ref{eq:Psi4_ddh_defn}) above]. To study how the different
multipole moments contribute to the recoil, we could plug
Eq.~(\ref{eq:psi4Ylmdef}) into Eq.~(\ref{dPdt_psi4}), as done, e.g.,
in Ref.~\cite{recoilPSU}. Here, we prefer to use the expression of the
linear momentum flux given in terms of the symmetric and trace-free
(STF) radiative mass and current multipole moments, as done in
Refs.~\cite{KT,BD,BDS,BS,JS}. 

Starting from Eq.~(4.20') in Ref.~\cite{KT}, we write the
linear momentum flux as  
\begin{widetext}
\begin{eqnarray}\label{thorne_420}
F_j \equiv \frac{dP_j}{dt} &=& \frac{G}{c^7}\,\sum_{\ell=2}^{\infty}
\left[\frac{2(\ell+2)(\ell+3)}{\ell(\ell+1)!(2\ell+3)!!} {}^{(\ell+2)}\mathbf{I}_{jA_\ell}
  {}^{(\ell+1)}\mathbf{I}_{A_\ell}\,\left (\frac{1}{c}\right)^{2(\ell-2)} 
+ \frac{8(\ell+3)}{(\ell+1)!(2\ell+3)!!}
  {}^{(\ell+2)}\mathbf{S}_{jA_\ell} {}^{(\ell+1)}\mathbf{S}_{A_\ell}\,\left (\frac{1}{c}\right)^{2(\ell-1)}
  \right. \nonumber\\
  & &\left. \hspace{1.4cm} +\frac{8(\ell+2)}{(\ell-1)(\ell+1)!(2\ell+1)!!} \epsilon_{jpq}
       {}^{(\ell+1)}\mathbf{I}_{pA_{\ell-1}} {}^{(\ell+1)}\mathbf{S}_{qA_{\ell-1}}\,
\left (\frac{1}{c}\right)^{2(\ell-2)} \right]\,,
\end{eqnarray}
\end{widetext}
where $\mathbf{I}_{A_\ell}$ ($\mathbf{S}_{A_\ell}$) are the
$\ell$-dimensional STF mass (current) tensors and
left-hand superscripts represent time derivatives. 
From these tensors, we can construct the radiative multipole moments
${\cal I}^{\ell m}$ and ${\cal S}^{\ell m}$ according to the normalization given by 
Eq.~(4.7) of Ref.~\cite{KT}:
\bse
\label{thorne_47}
\bea
{\cal I}^{\ell m} &=& \frac{16\pi}{(2\ell +1)!!}\nonumber\\
&& \cdot \left(\frac{(\ell +1)(\ell +2)}{2(\ell -1)\ell }\right)^{1/2} \mathbf{I}_{A_\ell }
\mathbf{Y}^{\ell m*}_{A_\ell }\,, \label{thorne_47_I} \\
{\cal S}^{\ell m} &=& \frac{-32\pi \ell }{(\ell +1)(2\ell +1)!!}\nonumber\\
&& \cdot \left(\frac{(\ell +1)(\ell +2)}{2(\ell -1)\ell }\right)^{1/2} \mathbf{S}_{A_\ell }
\mathbf{Y}^{\ell m*}_{A_\ell } \,, \label{thorne_47_S} 
\eea
\ese
where $\mathbf{Y}^{\ell m}_{A_\ell }$ are $\ell $-dimensional STF tensors that are
closely related to the usual scalar spherical harmonics by
\begin{equation}
Y_{\ell m}(\theta,\varphi) = \mathbf{Y}^{\ell m}_{i_1\cdots i_\ell } n^{i_1} \cdots
n^{i_\ell }\,,
\end{equation}
with $n^i = (\sin\theta \cos\varphi,\sin\theta
\sin\varphi,\cos\theta)^i$. Note that the radiative moments ${\cal
  I}^{\ell m}$ and ${\cal S}^{\ell m}$ are scalar quantities and have no
explicit spatial dependence. 
To simplify the notation below, we incorporate the ($\ell +1$) time
derivatives into the radiative multipole moments, and define 
\beq
\label{new_tensors}
I^{\ell m} \equiv {}^{(\ell +1)}{\cal I}^{\ell m}\,, 
\quad S^{\ell m} \equiv {}^{(\ell +1)}{\cal S}^{\ell m}\,.
\eeq

By combining Eqs.~(\ref{thorne_420}), (\ref{thorne_47}), and
(\ref{new_tensors}), we find that at leading order (in a $1/c$
expansion) the linear momentum flux is given by
\begin{widetext}
\bea
\label{dPdt_radmoments}
F^{(0)}_x+iF^{(0)}_y =
\frac{1}{336\pi}\hspace{-0.3cm}&&\left[ -14iS^{21}I^{22*}
+ \sqrt{14}I^{31}I^{22*} -\sqrt{210}I^{22}I^{33*} +
7i\sqrt{6}I^{20}S^{21*}-7i\sqrt{6}S^{20}I^{21*}
+ \right . \nonumber \\
&& \left. 14iI^{21}S^{22*} +\sqrt{42}I^{30}I^{21*} - 2\sqrt{21}I^{20}I^{31*} -
2\sqrt{35}I^{21}I^{32*} \right]\,,
\eea
and
\bea
\label{dPdtz_radmoments}
F^{(0)}_z=
\frac{1}{336\pi}\left[4\sqrt{14}\Re(I^{31}I^{21*})-14\Im(I^{21}S^{21*})
+2\sqrt{35}\Re(I^{22}I^{32*})-28\Im(I^{22}S^{22*})
+3\sqrt{7}I^{20}I^{30}\right]\,.
\eea
\end{widetext}
Note that Eq.~(\ref{dPdt_radmoments}) coincides with Eq.~(9) in
Ref.~\cite{DG} when we equate the radiative multipole moments with
the source moments~\cite{BD,BDS,BS,JS} and reduce
to a circular, non-spinning orbit in the $x$-$y$ plane. In this case
only the first three terms in Eq.~(\ref{dPdt_radmoments}) survive.

The next highest order terms ($1/c^2$ with respect to the leading terms) are proportional to the
mass octupoles $I^{3m}$, or current quadrupoles $S^{2m}$: 
\begin{widetext}
\bea
\label{dPdt_radmoments2}
F^{(1)}_x+iF^{(1)}_y =
\frac{1}{672\pi} \hspace{-0.3cm}&& \left[ -7i\sqrt{6}S^{32}I^{33*}  
-14\sqrt{6}I^{33}I^{44*} 
-4\sqrt{21}S^{20}S^{31*}-4\sqrt{35}S^{21}S^{32*}-2\sqrt{210}S^{22}S^{33*}
+ \right .\nonumber \\
&& \left. 2\sqrt{42}S^{30}S^{21*} + 14i\sqrt{3}I^{30}S^{31*}-14i\sqrt{3}S^{30}I^{31*}
+7i\sqrt{10}I^{31}S^{32*} -7i\sqrt{10}S^{31}I^{32*} -\right. \nonumber \\ 
&& \left . 2\sqrt{105}I^{30}I^{41*}+6\sqrt{7}I^{40}I^{31*}
-3\sqrt{70}I^{31}I^{42*} +3\sqrt{14}I^{41}I^{32*} -21\sqrt{2}I^{32}I^{43*} +\right . \nonumber \\
&& \left . 2\sqrt{14}S^{31}S^{22*}+ \sqrt{42}I^{42}I^{33*} +7i\sqrt{6}I^{32}S^{33*} \right ]\,,
\eea
and
\bea
\label{dPdtz_radmoments2}
F^{(1)}_z=
\frac{1}{336\pi} \hspace{-0.3cm}&& \left[
3\sqrt{7}S^{20}S^{30}
+4\sqrt{14}\Re(S^{21}S^{31*})+2\sqrt{35}\Re(S^{22}S^{32*}) 
-7\Im(I^{31}S^{31*})-14\Im(I^{32}S^{32*})-21\Im(I^{33}S^{33*}) 
+\right . \nonumber \\
&& \left . 2\sqrt{21}I^{30}I^{40}+3\sqrt{35}\Re(I^{31}I^{41*})
+6\sqrt{7}\Re(I^{32}I^{42*})+7\sqrt{3}\Re(I^{33}I^{43*}) \right ] \,.
\eea
\end{widetext}
Note that all of the terms in Eqs.~(\ref{dPdt_radmoments}) and
(\ref{dPdt_radmoments2}) contain products of multipoles with
$m'=m\pm1$, while the terms in Eqs.~(\ref{dPdtz_radmoments}) and
(\ref{dPdtz_radmoments2}) have $m'=m$, as with familiar
quantum-mechanical operators that involve similar $x_i$-weighted
integrations over the sphere. Also note that for mass-mass and
current-current terms, $\ell'=\ell\pm 1$, while for mass-current
terms, $\ell'=\ell$.

The above formulae (\ref{dPdt_radmoments})--(\ref{dPdtz_radmoments2}) 
are valid for completely general orbits, including eccentricity, 
spin terms and even for binary systems precessing out of the plane. However,
we can simplify them significantly by rotating into the frame where
the instantaneous orbital angular momentum is along the $z$-axis. 
Furthermore, by 
assuming that terms proportional to $\ddot{R}$ ($R$ being the binary radial 
separation) are negligible,
we find $I^{20}=I^{30}=S^{30}=I^{32}=I^{40}=I^{41}=I^{43}=0$. In the
approximation of $\ddot{R}=0$, the inclusion of terms linear in
$\dot{R}\ne 0$ adds no new multipole modes. 
In fact, one of the primary reasons the derivations above begin with
the mass and current tensors $\mathbf{A}_{A_\ell}$ and
$\mathbf{S}_{A_\ell}$ is to facilitate the calculation of the
individual radiative moments $I^{\ell m}$ and $S^{\ell m}$ and also
identify the contributions from $\dot{R}$ and $\ddot{R}$ terms from a
generalized binary orbit~\cite{LK}.
In the case of non-spinning BHs, the formulae 
(\ref{dPdt_radmoments})--(\ref{dPdtz_radmoments2}) can be additionally 
simplified by setting $S^{20}=I^{21}=S^{22}=S^{31}=S^{33}=0$. 
Quite interestingly, we obtain that the latter conditions 
are also valid in the special case of {\it non-precessing} BHs  
where the spins are aligned or anti-aligned with the orbital
angular momentum. Since these are the cases we consider in this paper,
we refer often to the following approximate formula for the
linear momentum flux:
\begin{widetext}
\beq
\label{flux_approx}
F_x+iF_y \simeq 
\frac{1}{672\pi} \left[ 
-28iS^{21}I^{22*} -2\sqrt{210}I^{22}I^{33*} 
-14\sqrt{6}I^{33}I^{44*}+2\sqrt{14}I^{31}I^{22*} -
7i\sqrt{6}S^{32}I^{33*} 
\right]\,, \quad F_z = 0. 
\eeq
\end{widetext}
As we will see below in Sec.~\ref{inspiral_phase}, the linear momentum
flux contributions from $I^{31}I^{22*}$ as well as other higher-$\ell$
modes are typically smaller by at least
an order of magnitude. When integrating Eq.~(\ref{flux_approx}) to get
the recoil velocity, we also find that (due in large part to the
relative phases between the modes) the contribution from
$S^{32}I^{33*}$ is rather minimal. Thus for most of the analysis that
follows, we will focus solely on the first three terms of
Eq.~(\ref{flux_approx}). 

In the following, sometimes we will use 
\beq
\mathbf{F} = \{F_x,F_y,F_z \}\,, \quad \quad \mathbf{\hat{F}} = \frac{\mathbf{F}}{|\mathbf{F}|}\,.
\eeq
All the non-precessing numerical simulations we will analyze have
$F_{z}=0$, so we can introduce a complex scalar flux
\beq
F \equiv F_x + i F_y\,.
\eeq

Since what we extract from the numerical simulations are the 
modes ${}_{-\!2}C_{\ell m}$ computed over the sphere surrounding the 
binary, we need to relate the ${}_{-\!2}C_{\ell m}$ to the 
radiative mass and current multipole moments defined above. 
From Eq.(4.3) of \cite{KT},
\begin{equation}
\label{eq:hmm}
h=\sum_{\ell m} ({}^{(\ell )}{\cal I}^{\ell m}T_{ab}^{E2,\ell
m}m^a m^b + {}^{(\ell )}{\cal S}^{\ell m}T_{ab}^{B2,\ell m}m^a m^b)\, ,
\end{equation}
where $h\equiv h_{ab} m^a m^b$ and $h_{ab}$ is the metric perturbation
$g_{ab}-\eta_{ab}$ in the transverse traceless gauge, which satisfies
Eq.~(\ref{eq:Riemann}), and $T_{ab}^{E2,\ell m}$ and 
$T_{ab}^{B2,\ell m}$ are the ``pure-spin'' harmonics of Thorne.
From Appendix A of \cite{martel05},
\bse
\begin{eqnarray}
\label{eq:TabE}
T_{ab}^{E2,\ell m}&=
&\frac{1}{\sqrt2}(\,{}_{-2}Y^{\ell m} m_a m_b +\,{}_{2}Y^{\ell m} m_a^*
m_b^*)\\
\label{eq:TabB}
T_{ab}^{B2,\ell m}&=
&\frac{-i}{\sqrt2}(\,{}_{-2}Y^{\ell m} m_a m_b -\,{}_{2}Y^{\ell m} m_a^* 
m_b^*).
\end{eqnarray}
\ese
Substituting Eqs.~(\ref{eq:TabE})--(\ref{eq:TabB}) into Eq.~(\ref{eq:hmm}) and recalling that $m^a m_a=0$ gives
\begin{equation}
h=\frac{1}{\sqrt{2}r}\sum_{\ell m}({}^{(\ell )}{\cal I}^{\ell m}+i {}^{(\ell )}{\cal
  S}^{\ell m})\,{}_{+2}Y^{\ell m}
\end{equation}
Now taking the complex conjugate and using the fact that
$_{+2}Y^{*\ell m}=(-1)^m{}_{-2}Y^{\ell -m}$ [note there is a typo in Eq.~(3.1)
of Ref.~\cite{goldberg}]
we obtain
\begin{eqnarray}
h^* &=&\frac{1}{\sqrt{2}r}\sum_{\ell m}(-1)^m({}^{(\ell )}{\cal I}^{\ell m*}-i {}^{(\ell )}{\cal S}^{\ell m*})\,{}_{-2}Y^{\ell -m} \nonumber\\
           &=&\frac{1}{\sqrt{2}r}\sum_{\ell m}(-1)^m({}^{(\ell )}{\cal
           I}^{\ell -m*}-i {}^{(\ell )}{\cal S}^{\ell
           -m*})\,{}_{-2}Y^{\ell m}\,. \nonumber \\
\end{eqnarray}
Using the tetrad choice of
Eqs.~(\ref{eq:tetrad})--(\ref{eq:Psi4_ddh_defn}), $\partial^2 
h^*/\partial t^2 =\ddot{h}_+-i\ddot{h}_\times=-\Psi_4$,
which decomposed into spin -2 weighted harmonics, gives
\begin{equation}
\frac{\partial^2 h^*}{\partial t^2}=-\frac{1}{Mr}\sum_{\ell m}{}_{-2}C_{\ell m}  \,{}_{-2}Y^{\ell m},
\end{equation}
allowing us to see term-by-term that
\begin{equation}
(-1)^m({}^{(\ell +2)}{\cal I}^{\ell -m*}-i {}^{(\ell +2)}{\cal
S}^{\ell -m*})=-\sqrt{2} {}_{-2}C_{\ell m} \, .
\end{equation}
Recall that  $(-1)^m{\cal I}^{\ell -m*} = {\cal I}^{\ell m}$ and $(-1)^m{\cal S}^{\ell -m*} = 
{\cal S}^{\ell m}$, which allows us to write
\begin{subequations}
\begin{eqnarray}
\label{Ilm_Clm}
{}^{(\ell +2)}{\cal I}^{\ell m} &=& -\frac1{\sqrt2}\left[{}_{-2}C_{\ell m}+\,
(-1)^m{}_{-2}C_{\ell-m}^*\right]\,,\\
\label{Slm_Clm}
{}^{(\ell +2)}{\cal S}^{\ell m} &=& -\frac{i}{\sqrt2}\left[{}_{-2}C_{\ell m}-
\,(-1)^m{}_{-2}C_{\ell-m}^*\right]\,.
\end{eqnarray}
\end{subequations}
Equations (\ref{dPdt_radmoments})--(\ref{flux_approx}) are expressed in terms 
of $I^{\ell m} \equiv {}^{(\ell +1)}{\cal I}^{\ell m}$ and $S^{\ell m} \equiv  
{}^{(\ell +1)}{\cal S}^{\ell m}$, which can be computed by integrating 
Eqs.~(\ref{Ilm_Clm}), (\ref{Slm_Clm}) once in time. To avoid the complication of an
undetermined constant of integration, we typically integrate
${}_{-2}C_{\ell m}(t)$ {\it backwards} in time, since in the numerical
data (and what we expect happens in reality) all the moments go
to zero exponentially after the merger. At early times on the other
hand, most of the modes are significantly non-zero and also include
a large amount of numerical noise due to the initial conditions. 

\subsection{Energy and angular momentum flux}

Unlike the equations for the linear momentum flux, which all involve
``beating'' between pairs of different modes, the energy and angular-momentum
flux expressions involve terms of the form $|I^{\ell m}|^2$,
allowing us to isolate the individual contributions 
from each mode. As
we will see below, for the comparable-mass binary systems that 
we analyse ($m_1$:$m_2$ = 1:1, 2:3, 1:2), the amplitude of the
mass quadrupole moment $I^{22}$ is roughly an order of magnitude
larger than the next largest mode. Thus it almost completely dominates
the energy and angular momentum fluxes, and we can write [see Eq.~(4.16) 
in Ref.~\cite{KT}]
\begin{equation}\label{dEdt}
\frac{dE}{dt} =\frac{1}{32\pi}\sum_{\ell m}
\left( |I^{\ell m}|^2 + |S^{\ell m}|^2 \right) \simeq \frac{1}{16\pi}|I^{22}|^2.
\end{equation}

The multipole expressions for angular momentum flux are somewhat more
complicated, but for the numerical
simulations considered in this paper, the only non-zero modes have
$\ell +m$ even for $I^{\ell m}$ and $\ell +m$ odd for $S^{\ell m}$, so we can neglect
the $(m,m\pm 1)$ cross-terms in Eq.~(4.23) of Ref.~\cite{KT}. These
cross-terms are responsible for angular momentum loss in the $x$-$y$
plane, so it is reasonable that they must be zero for non-precessing
planar orbits. In this case, where the angular momentum is solely
along the $\hat{\mathbf{z}}$-axis, we have 
\begin{eqnarray}\label{dJzdt}
\frac{dJ_z}{dt} &=& \frac{i}{32\pi}\sum_{\ell m}
m( {}^{(\ell )}{\cal I}^{\ell m*}\, {}^{(\ell +1)}{\cal I}^{\ell m} +
{}^{(\ell )}{\cal S}^{\ell m*}\, {}^{(\ell +1)}{\cal S}^{\ell m} ) \nonumber\\
& \simeq & -\frac{1}{8\pi}\, \Im\left[{}^{(2)}{\cal I}^{22*}\,
  {}^{(3)}{\cal I}^{22} \right]\, ,  
\end{eqnarray}
where we have restored the explicit time
derivatives as in Eq.~(\ref{new_tensors}).

\begin{center}
\begin{table*}
 \caption{Energy and angular momentum radiated in each of the dominant
 multipole modes. In parentheses we show the amount radiated only
 after the peak of GW energy flux. All units are normalized to $M=1$.}
\label{table:Elm}
 \begin{tabular}{c| c c c c c c c c c c }
  \hline  \hline
   Run & $E_{22}$ & $E_{21}$ & $E_{32}$ & $E_{33}$ & $E_{44}$ &
$J_{22}$ & $J_{21}$ & $J_{32}$ & $J_{33}$ & $J_{44}$ \\
& $(\times 10^{-2})$ & $(\times 10^{-4})$ & $(\times 10^{-4})$ & $(\times
10^{-4})$ & $(\times 10^{-4})$ & $(\times 10^{-1})$ & $(\times 10^{-4})$ &
$(\times 10^{-4})$ & $(\times 10^{-3})$ & $(\times 10^{-3})$  \\

  \hline 
   ${\rm EQ}_{+-}$ & 3.5 & $0.22$ 
   & $1.6$ & $0.04$ & $3.3$ & $2.2$  & $-0.70$ & $7.9$  & $-0.02$ & $1.9$ \\
   &($1.4$)&($0.17$)&($1.2$)&($0.02$)&($1.5$) &($0.50$)&($-0.46$)&($-2.0$)
   &($-0.01$)&($0.64$)
\smallskip\\
   ${\rm NE}^{2:3}_{00}$ & 3.1 & $0.61$ 
   & $0.90$ & $5.6$ & $2.9$
   & $2.2$             & $-2.1$ & $3.9$ 
   & $-3.1$ & $1.8$ \\
   &($1.1$)           &($0.40$)&($0.66$)
   &($2.8$)&($1.0$)
   &($0.45$)           &($-0.98$)&($2.5$)
   &($-1.1$)&($0.46$)
\smallskip\\
   ${\rm NE}^{1:2}_{00}$ &  2.5 & $1.4$ 
   & $0.47$ & $12.0$ & $2.7$
   & $1.8$             & $-4.8$ & $2.4$ 
   & $-6.9$ & $1.7$ \\
   &($0.87$)&($0.94$)&($0.30$)
   &($5.8$)&($0.73$)
   &($0.37$)           &($-2.4$)&($1.3$)
   &($-2.3$)&($0.30$)
\smallskip\\
   ${\rm NE}^{1:4}_{00}$ &  1.2 & $2.1$ 
   & $0.27$ & $16.0$ & $3.3$
   & $1.2$             & $-8.0$ & $1.6$ 
   & $-11.0$ & $2.4$ \\
   &($0.35$)&($1.4$)&($0.09$)
   &($6.6$)&($1.2$)
   &($0.16$)           &($-3.8$)&($0.27$)
   &($-2.9$)&($0.48$)
\smallskip\\
   ${\rm NE}^{2:3}_{+-}$ & 2.9 & $1.6$ 
   & $0.93$ & $5.2$ & $2.6$
   & $2.0$             & $-5.4$ & $2.1$ 
   & $-2.9$ & $1.6$ \\
   &($1.0$)           &($1.0$)&($0.67$)
   &($2.5$)&($0.82$)
   &($0.31$)           &($-2.9$)&($5.3$)
   &($-0.98$)&($0.33$)
\smallskip\\
   ${\rm NE}^{2:3}_{-+}$ & 3.3 & $0.14$ 
   & $1.1$ & $7.1$ & $2.9$
   & $2.3$             & $-0.50$ & $4.4$ 
   & $-3.9$ & $1.8$ \\
   &($1.1$)           &($0.09$)&($0.78$)
   &($3.4$)&($0.92$)
   &($0.44$)           &($-0.21$)&($3.1$)
   &($-1.3$)&($0.37$) \\
  \hline \hline
 \end{tabular}
\end{table*}
\end{center}

Integrating Eqs.~(\ref{dEdt}) and (\ref{dJzdt}) term-by-term, we can
calculate how much energy and angular momentum are radiated in each of
the dominant modes, similar to the approach of Ref.~\cite{berti07}. We
introduce the quantities $E_{\ell m}$ and $J_{\ell m}$ as the total energy
and angular momentum radiated in each ($\ell$, $m$) mode, computed by
integrating Eqs.~(\ref{dEdt}) and (\ref{dJzdt}) in time, term by
term (for conciseness, we combine both the $m$ and $-m$ terms into
$E_{\ell m}$ and $J_{\ell m}$ and restrict our notation to $m>0$). Note
that while $E_{\ell m}$ is always positive, $J_{\ell m}$ can also be
negative, corresponding to angular momentum in the $-\hat{z}$
direction. These results are shown in Table \ref{table:Elm}, along
with the contributions from just
the RD phase ($t>t_{\rm peak}$, where $t_{\rm peak}$ is the point at which
$|I^{22}|$ reaches its peak, closely corresponding to the peak in GW
energy emission). We will see below in Section
\ref{quasi-Newtonian} that these various energy contributions agree
closely with the Newtonian predictions for the relative
mass-scalings. For example, the energy $E_{22}$ in the inspiral phase
should scale as $\eta$, while the RD contribution should scale like
$\eta^2$. It is important to note that the different moments have
different scalings: $E_{33} \sim \eta^2 \delta m^2$, while the $I^{44}$
contribution has a much weaker dependence on mass ratio: $E_{44} \sim
\eta^2(1-3\eta)^2$. 

In the limit of very large initial
separation (small initial frequency), each of the $E_{\ell m}$ and
$J_{\ell m}$
should converge to a finite value, with the notable exception of
$J_{22}$. It is well-know that the angular momentum of a binary system
scales as $R^{1/2}$, and is thus unbound in the limit of $R\to\infty$,
but it is interesting to see that the higher-order contributions to
the angular momentum all converge at large $R$. This can be understood
directly from Eq.~(\ref{dJzdt}) in the Keplerian limit of
$R=M^{1/3}\omega^{-2/3}$. At leading order, radiation reaction follows the
relation $dt \sim \omega^{-11/3}d\omega$ so the angular momentum in
the inspiral is
\bea\label{J22insp}
J_{22} &=& \frac{1}{8\pi}\int_{t=-\infty}^{t_0} dt\,
\Im\left[{}^{(2)}{\cal I}^{22*}\, 
{}^{(3)}{\cal I}^{22}\right] \nonumber\\
&\sim& \int_{\omega=0}^{\omega_0} \omega^{2/3} \omega^{5/3}
\omega^{-11/3} d\omega \to \infty.
\eea
As we will see below in Section \ref{quasi-Newtonian}, 
for all the other energy and angular momentum modes, the fluxes from
Eqs.~(\ref{dEdt}),(\ref{dJzdt}) scale as $\omega^{10/3}$ or higher powers,
and thus converge when integrated over $\omega^{-11/3}d\omega$.

\section{Multipole analysis of the numerical simulations}
\label{multipolesNR}

In this Section we want to investigate how the different multipole moments 
evolve during the inspiral and ringdown phases of BH binary mergers.

\subsection{Inspiral phase}
\label{inspiral_phase}

As can be derived in PN theory~\cite{LB} and has been confirmed numerically 
in Refs.~\cite{CLMZ, Bakeretal1}, the $\ell=2, m=2$ 
mode in Eq.~(\ref{eq:psi4Ylmdef}) is circularly polarized to leading order
throughout the coalescence.
Because of this, Ref.~\cite{BCP} defined the 
(dominant) orbital angular frequency as
\begin{equation}
\label{dom}
\omega_{\rm D}^{\ell m} = -\frac{1}{m} \Im
\left(\frac{{}_{-2}\dot{C}_{\ell m}}{{}_{-2}C_{\ell m}}\right).
\end{equation}
Here, we extend Eq.~(\ref{dom}) by defining several 
(dominant) orbital angular frequencies, each of them 
being related to a specific multipole moment, $I^{\ell m}$ or $S^{\ell m}$, 
as
\begin{equation}\label{omega_lm}
\omega^{I\ell m}_{\rm D} = -\frac{1}{m} \Im
\left(\frac{\dot{I}^{\ell m}}{I^{\ell m}}\right)\,, \quad 
\omega^{S\ell m}_{\rm D} = -\frac{1}{m} \Im
\left(\frac{\dot{S}^{\ell m}}{S^{\ell m}}\right).
\end{equation}

\begin{figure*}
\includegraphics[width=0.48\textwidth,clip=true]{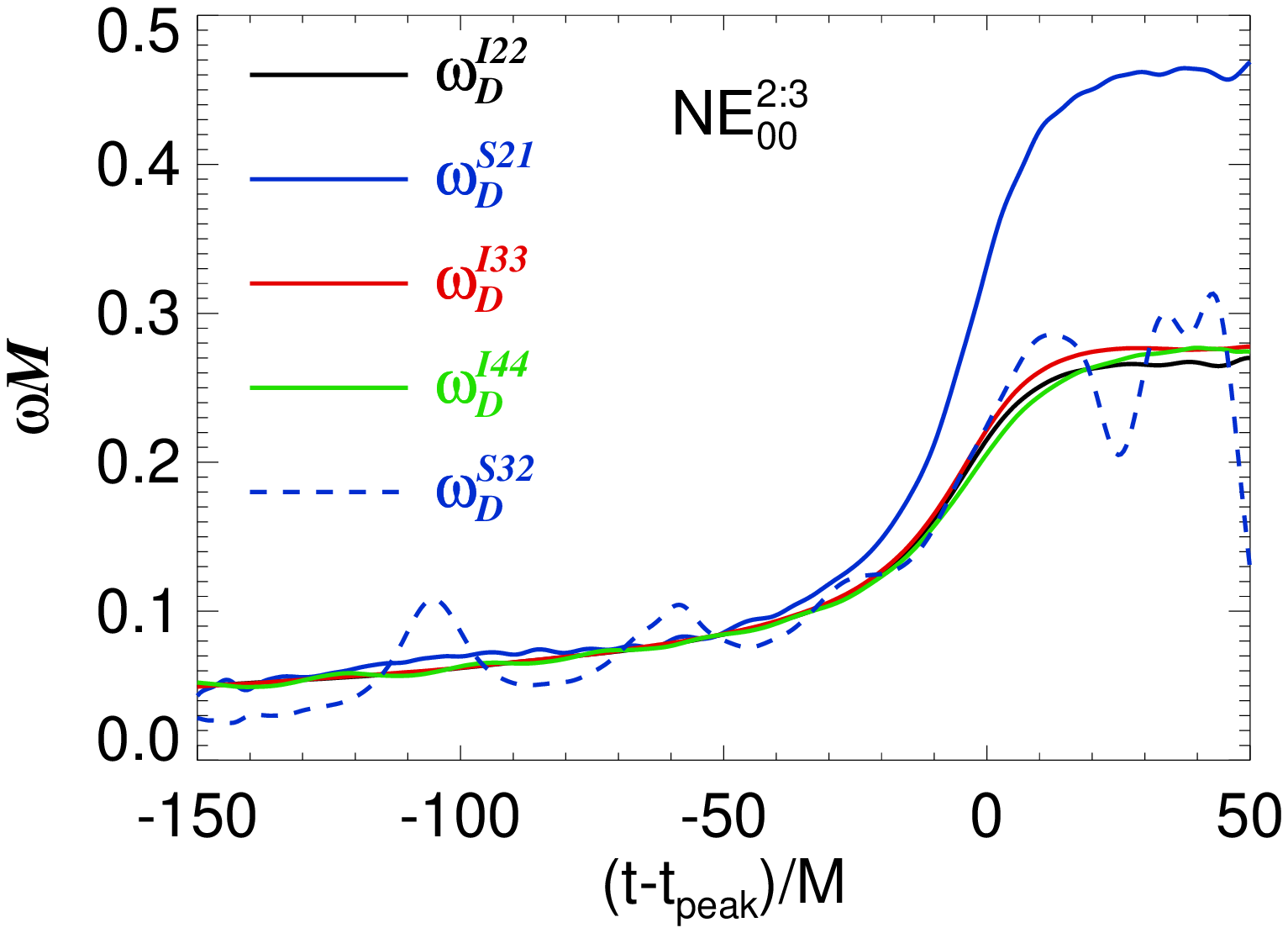}
\includegraphics[width=0.48\textwidth,clip=true]{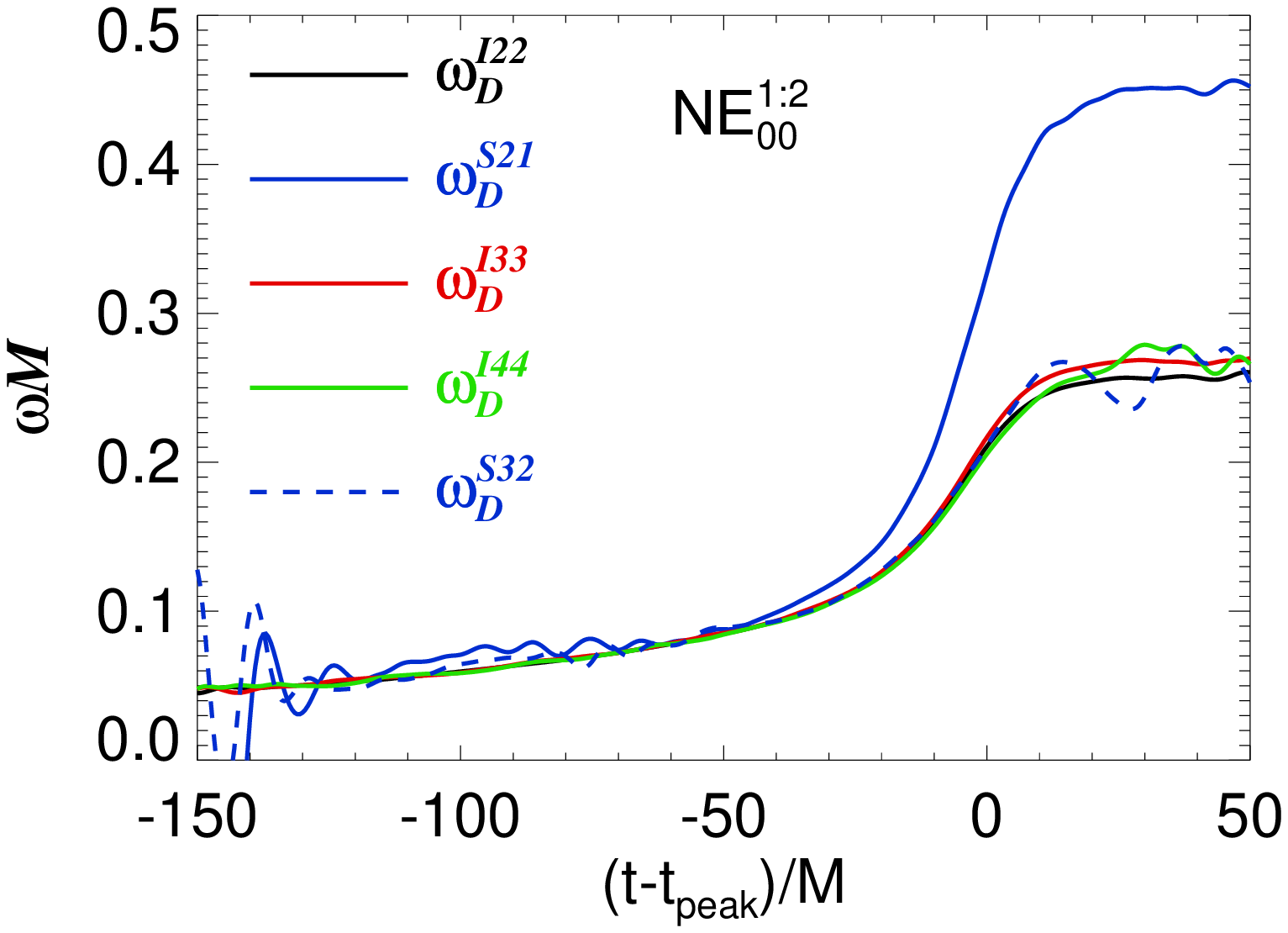}
\caption{\label{omega_modes} Dominant orbital angular frequency 
obtained from the individual radiative multipole moments, as determined by Eq.~
  (\ref{omega_lm}). The different frequencies with $\ell=m$ agree closely
throughout the inspiral and RD phases.
The frequency with $\ell=2,m=1$ decouples from the others at earlier time 
and reaches a much higher plateau. The left panel refers to the NE$_{00}^{2:3}$ run
and the right panel to the NE$_{00}^{1:2}$ run. We denote with $t_{\rm peak}$ the 
time at which $I^{22}$ reaches its maximum.}
\end{figure*} 

\begin{figure*}
\includegraphics[width=0.48\textwidth,clip=true]{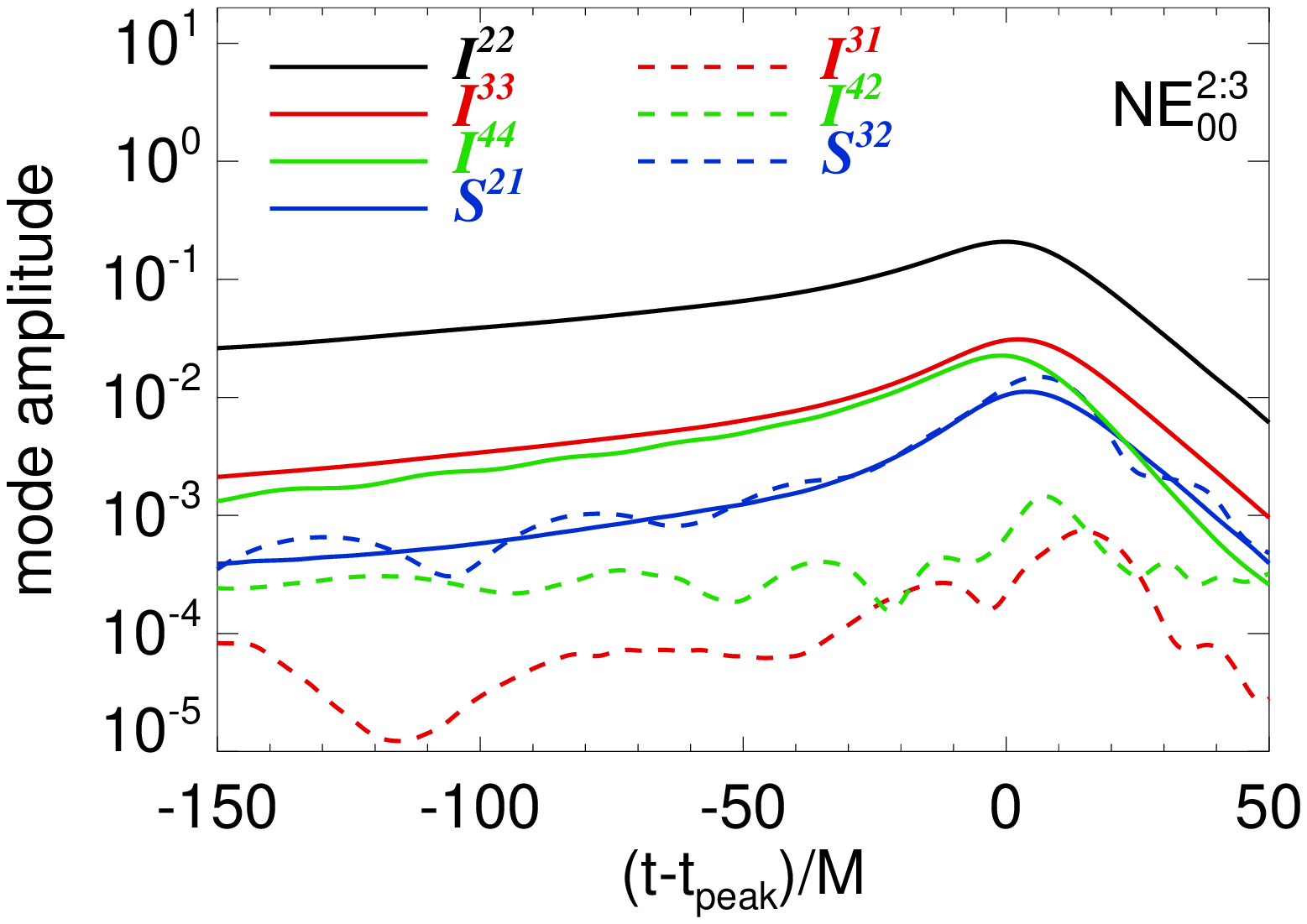}
\includegraphics[width=0.48\textwidth,clip=true]{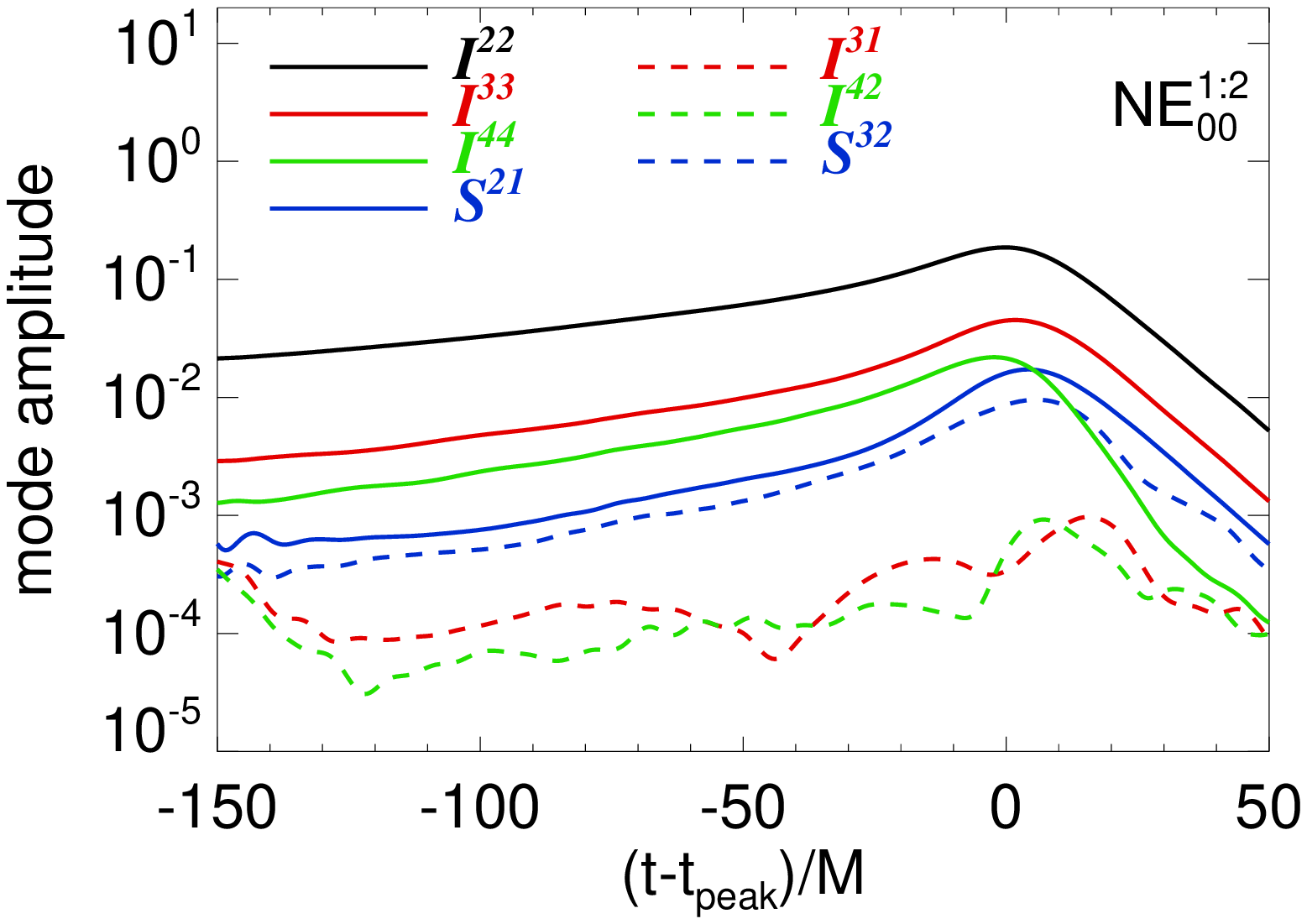}
\caption{\label{abs_lm} Amplitudes of the dominant radiative
  multipole moments.
On the left panel we show the modes for the NE$^{2:3}_{00}$ run, 
while on the right panel the modes for the NE$^{1:2}_{00}$ run.   
  The leading-order mass quadrupole $I^{22}$ is about an
  order of magnitude stronger than any other mode. The oscillating 
  behavior of the $S^{32}$ moment during RD is likely due to mode
  mixing with $I^{22}$. We denote with $t_{\rm peak}$ the 
time at which $I^{22}$ reaches its maximum.}
\end{figure*} 

We plot these frequencies in Fig.~\ref{omega_modes} for the dominant 
multipole moments $I^{22}$, $S^{21}$, $I^{33}$, $I^{44}$, and
$S^{32}$, for the NE$_{00}^{2:3}$ (left panel) and NE$_{00}^{1:2}$
(right panel) runs. The amplitudes of the $I^{31}$ and $I^{42}$ modes are too
weak and dominated by noise to extract a dominant frequency. In this
figure, as well as most shown in the rest of the paper, we plot the
time variable with respect to $t_{\rm peak}$. We notice
that the frequencies corresponding to the modes with $\ell=m$ agree quite
well throughout the inspiral and ringdown, but the frequency of the $S^{21}$ mode
decouples from the others approximately $50 M$ before the peak in  the
$I^{22}$ mode. As we shall see in Sec.~\ref{anatomy}, this is due to
the fact that, during the ringdown phase, the dominant angular frequency 
associated to the $S^{21}$ mode is almost twice as large as those of
the other leading modes~\cite{L85,E89,BCW}. This decoupling plays a
major role in determining the shape of the kick and anti-kick (see
Sec.~\ref{anatomy} below), and also suggests that the
transition to RD may begin long before the peak of the GW
flux. Similarly, the $S^{32}$ mode should converge to a higher RD
frequency ($\omega_{320}/2 \simeq 0.37/M_{\rm f}$ for these runs), but
may be limited by numerical noise here, as well as possible mode
mixing with the dominant $I^{22}$ moment.

\begin{figure*}
\includegraphics[width=0.48\textwidth,clip=true]{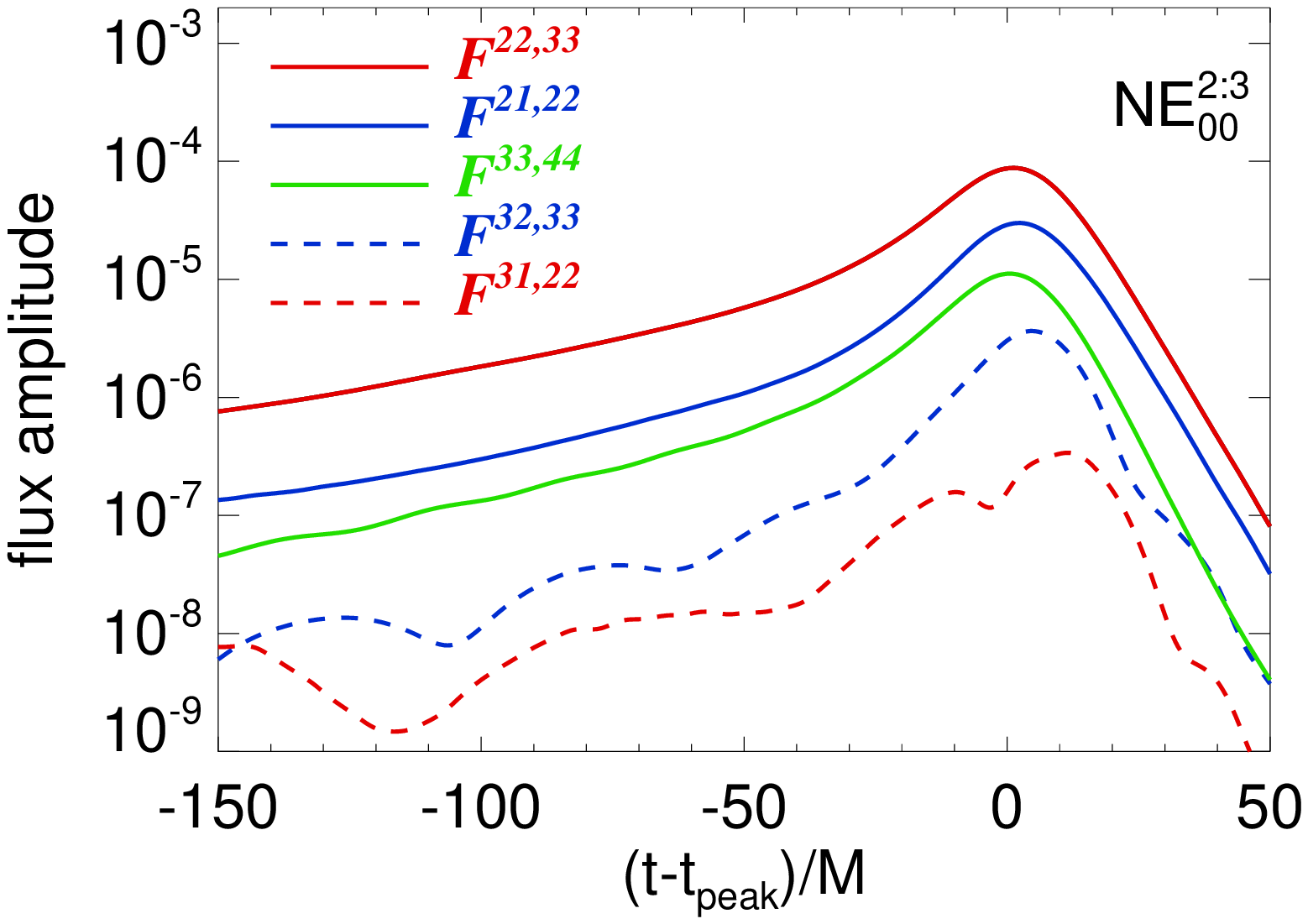}
\includegraphics[width=0.48\textwidth,clip=true]{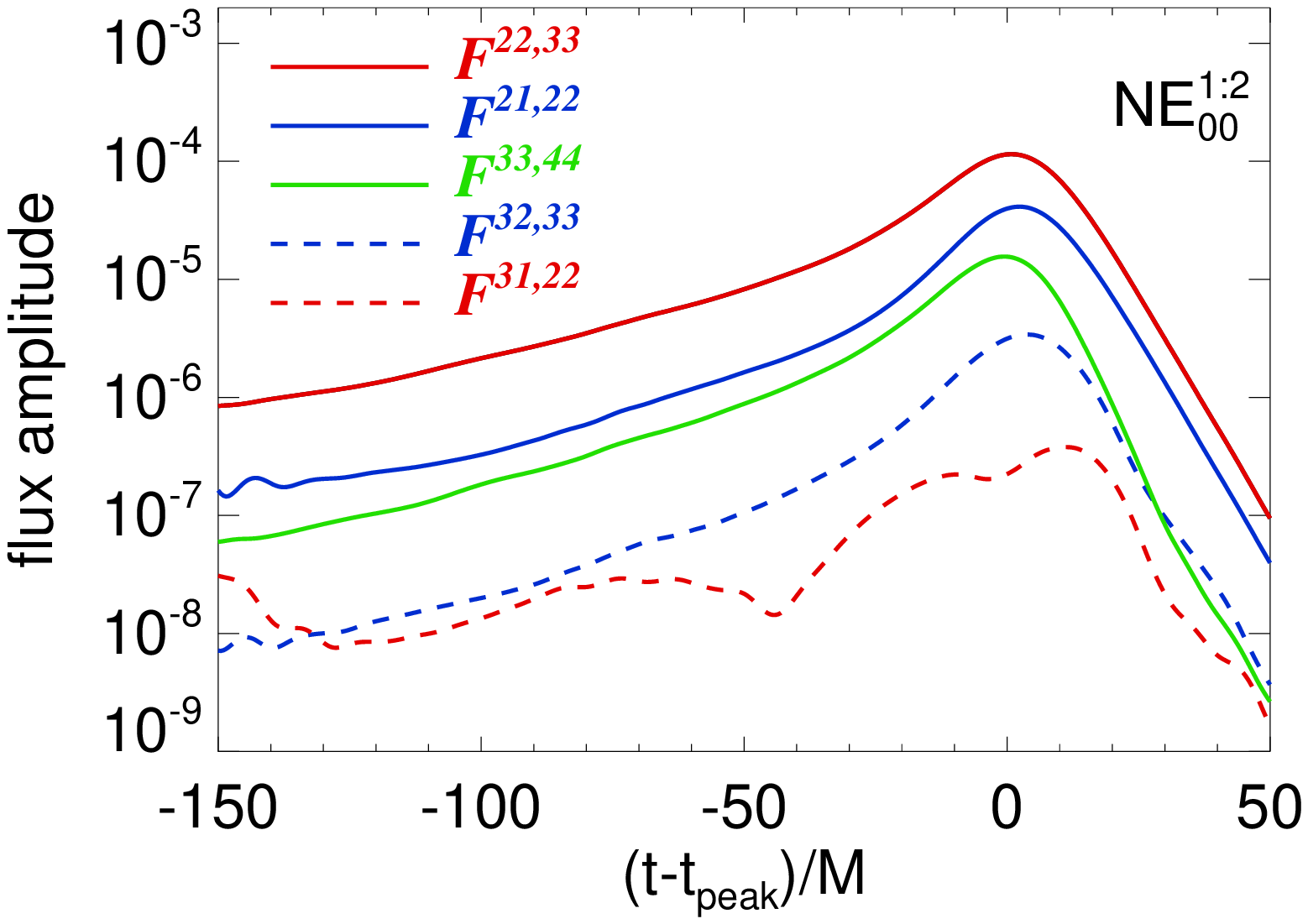}
\caption{\label{flux_lm} Linear momentum flux of the strongest radiative
  multipole moments, i.e., the ones in Eq.~(\ref{flux_approx}). 
On the left panel we show the modes for the NE$^{2:3}_{00}$ run, 
while on the right panel the modes for the NE$^{1:2}_{00}$ run.   
We denote with $t_{\rm peak}$ the time at which $I^{22}$ reaches its
  maximum.}
\end{figure*} 

In Fig.~\ref{abs_lm} we show the amplitudes of the multipole moments 
in Eq.~(\ref{flux_approx}).  Again, the left panel refers 
to the NE$^{2:3}_{00}$ run, while the right panel to the
NE$^{1:2}_{00}$ run. The mass-quadrupole moment $I^{22}$ clearly
dominates in both cases, while the $I^{31}$ and $I^{42}$ modes are so
weak as to be almost completely overwhelmed by numerical noise.
In addition to having dissimilar amplitudes, the different moments
also peak at slightly different times, which
may be related to the fact that RD modes are excited at different times. 
In particular, the modes mentioned above with $\ell \ne m$ tend to peak later
in time, perhaps due to a longer transition to the higher QNM frequency. 
As we shall see in Sec.~\ref{quasi-Newtonian}, as the mass ratio becomes more
extreme (i.e., decreasing $\eta$), the higher-order modes increase in relative amplitude, with
$I^{33}$ and $S^{21}$ both proportional to $\eta\, \delta m$. $I^{44}$ and
$S^{32}$, however, scale as $\eta(1-3\eta)$, so they increase
only slightly in the range of masses considered here.  

Next, in Fig.~\ref{flux_lm}, we show the amplitude of the linear
momentum flux from the mode-pairs included in
Eq.~(\ref{flux_approx}). Here we define the complex flux
$F^{21,22}=(-14i/336\pi)S^{21}I^{22*}$ and other $F^{\ell m,\ell' m'}$
analogously from Eq.~(\ref{flux_approx}). As in Fig.~\ref{abs_lm}, the
mass-quadrupole terms dominate,
with significantly smaller contributions from the $S^{32}$ and
$I^{31}$ modes. However, note the appreciable flux amplitude from the
$F^{33,44}\sim I^{33}I^{44*}$
term, which is formally a higher-order correction in a ($1/c$) expansion~\cite{BQW,DG}.
From Fig.~\ref{flux_lm}, we expect that the first three pairs of modes in 
Eq.~(\ref{flux_approx}) should contribute most significantly to the
recoil. Including the complex phase relations between the different
modes, we find this result will be supported further by the analysis in
Sec.~\ref{diff_mom}. 

\subsection{Ringdown phase}
\label{RD_phase}

We now extract the QNMs, notably the fundamental 
and the first two overtones, 
present in the most significant multipole moments during the RD phase.
We follow the procedure outlined in Ref.~\cite{BCP}. To avoid possible
constant offsets introduced by integrating
Eqs.~(\ref{Ilm_Clm}), (\ref{Slm_Clm}), we prefer to extract the QNMs
directly from the ${}_{-2}C_{\ell m}$ instead of using $I_{\ell m}$ or
$S_{\ell m}$. Additionally, from
Eqs.~(\ref{Ilm_Clm}), (\ref{Slm_Clm}), we see that ${}^{(1)}I^{\ell m}$ and
${}^{(1)}S^{\ell m}$ are made up of both ${}_{-2}C_{\ell m}$ {\it and}
${}_{-2}C_{\ell -m}$, which in general do not have the same QNM
frequencies, so it is more reliable to extract the RD modes from just
${}_{-2}C_{\ell m}$ (however, in practice we find that the RD phase is
dominated by modes with positive $m$). Following the approach of
Ref.~\cite{BCW}, we define the
complex frequencies $\sigma_{\ell mn}$:
\beq
\sigma_{\ell mn} \equiv \omega_{\ell mn} - i/\tau_{\ell mn},
\eeq
and each RD mode is proportional to $\exp(-i\sigma_{\ell mn}t)$. In this
notation, $\omega_{\ell mn}$ are the QNM oscillation frequencies [not to be
confused with the dominant frequencies of Eq.~(\ref{omega_lm})] and
$\tau_{\ell mn}$ are the mode decay times, all functions of the final black
hole mass and spin. The subscripts $\ell $ and $m$ are the same
spherical wavenumbers used above, and $n=0$ denotes the fundamental
mode, with $n=1,2,\cdots$, corresponding to the higher overtones. The
fundamental QNM frequencies $\sigma_{\ell m0}$ are listed in
Table~\ref{table:QNM_freqs} for the NR runs listed above. All
frequencies and decay times are measured in units of the final mass
$M_{\rm f}$.

\begin{center}
\begin{table*}
 \caption{Frequencies and decay times for the fundamental QNMs for
 each of the numerical simulations. $\omega_{\ell m 0}$ is in units of
 $M_{\rm f}^{-1}$ and $\tau_{\ell m 0}$ is in units of $M_{\rm f}$.}
\label{table:QNM_freqs}
 \begin{tabular}{c| c r r r r r r r r r r }
  \hline  \hline
   Run & $a_{\rm f}/M_{\rm f}$
   & \hspace{0.4cm} $\omega_{210}$ & $\tau_{210}$ 
   & \hspace{0.4cm} $\omega_{220}$ & $\tau_{220}$
   & \hspace{0.4cm} $\omega_{320}$ & $\tau_{320}$ 
   & \hspace{0.4cm} $\omega_{330}$ & $\tau_{330}$ 
   & \hspace{0.3cm} $\omega_{440}$ & $\tau_{440}$ \\
  \hline 
   ${\rm EQ}_{+-}$       & 0.697 & 0.454 & 12.2 & 0.531 & 12.4 
  & 0.758 & 11.9 & 0.841 & 12.0 & 1.14 & 11.8 \\
   ${\rm NE}^{2:3}_{00}$ & 0.675 & 0.450 & 12.1 & 0.521 & 12.2 
  & 0.749 & 11.7 & 0.827 & 11.9 & 1.12 & 11.7 \\
   ${\rm NE}^{1:2}_{00}$ & 0.633 & 0.442 & 11.9 & 0.505 & 12.1 
  & 0.734 & 11.6 & 0.803 & 11.7 & 1.09 & 11.5 \\
   ${\rm NE}^{1:4}_{00}$ & 0.423 & 0.411 & 11.5 & 0.445 & 11.5 
  & 0.674 & 11.1 & 0.711 & 11.1 & 0.963 & 10.9 \\
   ${\rm NE}^{2:3}_{+-}$ & 0.640 & 0.443 & 11.9 & 0.507 & 12.1 
  & 0.736 & 11.6 & 0.806 & 11.7 & 1.09 & 11.5 \\
   ${\rm NE}^{2:3}_{-+}$ & 0.704 & 0.456 & 12.2 & 0.533 & 12.4 
  & 0.760 & 11.9 & 0.845 & 12.1 & 1.14 & 11.9 \\
  \hline \hline
 \end{tabular}
\end{table*}
\end{center}

We present the RD analysis only for the NE$_{00}^{2:3}$ run, but the
others are qualitatively very similar. 
We have extracted the various QNM contributions to the
$_{-\!2}C_{\ell m}$ RD signal in the following way (see also Ref.~\cite{BCP}): 
We expect that at late times the $n=0$ QNM dominates.
We fit the signal after time $t_{\rm peak}+t_r$ to this single mode
using non-linear regression and choose $t_r$ to minimize the error in
the fit. We have four dimensionless parameters in this non-linear
fit: the QNM amplitude and phase, ${\cal C}_{\ell m 0}$ and $\phi_{\ell m 0}$, 
and the QNM frequency and decay time $M\omega_{\ell m 0}$ and
$\tau_{\ell m 0}/M$.  However, instead of fitting directly for these four
parameters, we treat $M\omega_{\ell m 0}$ and $\tau_{\ell m 0}/M$ as
functions of $a_{\rm f}/M_{\rm f}$ and $M_{\rm f}/M$ (which can be obtained via
interpolation from tabulated values given in Ref.~\cite{BCW}). 
The advantage of using $(a_{\rm f}/M_{\rm f},M_{\rm f}/M,{\cal C}_{\ell m 0},\phi_{\ell m 0})$ for
the set of fitting parameters comes when we fit to higher overtones.
As done in Ref.~\cite{BCP}, we extract the QNMs treating 
the real and imaginary parts of $_{-\!2}C_{\ell m}$ as independent.
Below we shall list results obtained from ${\rm Re}[_{-\!2}C_{\ell m}]$.

\begin{figure*}
\includegraphics[width=0.48\textwidth,clip=true]{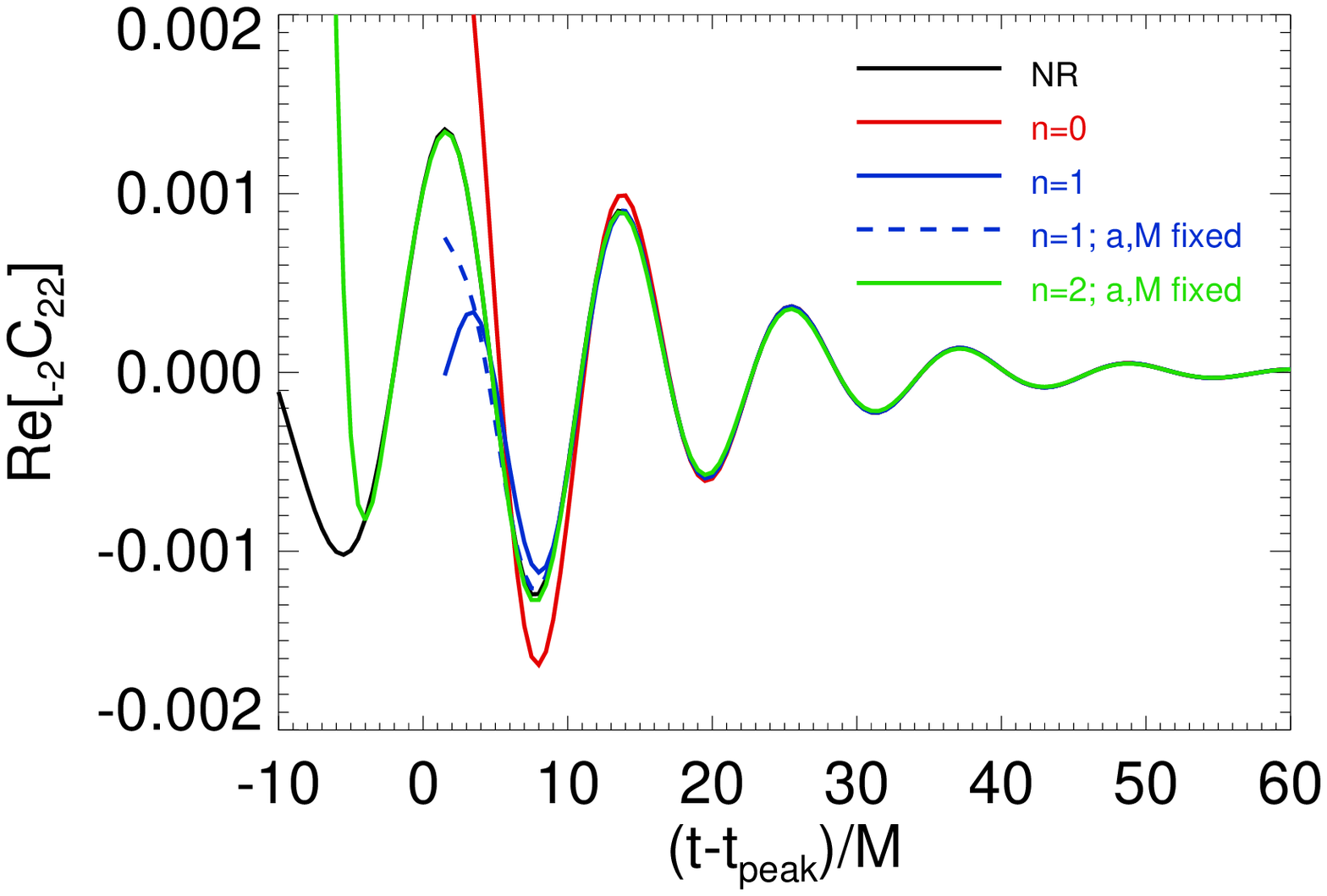}
\includegraphics[width=0.48\textwidth,clip=true]{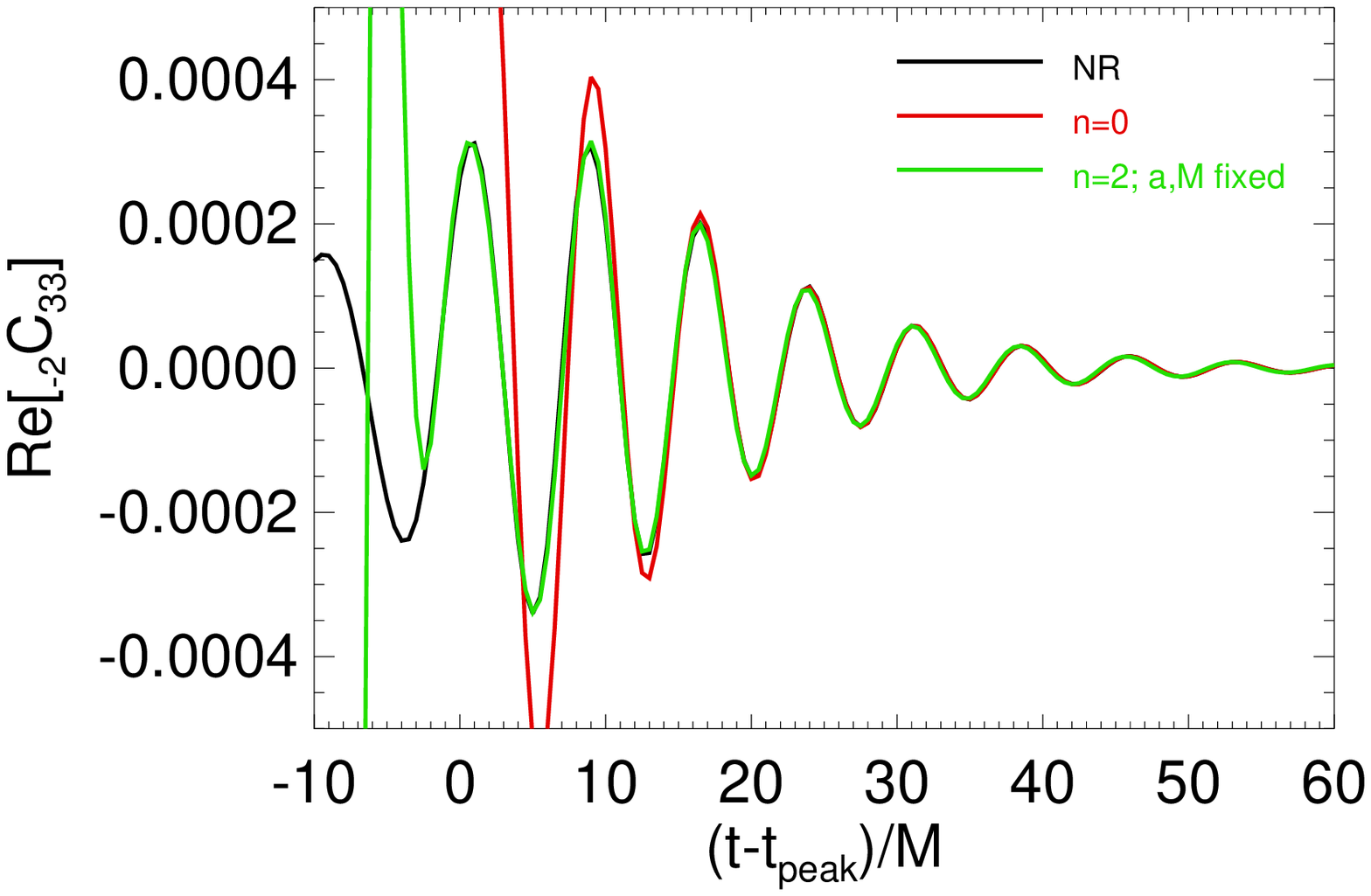}\\
\includegraphics[width=0.48\textwidth,clip=true]{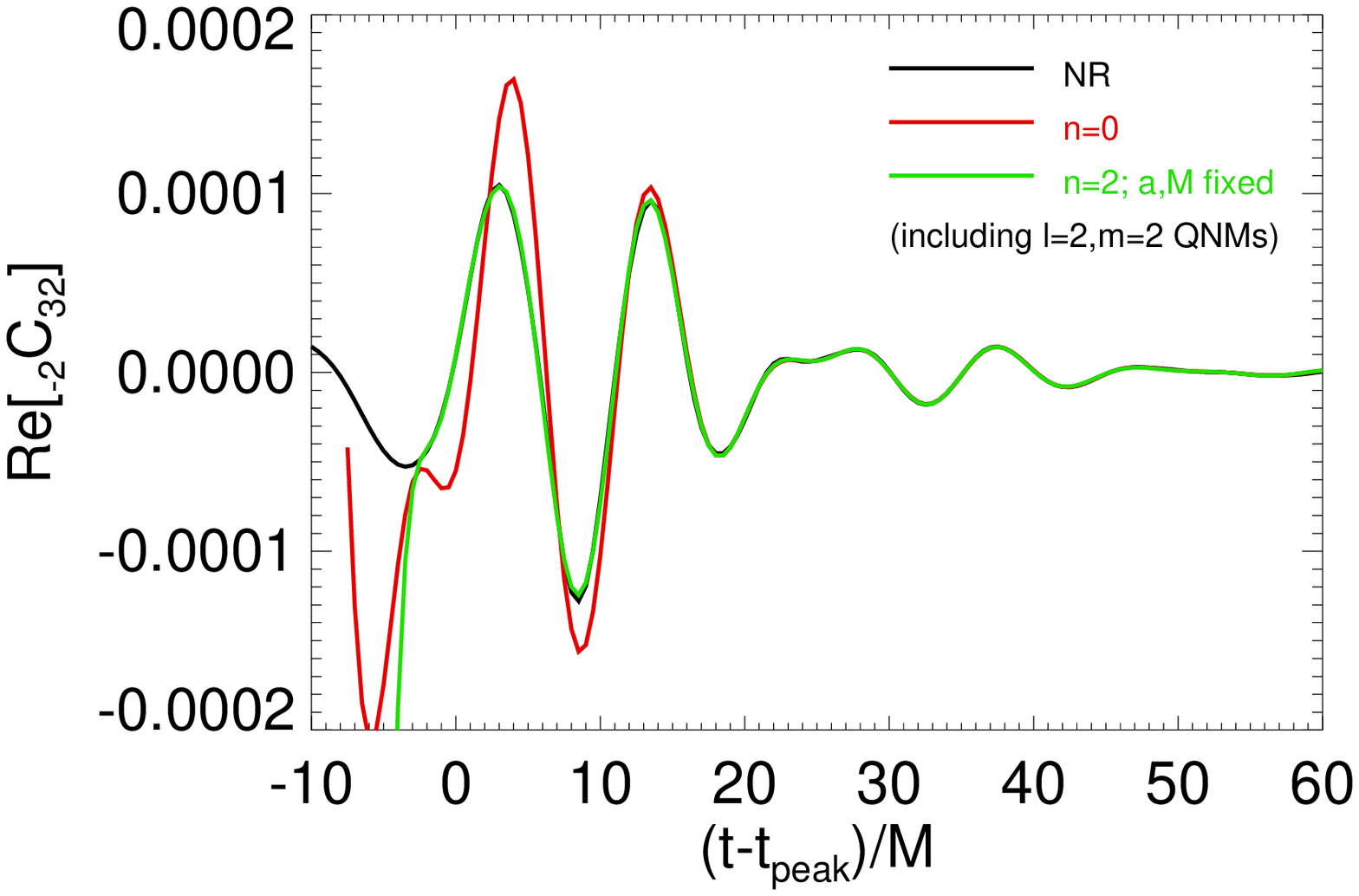}
\includegraphics[width=0.48\textwidth,clip=true]{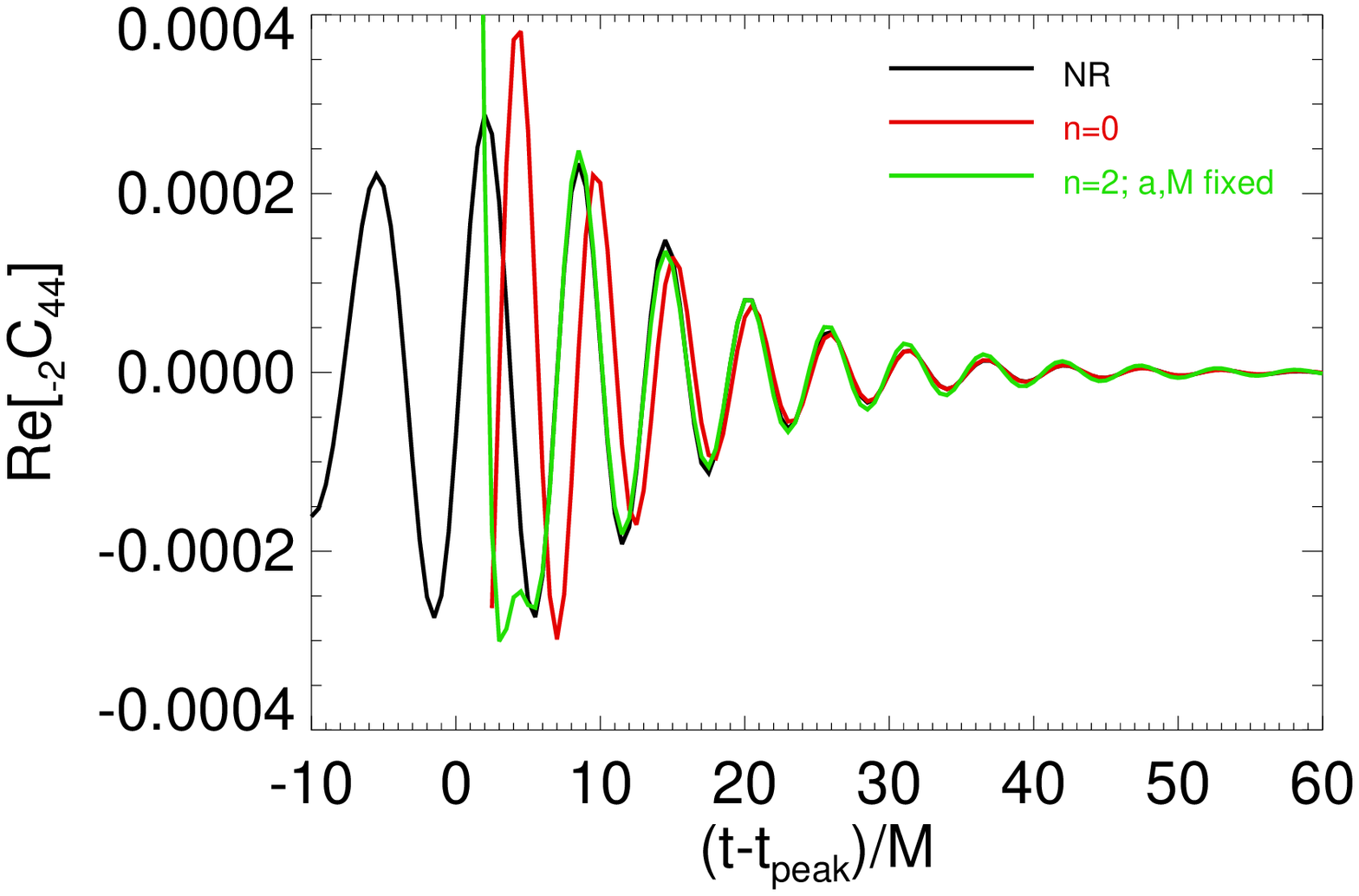}
\caption{\label{fig:rd22} Comparison of numerical and QNM 
waveforms for the NE$_{00}^{2:3}$ run. The dominant modes analyzed are
${}_{-2}C_{22}$ ({\it upper left}), ${}_{-2}C_{33}$
({\it upper right}), ${}_{-2}C_{32}$ ({\it lower left}), and
${}_{-2}C_{44}$ ({\it lower right}). Note that the ${}_{-2}C_{32}$
waveform includes contributions from the $\ell=2,m=2$ modes as well.
We denote with $t_{\rm peak}$ the time of the peak of $I^{22}$.}
\end{figure*}

By applying this procedure to the dominant mode, ${}_{-2}C_{22}$, 
we obtain $a_{\rm f}/M_{\rm f} = 0.669$ and $M/M_{\rm f} = 0.965$ 
together with the amplitude and phase of the fundamental 
QNM. We include additional overtones ($n>0$) successively. 
For each value of $n$, we refit the entire function, so for $n=0$ there are 4 parameters in the
fit, for $n=1$ there are 6, for $n=2$ there are 8, and so forth. 
Thus, applying a 6-parameter fit we successfully extract also the
first overtone simultaneously, obtaining 
slightly different values for  $a_{\rm f}/M_{\rm f} = 0.661$ 
and $M/M_{\rm f} = 0.958$. We find it impossible to extract, with a
single 8-parameter fit, also the second 
overtone. By contrast if we keep $a_{\rm f}/M_{\rm f}$ and $M/M_{\rm f}$ fixed 
and equal to the values obtained when extracting the fundamental 
QNM, we find that we can fit up to the second overtone. Moreover, 
quite interestingly, the fit provides waveforms that compare
very well with the NR waveforms up to the peak of $I_{22}$, 
as can be seen in the upper left panel of Fig.~\ref{fig:rd22}.

The remaining panels in Fig.~\ref{fig:rd22} show 
results for the other relevant modes ${}_{-2}C_{33}$, ${}_{-2}C_{44}$ and 
${}_{-2}C_{32}$. As obtained in Ref.~\cite{BCP}, 
we find a ``mode-mixing'' in ${}_{-2}C_{32}$, i.e., the RD waveform is a combination 
of $\ell=2,m=2$ and $\ell=3,m=2$ QNMs. This effect appears to be most
important between modes with the same $m$ value, and may possibly be
explained by the fact that the QNMs should really be expressed as {\it
  spheroidal}, not {\it spherical} harmonics \cite{BCW,BCP}. Including
both sets of modes means that the ${}_{-2}C_{32}$ is actually fit
using 14 parameters: the final mass and spin, and the amplitude and
phase of 6 QNMs.

By fitting the fundamental QNM for each ringdown waveform, 
we obtain $a_{\rm f}/M_{\rm f} = 0.671$ and $M/M_{\rm f} = 0.972$; 
$a_{\rm f}/M_{\rm f} = 0.527$ and $M/M_{\rm f} = 0.884$; $a_{\rm
f}/M_{\rm f} = 0.686$ and $M/M_{\rm f} = 0.981$, 
for ${}_{-2}C_{33}$, ${}_{-2}C_{44}$ and ${}_{-2}C_{32}$,
respectively. We also are able to extract the fundamental QNM for the
${}_{-2}C_{21}$ mode (not shown in Fig.~\ref{fig:rd22}) and find
$a_{\rm f}/M_{\rm f}=0.678$ and $M/M_{\rm f}=0.960$. 
All of these values for the inferred final BH spin and mass are 
rather consistent, except for ${}_{-2}C_{44}$. This discrepancy 
might be due to numerical resolution effects, and will be the object
of future investigations.

Thus we find that although we cannot simultaneously extract three QNMs 
(the fundamental and two overtones) and we are not able to 
clearly determine the onset of the RD phase, 
we {\it do} obtain that for $t > t_{\rm peak}$ 
the numerical waveforms can be well fitted by a superposition of 
three QNMs. This result explains why the simple matching procedure 
from inspiral to RD adopted in the EOB model~\cite{BD2,DG,BCP} can 
almost always work succesfully (see Ref.~\cite{EOB4PN} for some caveats). 
In Sec.~\ref{matching_RD} we shall adopt the same matching procedure of the EOB 
model when building the full waveform using the pseudo-analytic 
model of Sec.~\ref{quasi-Newtonian}.

\section{Effective Newtonian model}\label{quasi-Newtonian}

In an attempt to better understand the amplitudes and frequencies of
the various modes during the inspiral and merger phases, we present
here what we call the ``effective Newtonian'' (eN) model. It begins with
calculating the leading-order Newtonian formulae for each multipole
moment of the source, as a function of the BH masses, binary
separation $R$, and orbital phase $\phi$. To extend these formulae through
the end of the inspiral and into the merger phase, we introduce an
effective radial separation to absorb PN effects into
the leading-order multipole expressions. Each multipole moment is then
individually matched to a linear superposition of ringdown modes, as
is done in the effective-one-body model~\cite{BD2,DG,BCP}. Taken
together with the match to Kerr QNMs, this
eN model provides an excellent framework within which
we can understand the details of the linear momentum flux and
net recoil velocity.

\subsection{Newtonian Multipole Moments}\label{effective_radius}

Working at leading Newtonian order for each mode, we equate
the radiative multipole moments to the source multipole
moments. Restricting ourselves to circular, planar orbits, we find
that for non-spinning systems, the dominant modes are
\cite{BD,BDS,BS,JS}
\begin{subequations}\label{I_nospin}
\bea
\label{S21}
S_{\rm nospin}^{21} &=& -\frac{8}{3}i\sqrt{\frac{2\pi}{5}}\,\frac{\delta m}{M} \, \mu
\,R^3\, \omega^4 \,e^{-i\phi}, \\
\label{I22}
I_{\rm nospin}^{22} &=& 16i\sqrt{\frac{2\pi}{5}}\, \mu\, R^2\, \omega^3\, e^{-2i\phi}, \\
\label{I31}
I_{\rm nospin}^{31} &=& -\frac{2}{3}\sqrt{\frac{\pi}{35}}\, \frac{\delta m}{M} \, \mu\, 
R^3\, \omega^4\, e^{-i\phi}, \\
\label{S32}
S_{\rm nospin}^{32} &=& -\frac{16}{3}\sqrt{\frac{2\pi}{7}}\, \mu\,(1-3\eta)\, R^4\, \omega^5
\,e^{-2i\phi}, \\
\label{I33}
I_{\rm nospin}^{33} &=& 54\sqrt{\frac{\pi}{21}}\, \frac{\delta m}{M} \, \mu\,
R^3\, \omega^4\, e^{-3i\phi}, \\
\label{I42}
I_{\rm nospin}^{42} &=& \frac{16}{63}i\sqrt{2\pi}\,\mu\,(1-3\eta)\,R^4\,\omega^5\,
e^{-2i\phi}, \\
\label{I44}
I_{\rm nospin}^{44} &=& -\frac{256}{9}i\sqrt{\frac{2\pi}{7}}\,\mu\,(1-3\eta)\,
R^4\,\omega^5\, e^{-4i\phi},
\eea
\end{subequations}
where $R$ is the radial separation
and $\omega = \dot{\phi}$ is the binary orbital frequency. 
Considering only the mass quadrupole terms in the linear momentum flux
(i.e., the terms proportional to $S^{21}I^{22*}$, $I^{31}I^{22*}$,
and $I^{22}I^{33*}$),
we obtain the well-known result valid at Newtonian order~\cite{DG}: 
\begin{equation}
\label{flux_Newt0}
F^{(0)}  = -i \frac{464}{105}\frac{\delta m}{M} \, \mu^2\, R^5\, \omega^7\, e^{i\phi}.
\end{equation}
Including the next-highest order moments in
Eq.~(\ref{dPdt_radmoments2}), we get
\begin{equation}
\label{flux_Newt1}
F^{(1)} = -i \frac{11120}{1323}\frac{\delta m}{M} \, \mu^2\, (1-3\eta) R^7\, \omega^9\, e^{i\phi}.
\end{equation}
While there may also be next-to-leading order contributions from a PN
expansion of the multipole moments included in
Eq.~(\ref{dPdt_radmoments}) that would show up in
Eq.~(\ref{flux_Newt1}), we can effectively absorb those corrections
into the $R$ variable, as will be described below.

\begin{figure*}
\includegraphics[width=0.48\textwidth,clip=true]{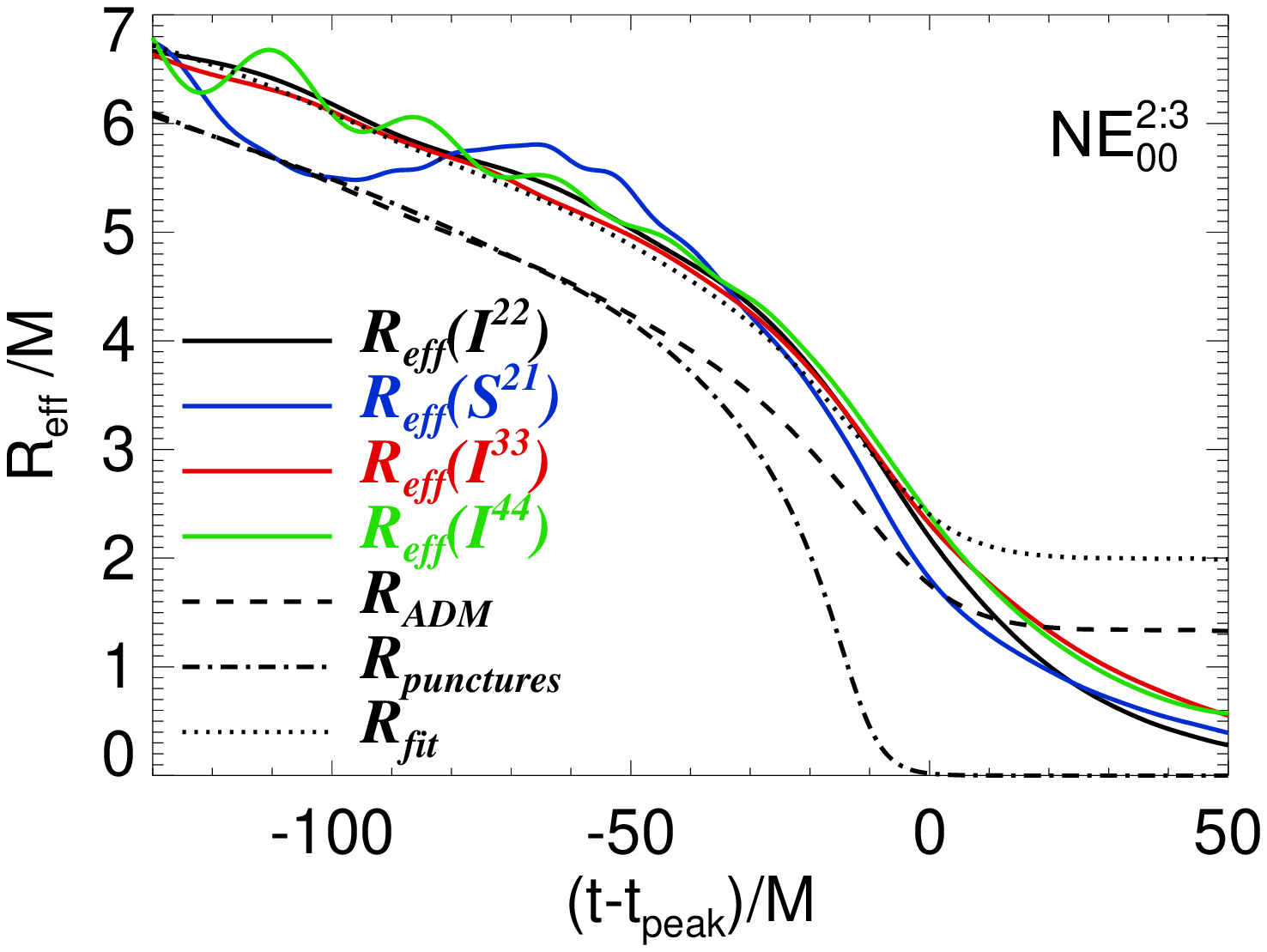}
\includegraphics[width=0.48\textwidth,clip=true]{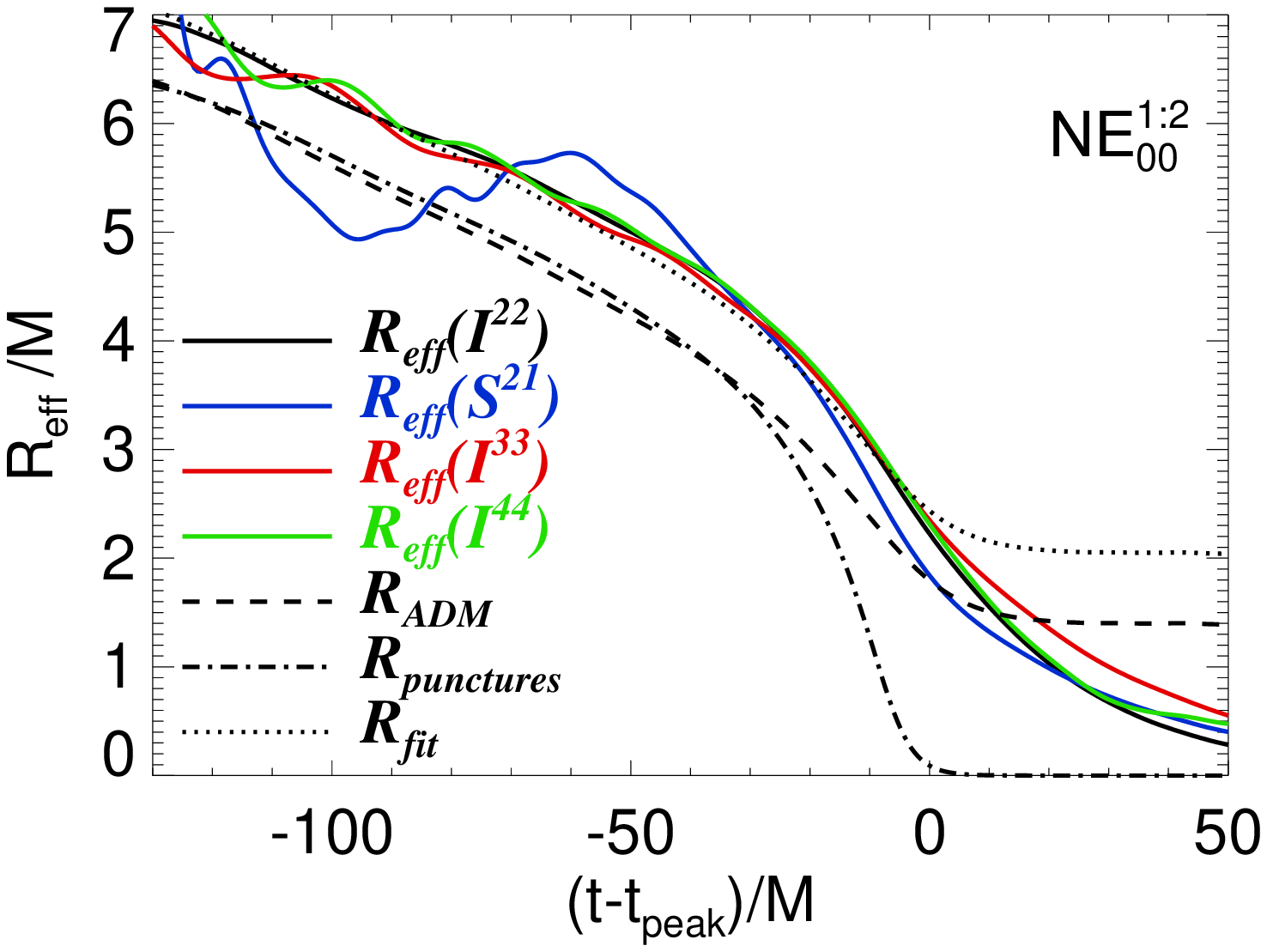}
\caption{\label{r_eff} Effective radius for different modes, derived
from Eqs.~(\ref{omega_lm}), (\ref{S21})--(\ref{I44}). 
The close agreement for the $R_{\rm eff}^{lm}$ suggests we can use a single
effective radius $R_{\rm eff}(t)$ for the Newtonian expressions. We believe
that the large oscillations in $R_{\rm eff}^{21}$ are due to initial
eccentricity at early times. Also plotted is the ADM radius (dashed
curves) derived from the orbital frequency via Eq.~(\ref{eqn:R_ADM}),
the coordinate separation of the BH punctures (dot-dashed curves), and
the empirical fit $R_{\rm fit}$ (dotted curves) obtained by shifting 
$R_{\rm ADM}$ by 0.65. The results correspond to the
NE$_{00}^{2:3}$ (left panel) and NE$_{00}^{1:2}$ (right panel)
runs. We denote with $t_{\rm peak}$ the time at which $I^{22}$ reaches its maximum.}
\end{figure*} 

Combining Eqs.~(\ref{flux_Newt0}) and (\ref{flux_Newt1}) we find the
linear momentum flux scales like
\bea\label{jena_formula}
|F^{(0)}+F^{(1)}| &\propto & \frac{\delta m}{M}\, \mu^2
\left[1+\frac{3475}{1827}(1-3\eta)R^2 \omega^2\right] \nonumber\\
&\approx& \frac{3}{2}\frac{\delta m}{M}\, \mu^2 (1-0.9\eta),
\eea
which is remarkably similar to the result found in
Ref.~\cite{recoilJena}. Here we have used $R^2\omega^2\approx 0.23-0.25$ at
the peak of the energy flux, which seems to be quite robust for a
range of mass ratios. However, the extremely close agreement with
Ref.~\cite{recoilJena} is probably to some degree a coincidence, since
this simple Newtonian formula does not include any details of the
phase relations between different modes, which become especially
important during the transition from inspiral to ringdown (see 
Sec.~\ref{transition} below). Since Eq.~(\ref{jena_formula}) really
only applies to the inspiral portion, if anything, it should be a
predictor of how the {\it peak} recoil velocity scales. This is not
necessarily the same as the {\it final} recoil, since
we find that more extreme-mass-ratio BH binaries have a
relatively smaller anti-kick, which should also play an important role
in the scaling relation of Ref.~\cite{recoilJena}.

If we compute the above multipole moments (\ref{S21})--(\ref{I44}) 
using $\omega$ as given by Eq.~(\ref{omega_lm}) and $R$  as obtained from the puncture trajectories, 
we do not find a very good agreement with the numerical results. This is not 
surprising since there is no reason to believe that the Newtonian 
approximation should work well all along the inspiral phase. We 
should expect that higher-order PN corrections become important as 
we approach the merger. Furthermore, $R$ is a
coordinate-dependent quantity, and thus does not necessarily have the
same meaning in a PN expression as in NR. Since our scope is limited to a diagnostic 
of the NR results, and not to a precise comparison with PN calculations, 
instead of including PN corrections in
Eqs.~(\ref{I_nospin})-(\ref{flux_Newt1}), we investigate whether by
properly scaling the  
Newtonian expressions we can get a better agreement until the merger. We can also think 
of this normalization as a way of resumming the PN expansion.

Quite interestingly, if we compute the amplitudes 
$|I^{\ell m}|$ or $|S^{\ell m}|$ from the numerical data, 
and the angular frequency $\omega$ from Eq.~(\ref{omega_lm}), 
we find that the radii $R^{\ell m}$ which appear in the RHS of 
Eqs.~(\ref{S21})--(\ref{I44}) are rather 
independent of the multipole moments $\ell $ and $m$, as Fig.~\ref{r_eff} 
shows. We denote the radii $R^{\ell m}$ computed numerically as {\it effective} 
radii $R_{\rm eff}^{\ell m}$. The close agreement between the frequencies (see 
Fig.~\ref{omega_modes}) and effective radii for each mode suggests we
can use the Newtonian expressions and a single $R_{\rm eff}(t)$ and
orbital frequency $\omega(t)$, 
e.g., $R_{\rm eff}^{22}(t)$ and $\omega_{\rm D}^{I22}$for all modes with a high degree of accuracy for 
the entire inspiral phase and even during the transition to merger.

For comparison we also show in Fig.~\ref{r_eff} the radius
from the puncture trajectory (dot-dashed curves) and the radius
computed using the Arnowitt-Deser-Misner transverse-traceless gauge
(dashed curves), given as a function of frequency through 3PN order
by~\cite{BI}
\begin{widetext}
\beq\label{eqn:R_ADM}
R_{\rm ADM} = M^{1/3}\, \omega^{-2/3}\,\left [ 1 + \omega^{2/3}\,\left (-1 + \frac{\eta}{3} \right ) 
+ \omega^{4/3}\,\left ( -\frac{1}{4} + \frac{9}{8}\,\eta + \frac{\eta^2}{9} \right ) 
+ \omega^2\,\left ( -\frac{1}{4} - \frac{1625}{144}\,\eta + \frac{167}{192}\,\eta\, \pi^2 
- \frac{3}{2}\,\eta^2 + \frac{2}{81}\,\eta^3\right ) \right ]\,.
\eeq
\end{widetext}
Here we use the orbital frequency $\omega$ derived from the $I^{22}$
mode via Eqn.~(\ref{omega_lm}), giving a constant value during
the RD phase when the orbital frequency is meaningless. 
Fig.~\ref{r_eff} shows interesting agreement between $R_{\rm ADM}$ and 
the radius from the puncture trajectory, and a constant offset 
between $R_{\rm ADM}$ and $R_{\rm eff}$. The latter is due to the fact 
that the amplitude of the multipole moments computed at
leading Newtonian order does not reproduce the numerical 
relativity amplitude~\cite{BCP,Baker:2006kr}, and higher order 
PN corrections need to be included. Motivated by this similarity
between $R_{\rm ADM}$ and $R_{\rm eff}$, we attempt to fit empirically
the $R_{\rm eff}$ curves in Fig.~\ref{r_eff} by simply shifting 
$R_{\rm ADM}$ by  $0.65$. The fit curve is included as 
a dotted curve in Fig.~\ref{r_eff}. As we accumulate longer and 
more accurate NR data for a wider range of $\eta$ values, and 
study possible analytic resummation  of higher-order PN 
amplitude corrections, we should be able to work out 
a widely applicable amplitude-scaling factor to be included 
in leading-order analytic waveforms~\cite{EOB4PN}.

In the next section, we shall investigate how 
this simple eN model can be combined 
with a superposition of QNMs, as described in Sec.~\ref{RD_phase}, 
giving a good representation of the NR results. 

\subsection{Matching to ringdown}
\label{matching_RD}

We now match the inspiral and RD 
waveforms in a mode-by-mode fashion following the philosophy of the 
EOB approach~\cite{BD2}. Note this is not the same
analysis of Section \ref{RD_phase}, where we {\it fit} the numerical
data throughout the RD phase with a superposition of QNMs. Here we
{\it match} the data at a single point at the transition from inspiral
to RD and see how well it agrees with the rest of the RD phase. A
similar attempt was followed in Ref.~\cite{DG},
where for simplicity the authors performed the matching to the 
Schwarzschild QNM frequencies, while we use the Kerr QNM frequencies 
and match to the fundamental QNM frequency and the first 
two overtones, as done in Ref.~\cite{BCP}. 
We obtain the QNM frequencies and decay times from Ref.~\cite{BCW} as
a function of $a_{\rm f}/M_{\rm f}$ (taken from
Table~\ref{table:idparams} above).
For the fundamental and two overtone QNMs, we can match a given multipole 
mode by equating it and two time derivatives to a linear combination
of QNMs.

We write
\beq
\label{RD}
I^{\ell m}(t) = A(t)\,e^{-i\phi(t)} = \sum_{n=0}^{\infty} A_{\ell mn}\,
e^{-i\sigma_{\ell mn} (t-t_{\rm match})},
\eeq
where the complex QNM frequencies are known functions of the final BH
mass and spin, and we must solve for the complex amplitudes
$A_{\ell mn}$. Matching three QNMs we get
\bse
\begin{eqnarray}
I^{\ell m}(t_{\rm match}) &=& \sum_{n=0}^{2} A_{\ell mn}, \\
\frac{d}{dt}I^{\ell m}(t_{\rm match}) &=& -i \,\sum_{n=0}^{2} \sigma_{\ell mn}
A_{\ell mn}, \\
\frac{d^2}{dt^2}I^{\ell m}(t_{\rm match}) &=& - \sum_{n=0}^{2} \sigma_{\ell mn}^2 \,A_{\ell mn},
\end{eqnarray}
\ese
or as a simple matrix equation
\begin{equation}
\left(\begin{array}{ccc}
1 & 1 & 1 \\
-i\sigma_{\ell m0} & -i\sigma_{\ell m1} & -i\sigma_{\ell m2} \\
-\sigma_{\ell m0}^2 & -\sigma_{\ell m1}^2 & -\sigma_{\ell m2}^2
\end{array}\right) \left(\begin{array}{c} A_{\ell m0} \\
A_{\ell m1} \\ A_{\ell m2} \end{array}\right) =
\left(\begin{array}{c} I^{\ell m} \\
\dot{I}^{\ell m} \\ \ddot{I}^{\ell m} \end{array}\right).
\end{equation}

\begin{figure*}
\includegraphics[width=0.48\textwidth,clip=true]{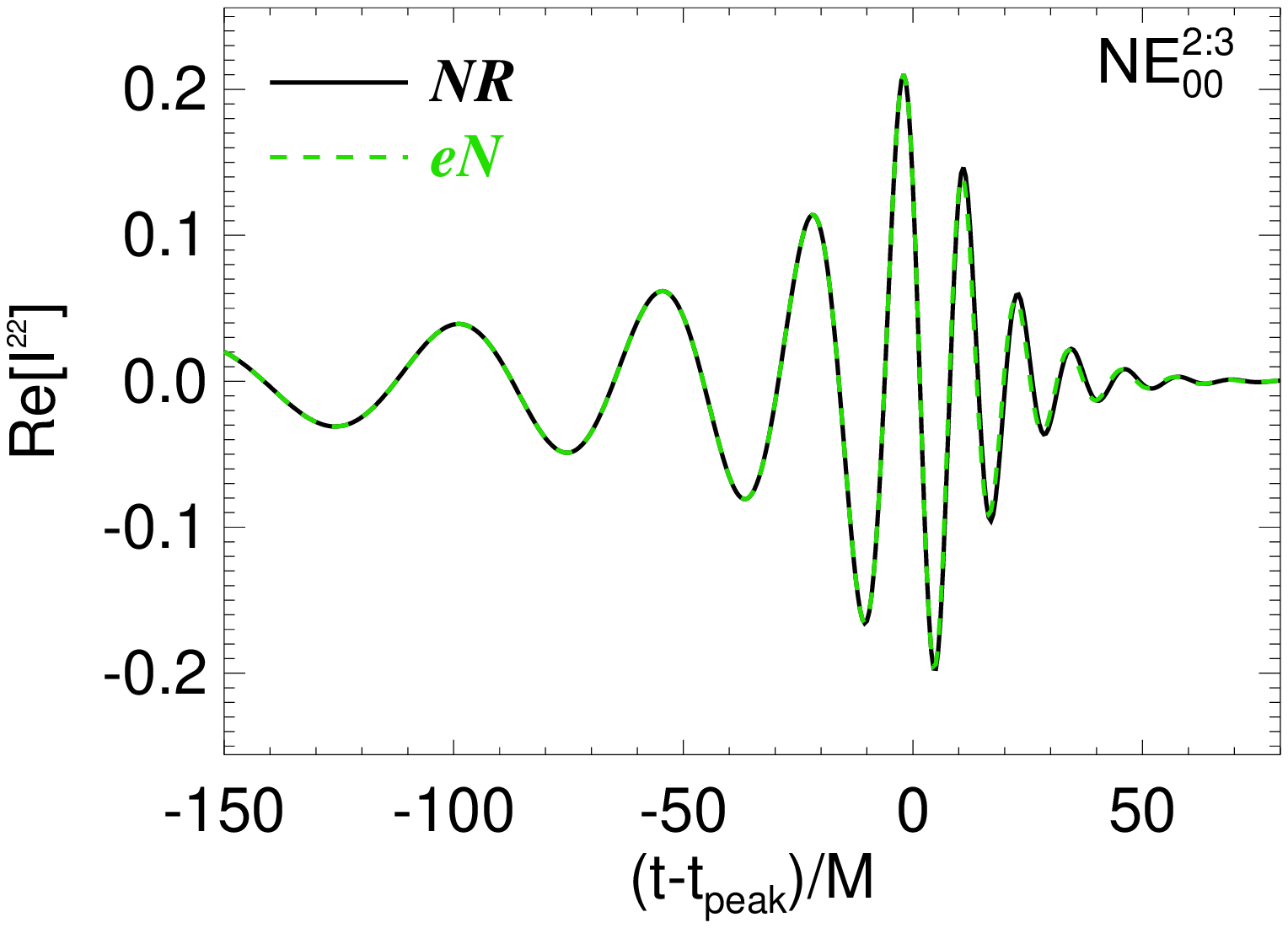}
\includegraphics[width=0.48\textwidth,clip=true]{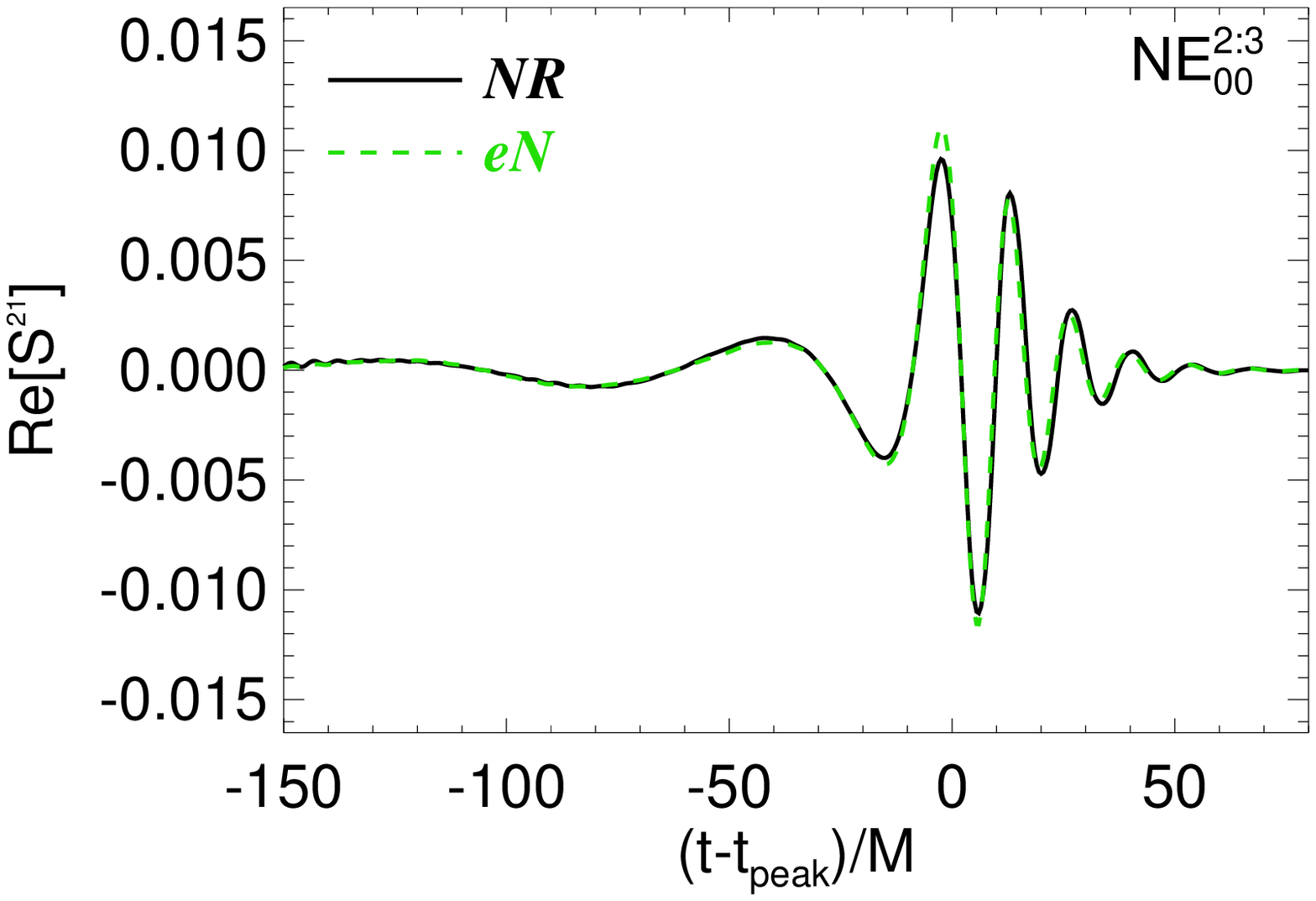}\\
\includegraphics[width=0.48\textwidth,clip=true]{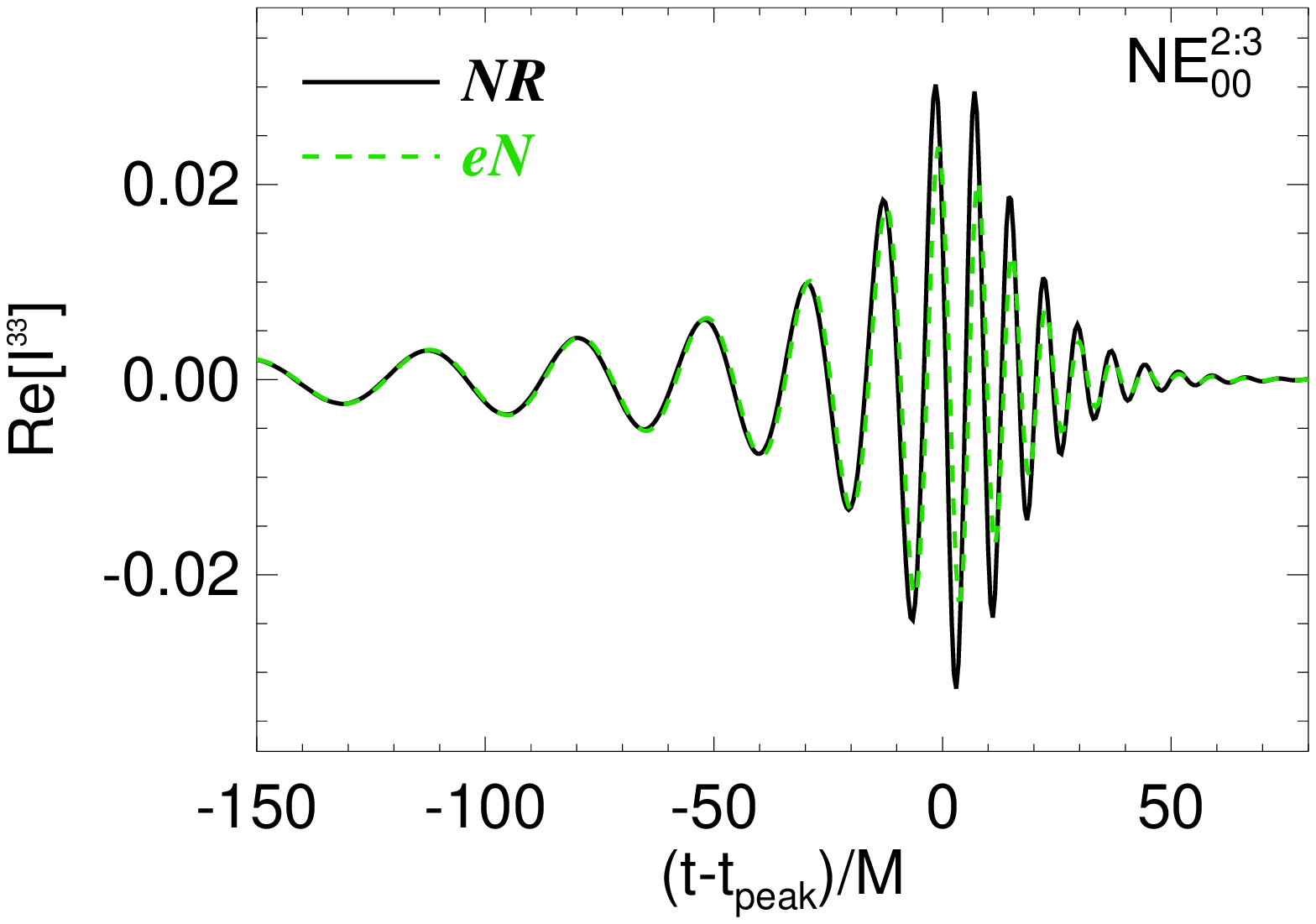}
\includegraphics[width=0.48\textwidth,clip=true]{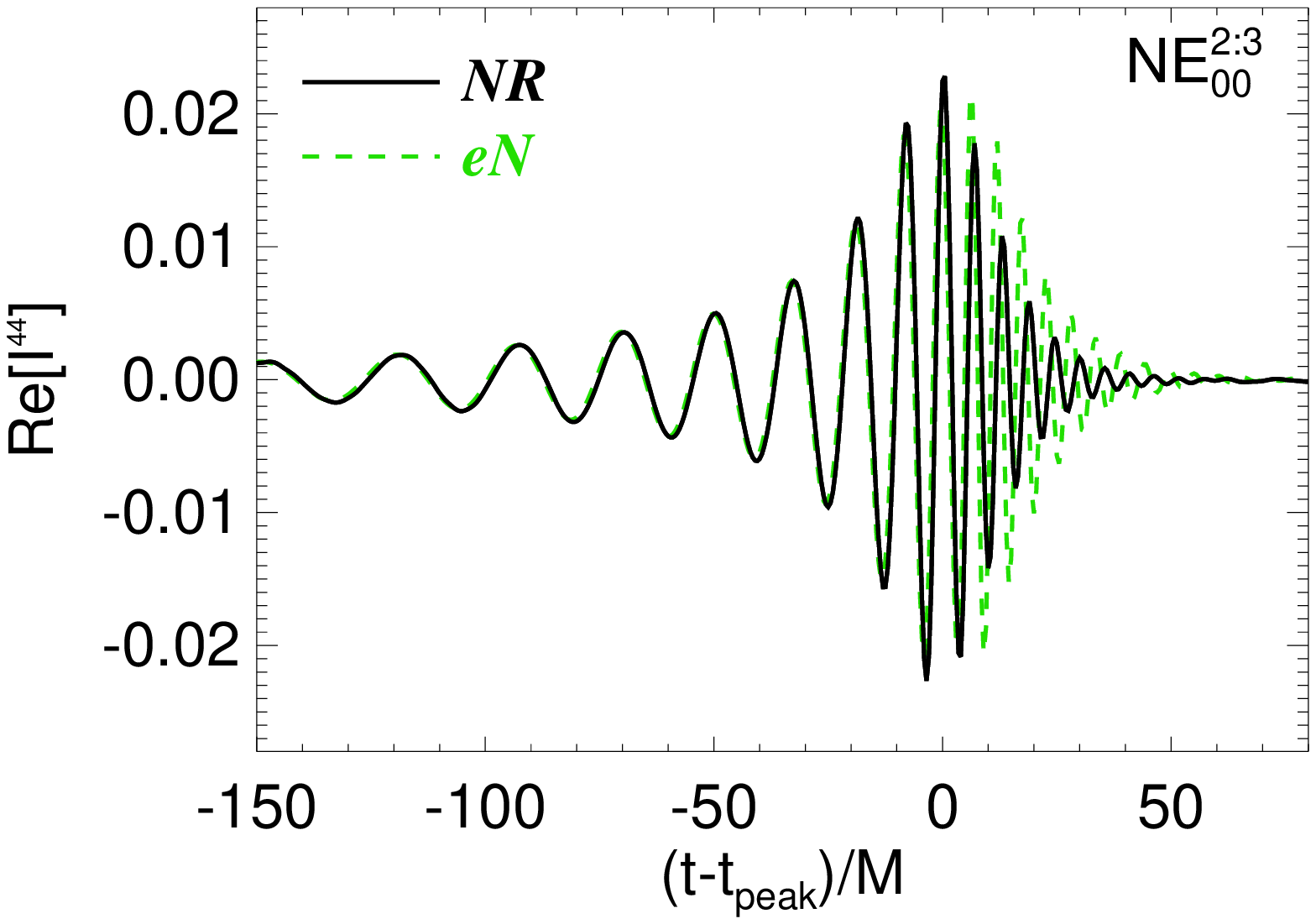}
\caption{\label{compare_ring} Comparison of the effective Newtonian
  and NR radiative modes during inspiral, merger and RD phases. The
  data refer to the NE$_{00}^{1:2}$ run. We denote with $t_{\rm peak}$
  the time at which $I^{22}$ reaches its maximum.}
\end{figure*} 

\begin{figure}
\includegraphics[width=0.5\textwidth,clip=true]{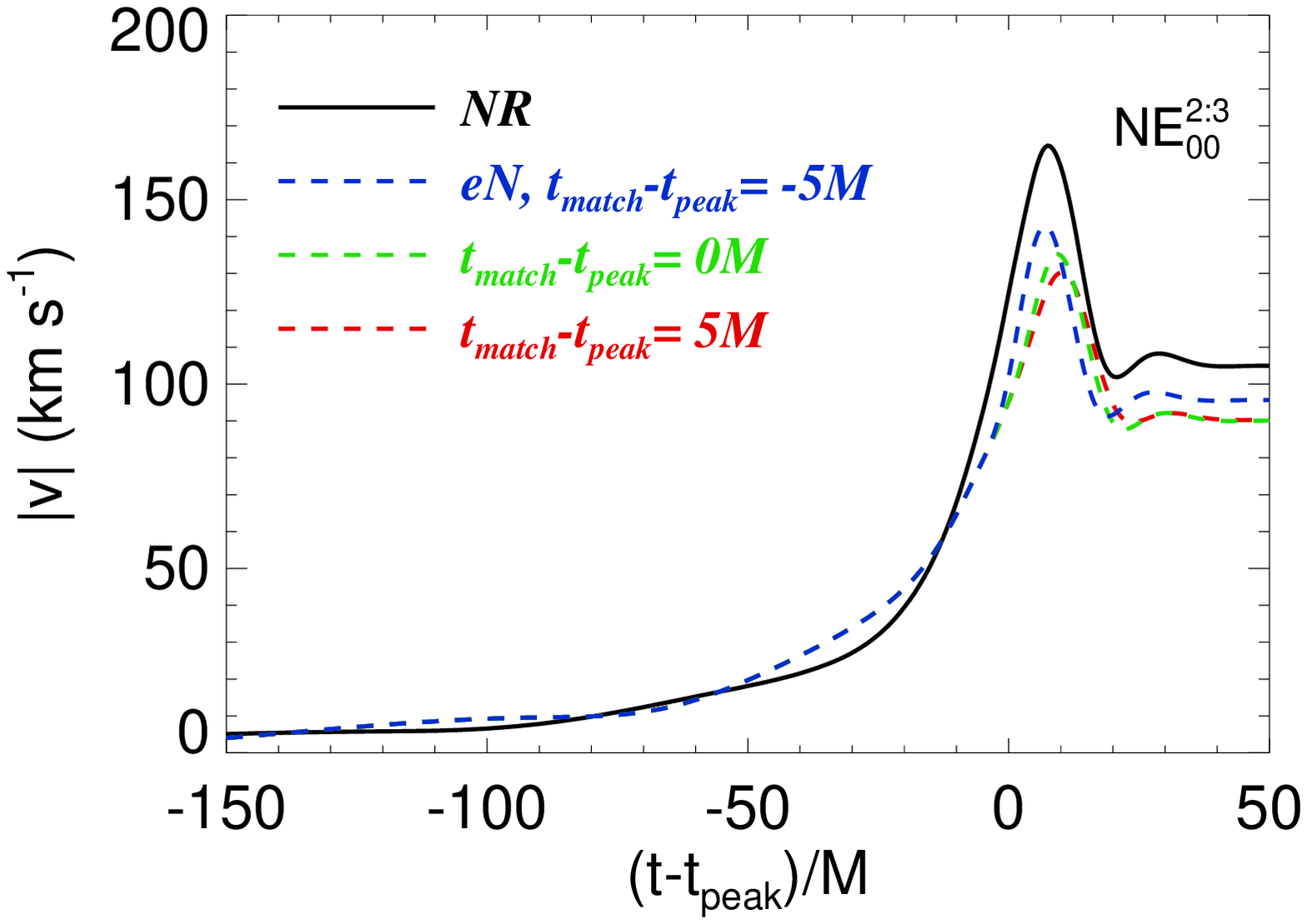}
\caption{\label{kick_match} Comparison of the effective Newtonian model and NR predictions for the recoil velocity 
for a range of inspiral-RD matching points. We denote with $t_{\rm peak}$ the 
time at which $I^{22}$ reaches its maximum. The data refer to the NE$_{00}^{2:3}$ run.}
\end{figure} 

In Fig.~\ref{compare_ring}, we compare the NR modes to  
the modes obtained by the effective Newtonian model described in 
Sec.~\ref{effective_radius} until $t_{\rm match}$ and by the superposition 
of three QNMs for $t > t_{\rm match}$. During the
inspiral, the different moments are calculated according to
Eqs.~(\ref{S21})-(\ref{I44}), using a single $R_{\rm eff}$ and
$\omega_{\rm D}$ determined from the $I^{22}$ mode, with the
exception of the $S^{21}$ mode, where we instead use the higher
frequency $\omega_{\rm D}^{S21}$ (but same $R_{\rm eff}$). 
We treat $t_{\rm match}$ as a free parameter: if we stop the inspiral
too early, the eN mode amplitudes are still growing,
so the sudden transition to decaying RD modes prematurely reduces
them. On the other hand, if the inspiral is continued too long, we
tend to lose the important phase shifts 
between the modes that only begin during the transition to
RD. This is particularly evident in the $I^{44}$
mode, which undergoes an unexplained phase-shift around
the transition to RD, and also decays at somewhat different rate
than is predicted from QNM theory (see above, Sec.~\ref{RD_phase}).
Motivated by the results of Sec.~\ref{RD_phase}, notably by the fact
that a superposition of three QNMs can fit very well the NR waveforms
starting from the peak of the energy flux, we choose as best-matching
point the peak of the energy flux. 

Having shown a reasonably close match for each of the
radiative multipoles between the effective Newtonian model and the
numerical data, it stands to reason that the total recoil calculated
with this model should agree as well. This is shown in Fig.~\ref{kick_match}, 
where we have also varied the matching point around $t_{\rm
  peak}$. We first note the close agreement between the eN
models with varying $t_{\rm match}$, suggesting the
inspiral-to-ringdown matching method described above is relatively
robust. Not surprisingly, since the individual modes agree, we also
find reasonable agreement between the NR data and the eN predictions
for the recoil.

However, this agreement may be partially fortuitous, since the eN model
cannot predict the mode phase shifts
around $t=t_{\rm peak}$, most notably that of the $I^{44}$ mode
described above. 
In Section \ref{transition} below, we will examine
this phasing in greater detail and show how it affects the overall
kick. At this point, we unfortunately do not have a clear
understanding of the underlying cause of the phase shift, but it may
well be related to the slightly different times of transition from
inspiral to ringdown for the different modes. Preliminary results
also suggest that this de-phasing effect is reduced in more
extreme-mass-ratio systems, as we shall see in Appendix~\ref{app}.

\section{Anatomy of the kick}
\label{anatomy}

In the above Sections, we have laid the groundwork for a multipolar
analysis of the gravitational recoil, describing the momentum flux as
a combination of radiative multipole modes. Along with the
psuedo-analytic models for the inspiral and ringdown phases, we can
now give a detailed description of the ``anatomy'' of the kick, namely
the way the different modes combine to produce a peak recoil velocity,
followed by a characteristic anti-kick and then asymptotic approach to
the final value of the BH recoil.

\subsection{Contribution from different moments}
\label{diff_mom}

In Sec.~\ref{linear_flux}, we showed how the radiative multipole
moments contribute to the linear momentum flux through the integral
of the $\Psi_4$ scalar
[Eqs.~(\ref{eq:psi4Ylmdef}),(\ref{dPdt_psi4})]. Here, we want to  
determine exactly which modes we need to include in the multipole expansion 
Eq.~(\ref{thorne_420}) to get a good representation of the full recoil, and 
which are the pairs of modes in Eq.~(\ref{flux_approx}) that 
contribute most.

\begin{figure*}
\includegraphics[width=0.48\textwidth,clip=true]{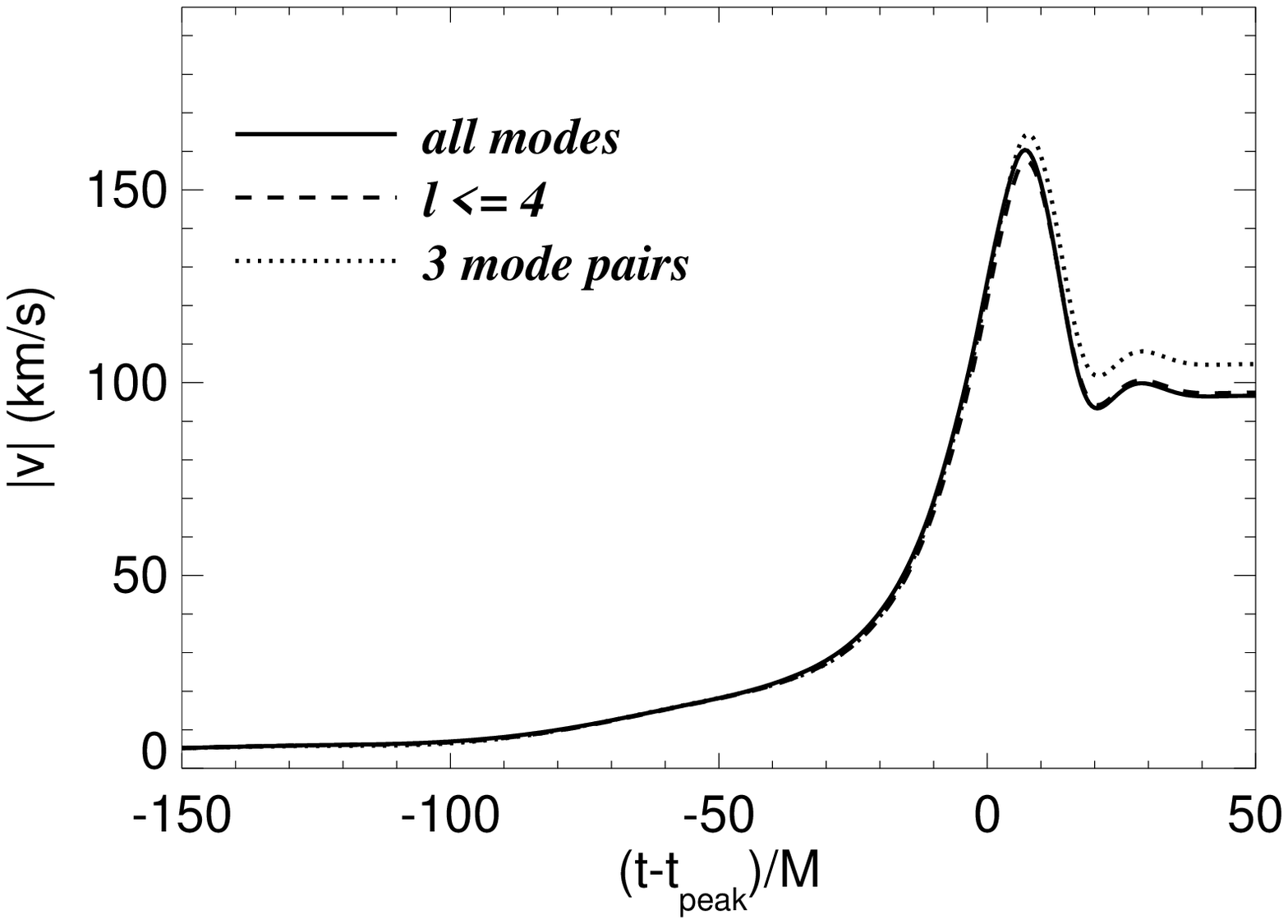}
\includegraphics[width=0.48\textwidth,clip=true]{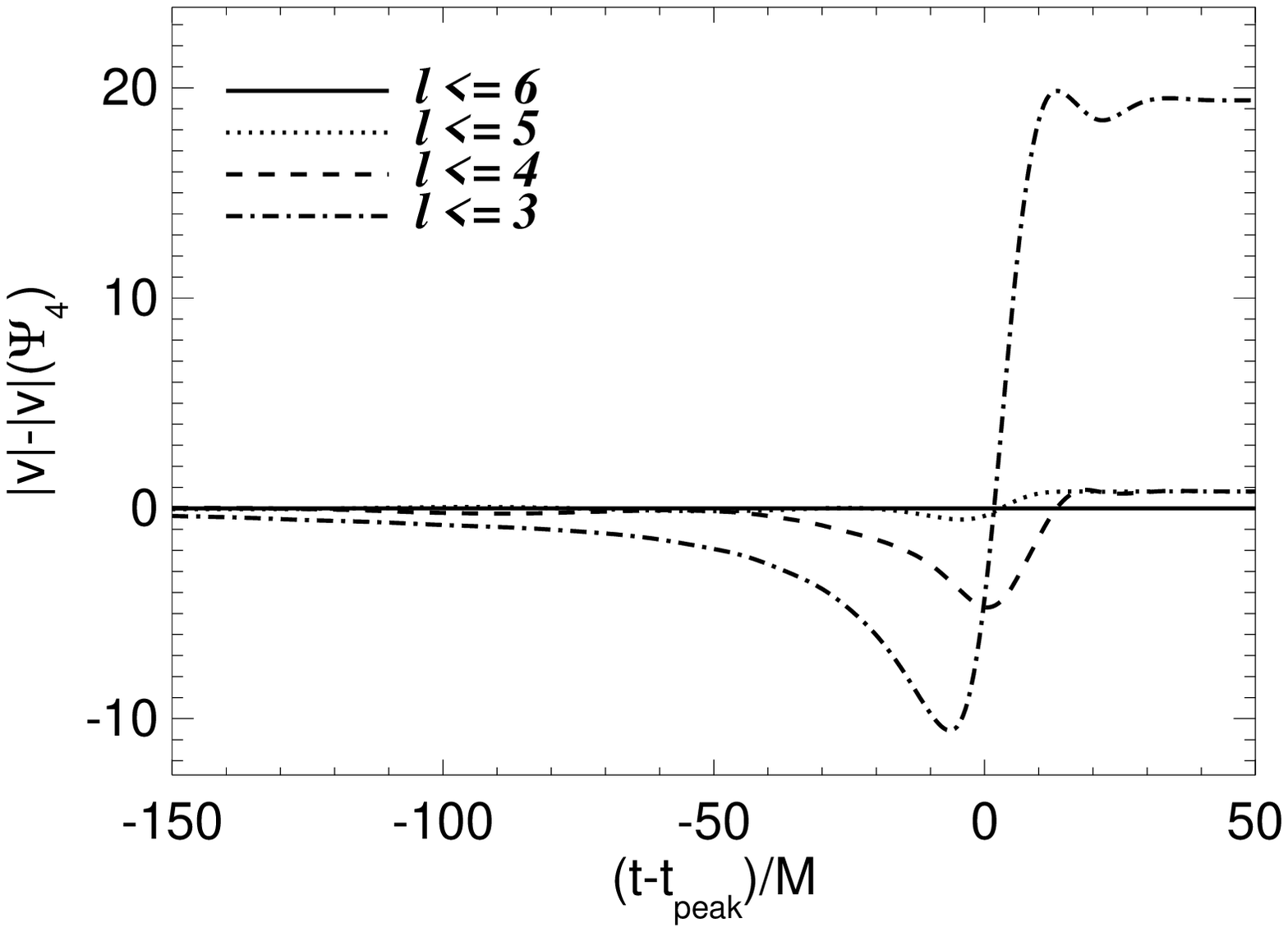} 
\caption{\label{v_tot1} In the left  panel we show the net recoil kick, integrated from the
  linear momentum flux via Eq.~(\ref{dPdt_psi4}) (solid curve), 
  from all modes with $\ell\le 4$ (dashed curve) and also limiting the
  modal composition of $\Psi_4$ to just the three dominant mode pairs 
  in Eq.~(\ref{flux_approx}) (dotted curve). In the right panel 
  we show the difference between the exact result and the $\Psi_4$
  expansion Eq.~(\ref{eq:psi4Ylmdef}), limited to $\ell\leq 3,4,5,6$. 
The data refer to the NE$_{00}^{2:3}$ run. 
We denote with $t_{\rm peak}$ the time at which $I^{22}$ reaches its maximum.}
\end{figure*}

\begin{figure*}
\includegraphics[width=0.48\textwidth,clip=true]{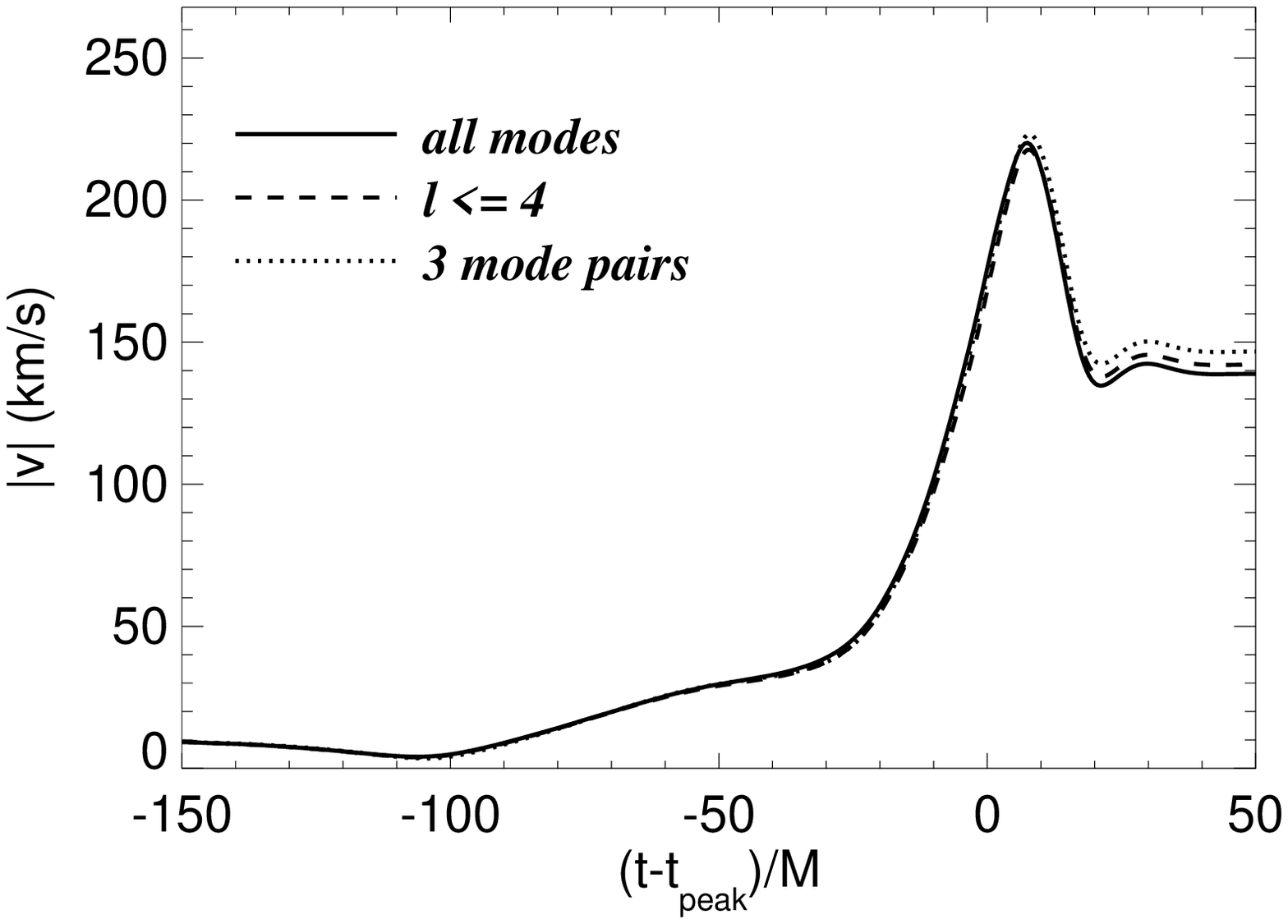}
\includegraphics[width=0.48\textwidth,clip=true]{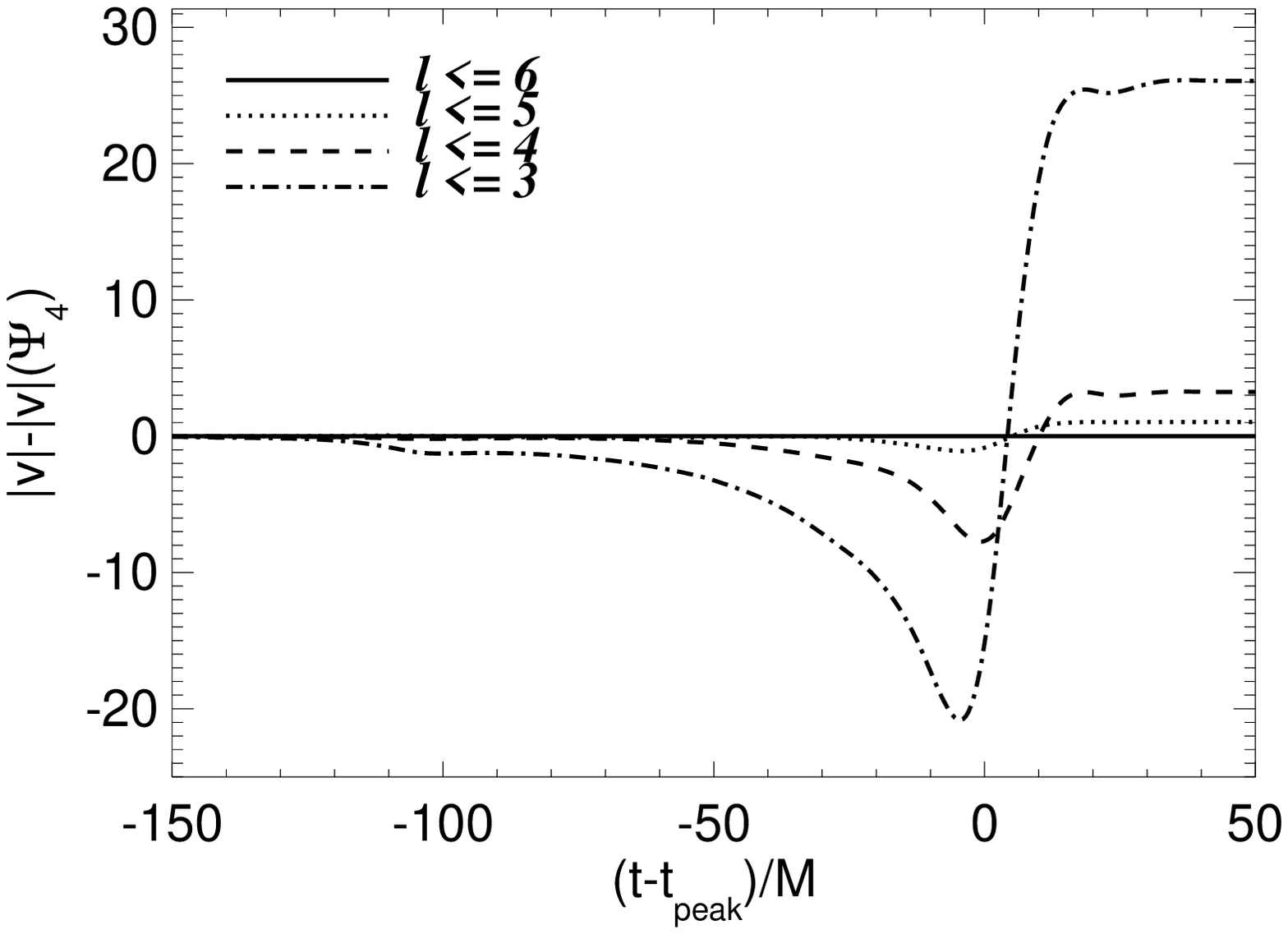}
\caption{\label{v_tot2} Same as Fig.~\ref{v_tot1}, but for the
  NE$_{00}^{1:2}$ run.}
\end{figure*}

By including only a select choice of terms in the $\psi_4$ expansion
Eq.~(\ref{eq:psi4Ylmdef}), we can calculate the linear momentum flux
by direct integration of Eq.~(\ref{dPdt_psi4}) and compare it with the
predictions of Eqs.~(\ref{dPdt_radmoments})-(\ref{flux_approx}), in
each case including only the appropriate moments. This is a good
way of double-checking those lengthy equations term-by-term, and in
practice we find excellent agreement, limited only by the numerical
accuracy of the simulations. Similarly, we can use this method of
truncated expansion to determine which modes are necessary for
calculating the recoil up to a given accuracy.
The results of using higher and higher order multipolar moments are
shown in Figs.~\ref{v_tot1} and \ref{v_tot2} for the NE$_{00}^{2:3}$ and
NE$_{00}^{1:2}$ runs, respectively.

In the left panels of Figs.~\ref{v_tot1} and \ref{v_tot2} we show with a
solid curve the exact recoil velocity from Eq.~(\ref{dPdt_psi4}),
with a dashed curve the contribution from terms up to $\ell=4$, i.e., those
obtained from Eq.~(\ref{dPdt_radmoments}) and
  (\ref{dPdt_radmoments2}), and with a dotted curve the contribution
from just the three leading terms in Eq.~(\ref{flux_approx}), valid
for non-precessing BHs with kicks in the orbital plane.
We conclude that the linear momentum flux is dominated by the
$I^{33}I^{22*}$, $I^{33}I^{44*}$, and $S^{21}I^{22*}$ terms, which
combine to produce 
the primary kick and anti-kick agreeing with the exact result
within $\lesssim 10\%$ throughout the entire merger. Note that the
flux from the $S^{32}I^{33*}$ term, while not insignificant in
Fig.~\ref{flux_lm}, contributes almost nothing to the net recoil
velocity. This is largely due to phase relations between the various
modes during the transition from inspiral to ringdown, described below
in Sec.~\ref{transition}.

In the right panels of Figs.~\ref{v_tot1} and \ref{v_tot2} we 
show the difference between the calculation obtained including terms up 
to $\ell=3,4,5,6,$ and the exact result. It seems clear that we need modes 
up to and including $\ell=4$ to get an accurate
estimate of the recoil velocity. For more extreme mass ratios,
higher-order moments become relatively more important, but remain
strongly sub-dominant to the $\ell\le 4$ modes \cite{recoilAEI,berti07}.

\begin{figure}
\includegraphics[width=0.48\textwidth,clip=true]{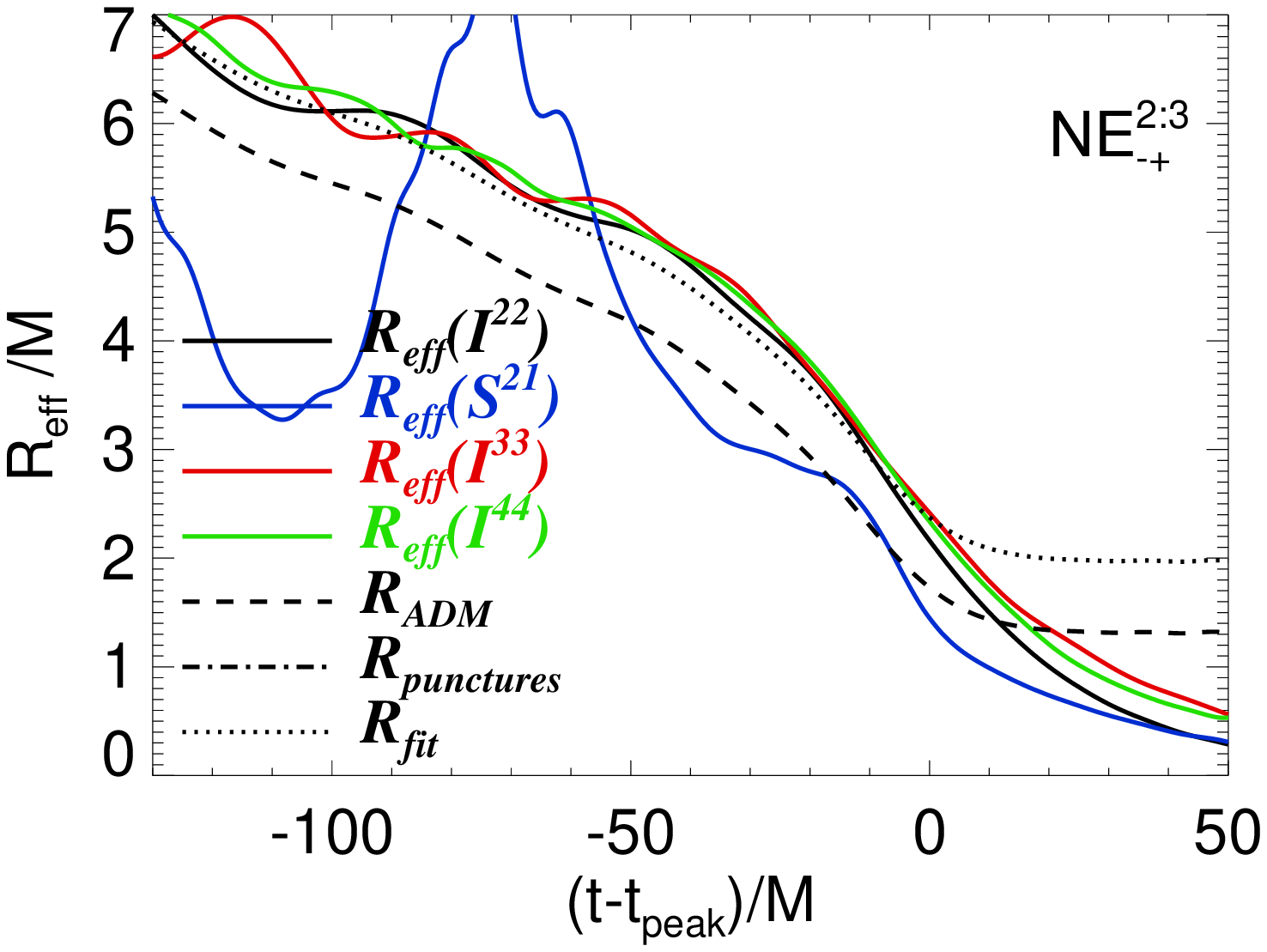}
\caption{\label{r_eff_mp} $R_{\rm eff}$ derived from different
  multipole modes, as in Fig.~\ref{r_eff}, for the NE$_{-+}^{2:3}$
  run. The $S^{21}$ mode for this run has comparable contributions
  from $\delta m$ and $\Delta^z$, making it difficult to derive a
  reasonable $R_{\rm eff}(S^{21})$.}
\end{figure} 

To understand more clearly the relative contributions of the different modes
to the total recoil, we will include analysis of a few
more simulations including non-precessing spins. As mentioned above in
Sec.~\ref{linear_flux}, non-precessing spins do not introduce any
additional moments compared to the non-spinning simulations, but
simply modify the relative amplitudes of the different modes in
Eq.~(\ref{flux_approx}) by adding the spin terms. Thus, once we 
determine how the spins modify the individual modes, we 
can use the same analysis for the spinning and non-spinning cases. 

Again equating the radiative multipole moments with the source moments, 
we get the leading order spin-orbit modifications
to Eqs.~(\ref{S21})--(\ref{I44}) [see Eqs.~(3.14),(3.20) in
  Ref.~\cite{LK} and Eq.~(5.5) in Ref.~\cite{BBF}]:
\begin{subequations}\label{spinorbit_moments}
\begin{eqnarray}
S^{21}_{\rm SO} &=& -4i\sqrt{\frac{2\pi}{5}}\,\eta\, R\, \omega^3 \,e^{-i\phi}
\Delta^z, \\
I^{22}_{\rm SO} &=& \frac{64}{3}i\sqrt{\frac{2\pi}{5}}\,\eta\, R^2\,
\omega^4\, e^{-2i\phi}\, \xi^z\, \\
S^{32}_{\rm SO} &=& -\frac{32}{3}\sqrt{\frac{2\pi}{7}}\,\eta\, R^2\,
\omega^4\, e^{-2i\phi}\, \xi^z, \\
I^{31}_{\rm SO} &=& -\frac{2}{3}\sqrt{\frac{\pi}{35}}\, \eta\, 
R^3\, \omega^5\, e^{-i\phi}\, \Sigma_{31}^z, \\
I^{33}_{\rm SO} &=& 54\sqrt{\frac{\pi}{21}}\, \eta\,
R^3\, \omega^5\, e^{-3i\phi}\, \Sigma_{33}^z, 
\end{eqnarray}
\end{subequations}
where we have introduced the spin vectors 
\begin{subequations}\label{sigma_z}
\bea
\Sigma_{31} &\equiv& \frac{11}{2}\frac{\delta m}{M}\, \mathbf{S} +
\frac{1}{2}(11-39\eta)\mathbf{\Delta}, \\
\Sigma_{33} &\equiv& \frac{3}{2}\frac{\delta m}{M}\, \mathbf{S} +
\frac{3}{2}(1-5\eta)\mathbf{\Delta}.
\eea
\end{subequations}

\begin{figure*}
\includegraphics[width=0.48\textwidth,clip=true]{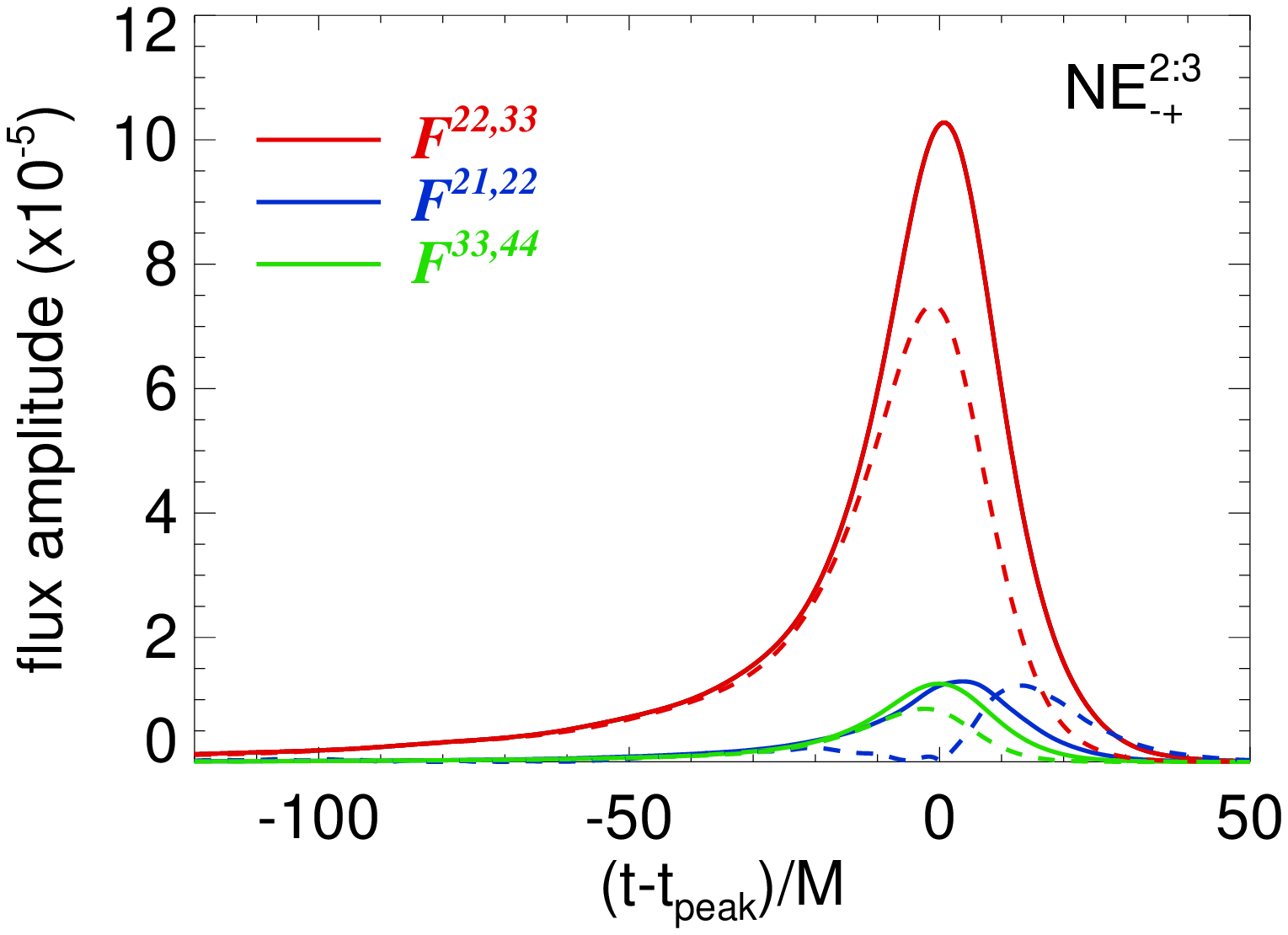}
\includegraphics[width=0.48\textwidth,clip=true]{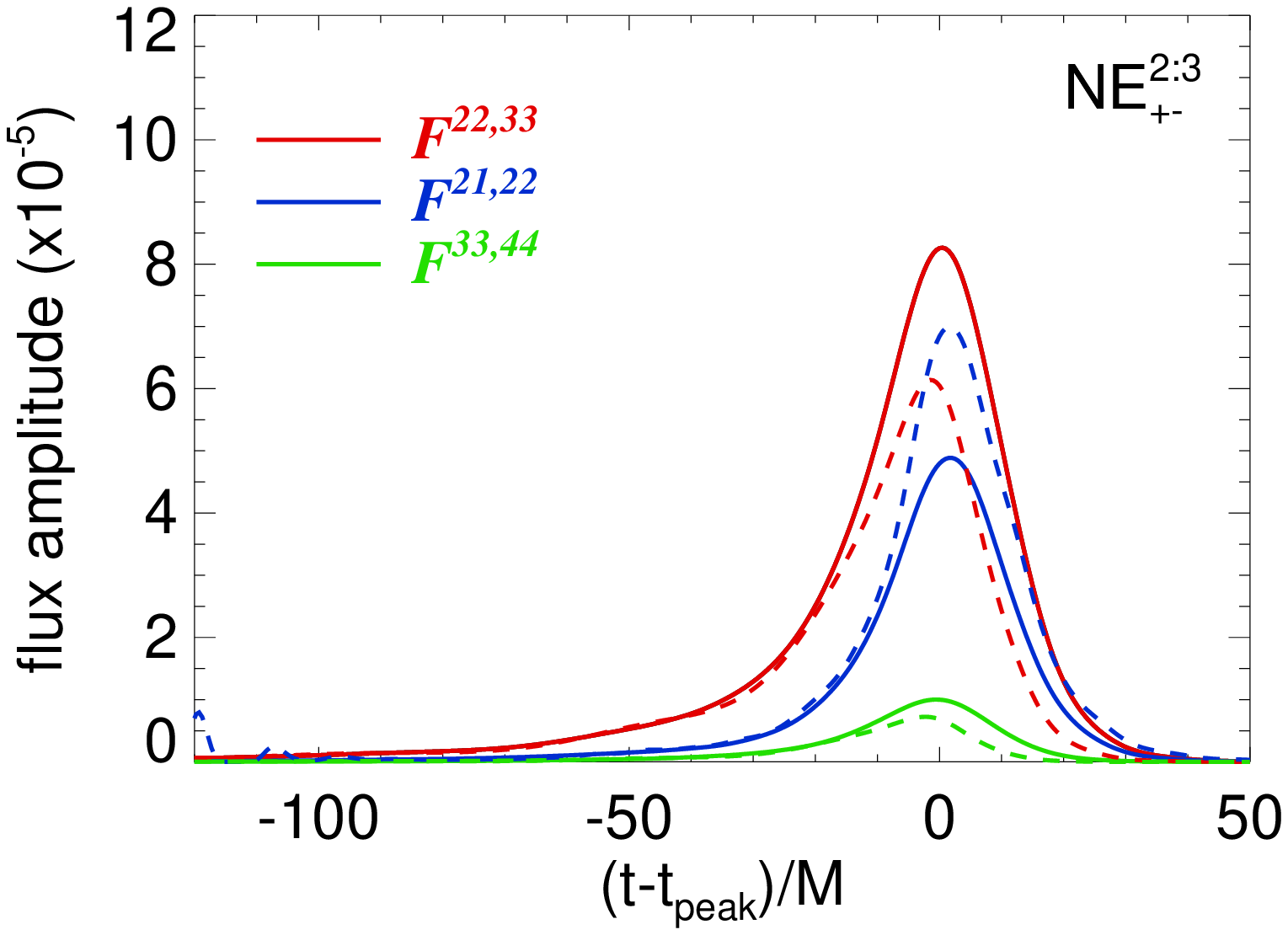}\\
\includegraphics[width=0.48\textwidth,clip=true]{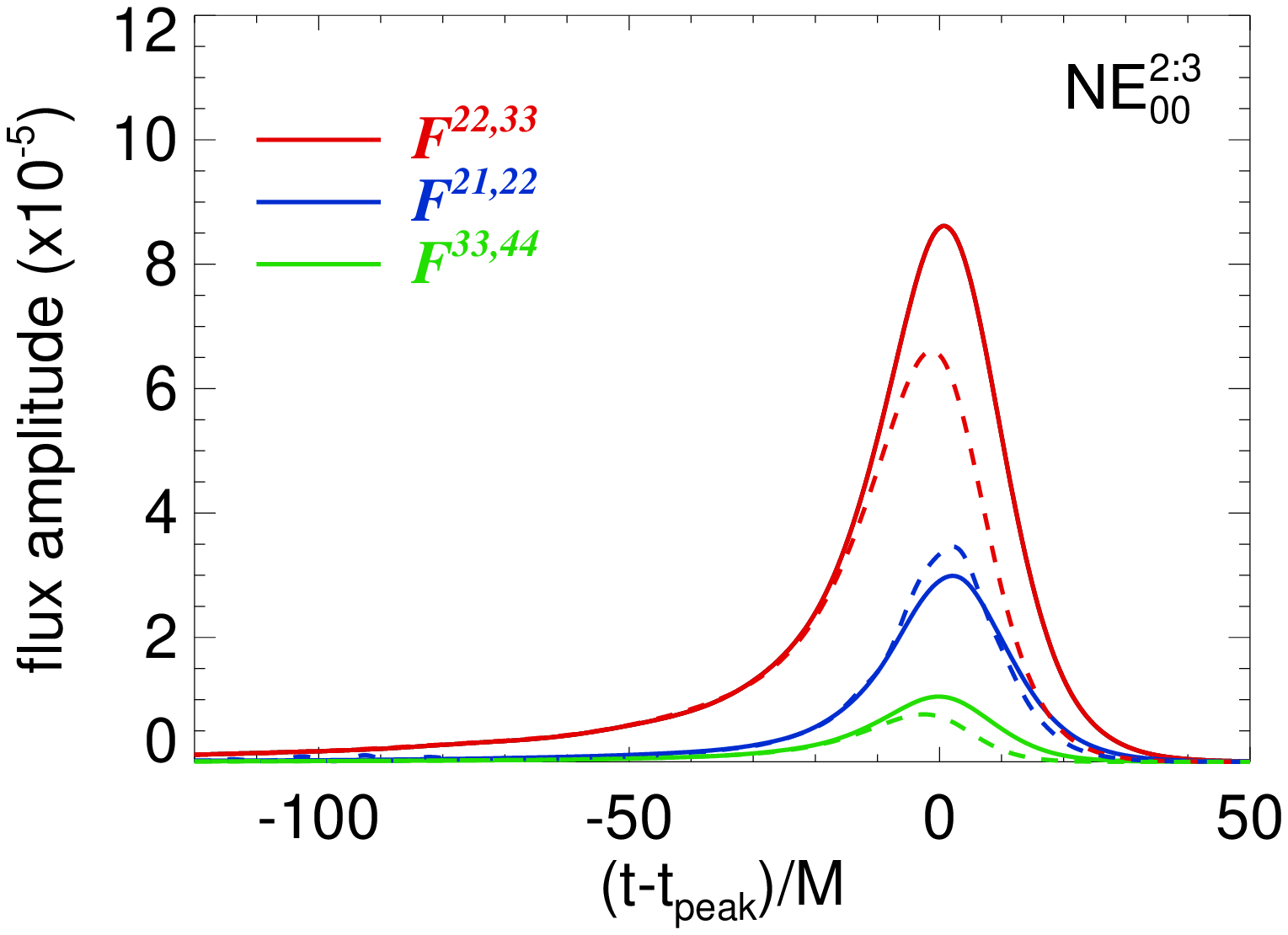}
\includegraphics[width=0.48\textwidth,clip=true]{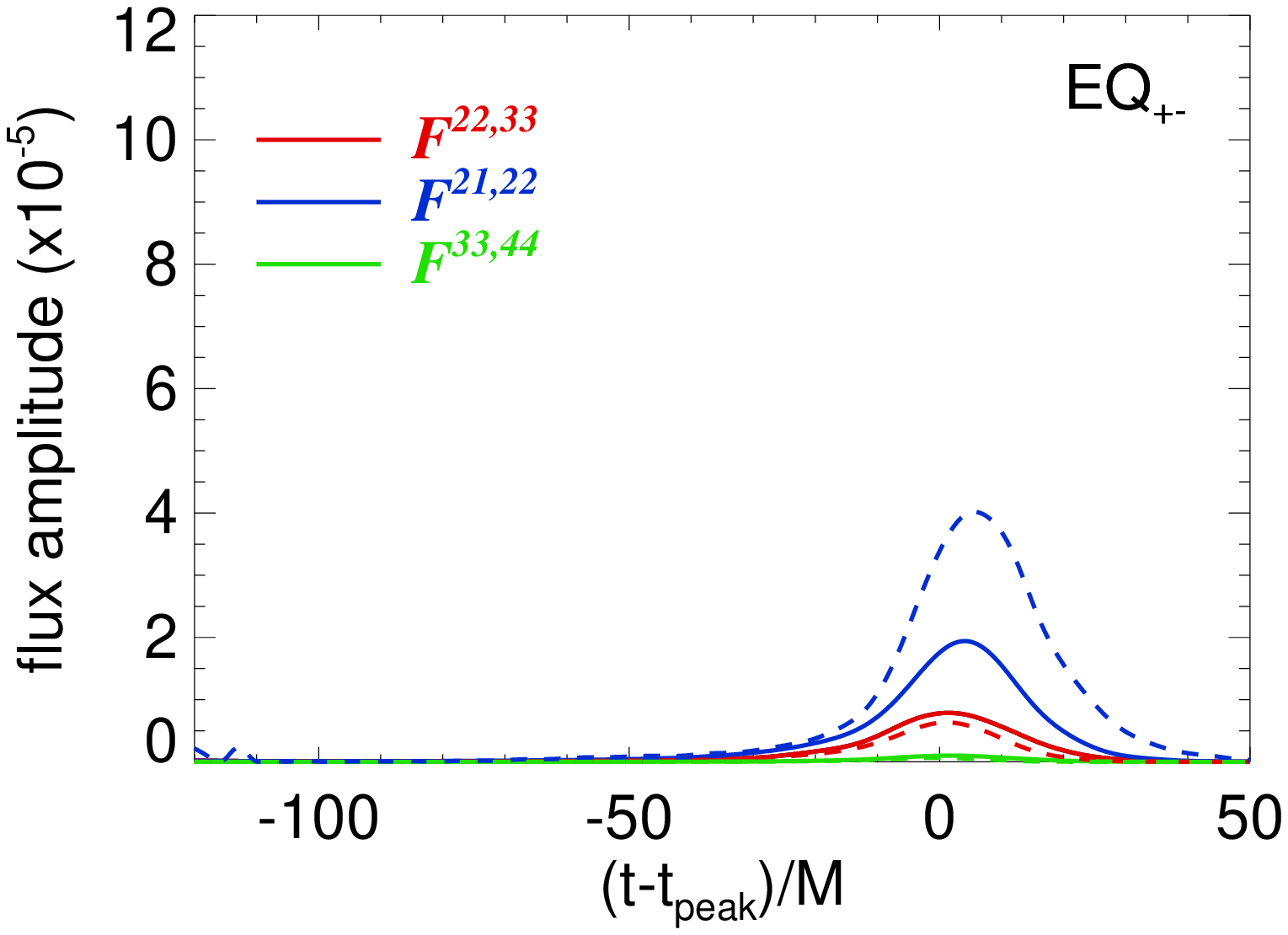}
\caption{\label{ne_modeamps} Relative amplitudes of the dominant 
multipole mode-pairs in the linear momentum flux. Also shown in the dashed
  curves are the eN model predictions for the flux amplitudes.
We denote with $t_{\rm peak}$ the time at which $I^{22}$ reaches its maximum.} 
\end{figure*} 

In all of the simulations considered here, the dimensionless spins are equal
($|a_1|/m_1=|a_2|/m_2$) and point in opposite directions, $\xi^z=0$, 
so for the leading-order terms in Eqn.~(\ref{flux_approx}) we are left
only with the 
modifications of $S^{21}$ and $I^{33}$, due to $\Delta^z$ and 
$\Sigma_{33}^z$, respectively. Then Eqs.~(\ref{I_nospin}) and
(\ref{spinorbit_moments}) give the linear momentum flux 
during the inspiral for each of the first three dominant terms in
Eq.~(\ref{flux_approx}):
\begin{widetext}
\begin{subequations}\label{newtonian_flux}
\bea 
F_{\rm insp}^{21,22} &=& \frac{16}{45}i \frac{\mu^2}{M}\, R^3\, \omega^6\, 
(2\delta m \, R^2 \omega +3\Delta^z)\, e^{i\phi}, 
\label{newtonian_flux1} \\
F_{\rm insp}^{22,33} &=& -\frac{36}{7}i \frac{\mu^2}{M}\,
R^5\,\omega^7\, (\delta m + \omega\, \Sigma_{33}^z)
\,e^{i\phi},
\label{newtonian_flux2} \\ 
F_{\rm insp}^{33,44} &=& -\frac{64}{7}i \frac{\mu^2}{M}\, (1-3\eta)\,
R^7\,\omega^9\, (\delta m + \omega\, \Sigma_{33}^z)\, e^{i\phi}
\label{newtonian_flux3}.
\eea
\end{subequations}
\end{widetext}
While these flux formulae contain terms of various orders in
$\omega$, we expect that the effective Newtonian scaling of $R$ 
ensures that we are including all relevant PN terms, at least in the
cases where the $\delta m$ terms dominate over the spin
corrections. When the spin terms begin to dominate, we find that it
becomes more difficult to use a single effective $R$ for all
modes. This can be seen in Fig.~\ref{r_eff_mp}, which plots $R_{\rm
  eff}$ as in Fig.~\ref{r_eff}, but for the NE$_{-+}^{2:3}$ run, where
the $\Delta^z$ and $\delta m$ terms in Eq.~(\ref{newtonian_flux1})
are comparable, making it difficult to derive a reasonable $R_{\rm
  eff}(S^{21})$. 

Even for non-spinning runs, in order to get reasonable agreement with
the NR data, we find that one 
must be careful towards the end of the inspiral to distinguish between
$\omega_{\rm D}^{I22}$ and
$\omega_{\rm D}^{S21}$ in Eq.~(\ref{newtonian_flux1}):
\begin{equation}
F_{\rm insp}^{21,22} \propto (\mu^2/M) R^3\, (\omega_{\rm D}^{I22})^3\,
(\omega_{\rm D}^{S21})^3\, (2\delta m \, R^2 \omega_{\rm D}^{S21}
+3\Delta^z)\, .
\end{equation}

The amplitudes of these fluxes are plotted in Fig.~\ref{ne_modeamps} 
for the four runs NE$_{-+}^{2:3}$, NE$_{+-}^{2:3}$, NE$_{00}^{2:3}$,
and EQ$_{+-}$. As seen in Table~\ref{table:idparams}, the
NE$_{-+}^{2:3}$ run has $\Delta^z = 0.2M^2$, while the
NE$_{+-}^{2:3}$ run has $\Delta^z = -0.2M^2$, respectively adding 
destructively and constructively with the $\delta m$ term in
Eq.~(\ref{newtonian_flux1}). This difference is clearly seen in the
blue curves in the top two panels of Fig.~\ref{ne_modeamps}. Also notable
in these plots is the somewhat smaller difference in the amplitudes of
$F^{22,33}$, due to a similar
effect from the constructive/destructive additions of $\delta m$ and
$\Sigma_{33}^z$ in Eq.~(\ref{newtonian_flux2}).
As we see in Fig.~\ref{ne_modeamps}, NE$_{00}^{2:3}$ appears to be the
average of NE$_{+-}^{2:3}$ and NE$_{-+}^{2:3}$, while the flux from
EQ$_{+-}$ is strongly suppressed due to the $\delta m=0$ terms in
Eq.~(\ref{newtonian_flux}), leaving only the flux from the terms
proportional to $\Delta^z=-0.2M^2$ and $\Sigma_{33}^z=0.075$. However,
as noted above, when the spin terms dominate the flux, as in the
case of equal-mass BHs, the eN model with a single $R_{\rm eff}$
begins to break down. Yet even in this situation,
Eqs.~(\ref{newtonian_flux1})-(\ref{newtonian_flux3}) still have
qualitative (if not quantitative) predictive value, including the
relative phases between the different mode-pair fluxes during the
inspiral. 

In each panel of Fig.~\ref{ne_modeamps}, we also plot
with dashed lines the eN prediction for the various flux
amplitudes. In almost all cases, the eN flux is quite close to the
NR results up to about $10M$ before $t_{\rm peak}$, when the
eN model begins to break down, especially for the spinning runs.
The amplitude differences near the peaks are comparable to those seen in
Fig.~\ref{compare_ring} for the NE$_{00}^{2:3}$ run. The notable
exception is the $F^{21,22}$ flux from the NE$_{-+}^{2:3}$ and
EQ$_{+-}$ runs, where the spin terms dominate over the $\delta m$
terms.

\subsection{Transition to ringdown and the de-phasing of the multipole
modes}\label{transition}

Since the flux vectors defined by Eq.~(\ref{newtonian_flux})
will not generally be co-linear, to understand the time evolution of the recoil velocity,
we must first understand the phase relations between the different
modes. From Eqs.~(\ref{I_nospin}), (\ref{spinorbit_moments}), and
(\ref{newtonian_flux}),
we see that during the inspiral phase, the individual moments and the resulting
flux vectors evolve according to a single orbital phase $\phi$, with
$F_{\rm insp}^{21,22}$ pointing 
in the opposite direction to $F_{\rm insp}^{22,33}$ and 
$F_{\rm insp}^{33,44}$. However, as we can see from
Fig.~\ref{omega_modes}, as the binary evolves from inspiral to 
RD, the frequency (and thus phase) of the $S^{21}$ mode decouples 
from the other dominant modes. Upon closer inspection, we find that 
even the $I^{22}, I^{33}$ and $I^{44}$ modes deviate from each other 
enough to undergo a significant phase shift at the inspiral-RD
transition. 

To quantify these effects, we define the following phase differences:
\begin{subequations}
\bea
\label{phase_shifts1}
\cos\psi^{2-3} &=& \mathbf{\hat{F}}_{\rm insp}^{21,22} \cdot 
\mathbf{\hat{F}}_{\rm insp}^{22,33}, \\
\label{phase_shifts2}
\cos\psi^{2-4} &=& \hat{\mathbf{F}}_{\rm insp}^{21,22}\cdot
\mathbf{\hat{F}}_{\rm insp}^{33,44}, \\
\label{phase_shifts3}
\cos\psi^{3-4} &=& \hat{\mathbf{F}}_{\rm insp}^{22,33}\cdot
\mathbf{\hat{F}}_{\rm insp}^{33,44}.
\eea
\end{subequations}
Here we use the notation $\psi^{m-m'}$ to describe
the phase difference between two complex flux vectors, where $m$ and
$m'$ correspond to the {\it larger} $m$-values of each mode pair
that makes up the flux. These definitions are valid throughout the
inspiral, merger, and ringdown phases.
In the inspiral phase, we can see that for the unequal-mass runs where
$\delta m$ dominates with respect to the spin terms in 
Eqs.~(\ref{newtonian_flux1})--(\ref{newtonian_flux3}), we have
\beq\label{phase_shifts_insp}
\cos\psi_{\rm insp}^{2-3}=\cos\psi_{\rm insp}^{2-4}=-1\,, \quad \cos\psi_{\rm insp}^{3-4}=1\,.
\eeq
For the EQ$_{+-}$ run with $\delta m=0$, Eq.~(\ref{newtonian_flux})
predicts that all phases have $\cos\psi_{\rm insp}=1$ during
the inspiral (as shown in Table~\ref{table:idparams}, $\Delta^z$ and
$\Sigma^z_{33}$ have opposite signs, so all the flux vectors in
Eq.~(\ref{newtonian_flux}) are parallel).
During the RD phase, using Eq.~(\ref{RD}), we can approximate the flux
vectors and phase evolution in terms of the fundamental QNM
frequencies $\sigma_{\ell m0}$:
\begin{widetext}
\begin{subequations}\label{flux_RD}
\bea
\label{flux_RD1}
F_{\rm RD}^{21,22} &\simeq& F_{\rm match}^{21,22}
\exp[-i(\sigma_{210}-\sigma_{220}^*)(t-t_{\rm match})]\, ,\\
\label{flux_RD2}
F_{\rm RD}^{22,33} &\simeq& F_{\rm match}^{22,33}
\exp[-i(\sigma_{220}-\sigma_{330}^*)(t-t_{\rm match})]\, ,\\
\label{flux_RD3} 
F_{\rm RD}^{33,44} &\simeq& F_{\rm match}^{33,44}
\exp[-i(\sigma_{330}-\sigma_{440}^*)(t-t_{\rm match})]\, ,
\eea
\end{subequations}
where the $F^{\ell m,\ell' m'}_{\rm match}$ fluxes include complex
phase information at the matching point. Taking the phase differences
between these RD modes gives
\begin{subequations}\label{phase_shifts_RD}
\bea
\label{phase_shifts_RD1}
\cos\psi_{\rm RD}^{2-3} &\simeq&
\cos[(\omega_{210}-2\omega_{220}+\omega_{330})(t-t_{\rm
    match})+\Phi_{\rm match}^{2-3}]\,, \\ 
\label{phase_shifts_RD2}
\cos\psi_{\rm RD}^{2-4} &\simeq&
\cos[(\omega_{210}-\omega_{220}-\omega_{330}+\omega_{440})(t-t_{\rm
    match})+\Phi_{\rm match}^{2-4}]\,,\\ 
\label{phase_shifts_RD3} 
\cos\psi_{\rm RD}^{3-4} &\simeq&
\cos[(\omega_{220}-2\omega_{330}+\omega_{440})(t-t_{\rm
    match})+\Phi_{\rm match}^{3-4}]\,. 
\eea
\end{subequations}
\end{widetext}
Here $\Phi_{\rm match}$ is a phase offset determined at the transition
from inspiral to ringdown.
Quite interestingly, we find that for the range of final BH spin parameters
$0.5\le a_{\rm f}/M_{\rm f} \le 0.8$, the linear combinations of frequencies 
in Eqs.~(\ref{phase_shifts_RD1})--(\ref{phase_shifts_RD3}) vary by
less than $\sim 5\%$. Thus, if we compute the above expressions 
for the $\omega_{lm0}$ corresponding to $a_{\rm f}/M_{\rm f}=0.7$, we
have \cite{BCW}
\begin{subequations}\label{phase_shifts_RDe}
\bea
\label{phase_shifts_RD1e}
\cos\psi_{\rm RD}^{2-3} &\simeq & \cos \left [
  \frac{0.23}{M_{\rm f}} (t- t_{\rm match}) +
\Phi_{\rm match}^{2-3} \right ]\,,\\
\label{phase_shifts_RD2e}
\cos\psi_{\rm RD}^{2-4} &\simeq &\cos\left [
  \frac{0.22}{M_{\rm f}}(t-t_{\rm match}) +\Phi_{\rm match}^{2-4}
  \right ]\,,\\ 
\label{phase_shifts_RD3e}
\cos\psi_{\rm RD}^{3-4} &\simeq &\cos \left [
  \frac{0.012}{M_{\rm f}}(t-t_{\rm match})+\Phi_{\rm match}^{3-4}
  \right ].
\eea
\end{subequations}
\begin{figure*}
\includegraphics[width=0.48\textwidth,clip=true]{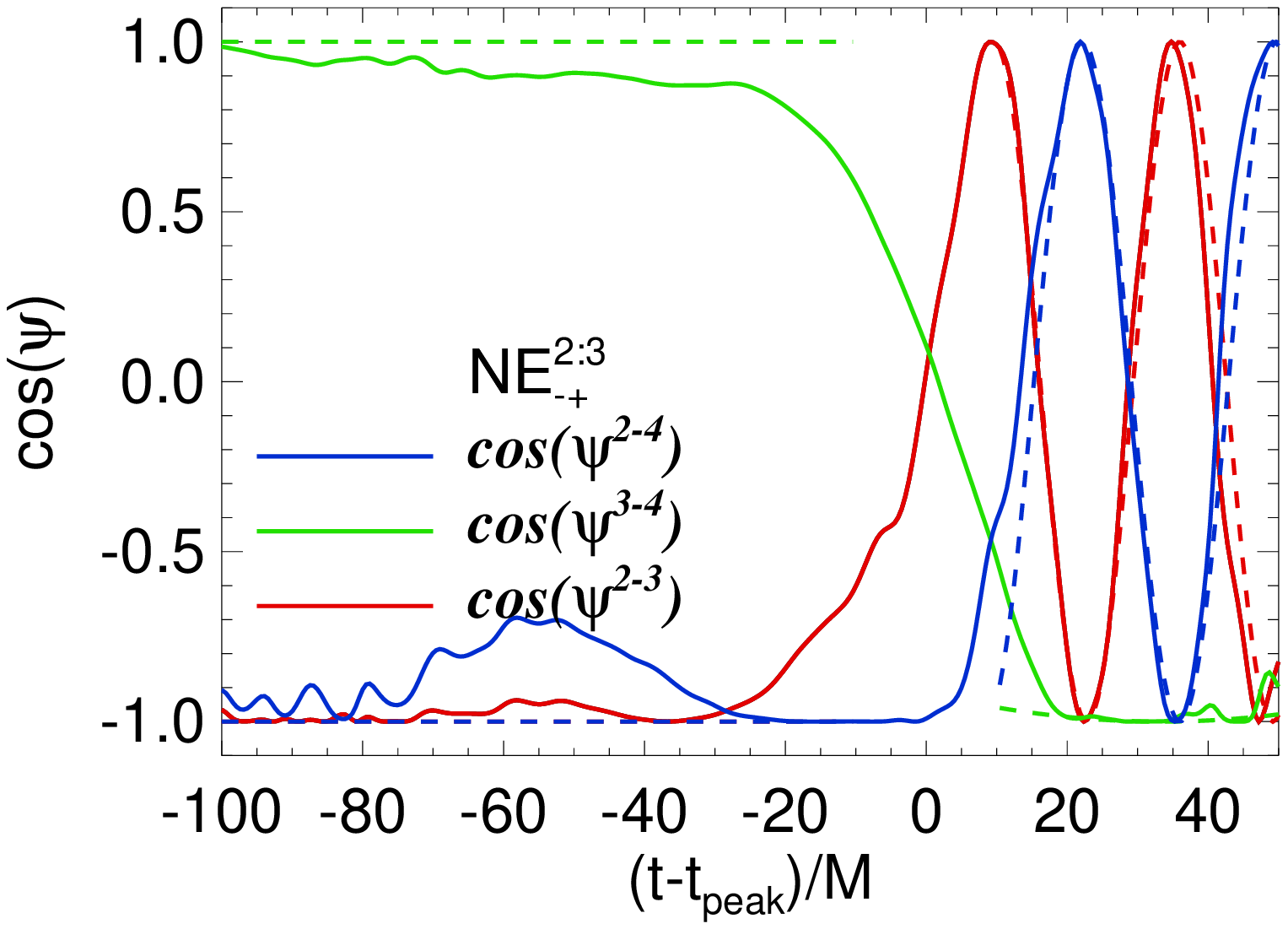}
\includegraphics[width=0.48\textwidth,clip=true]{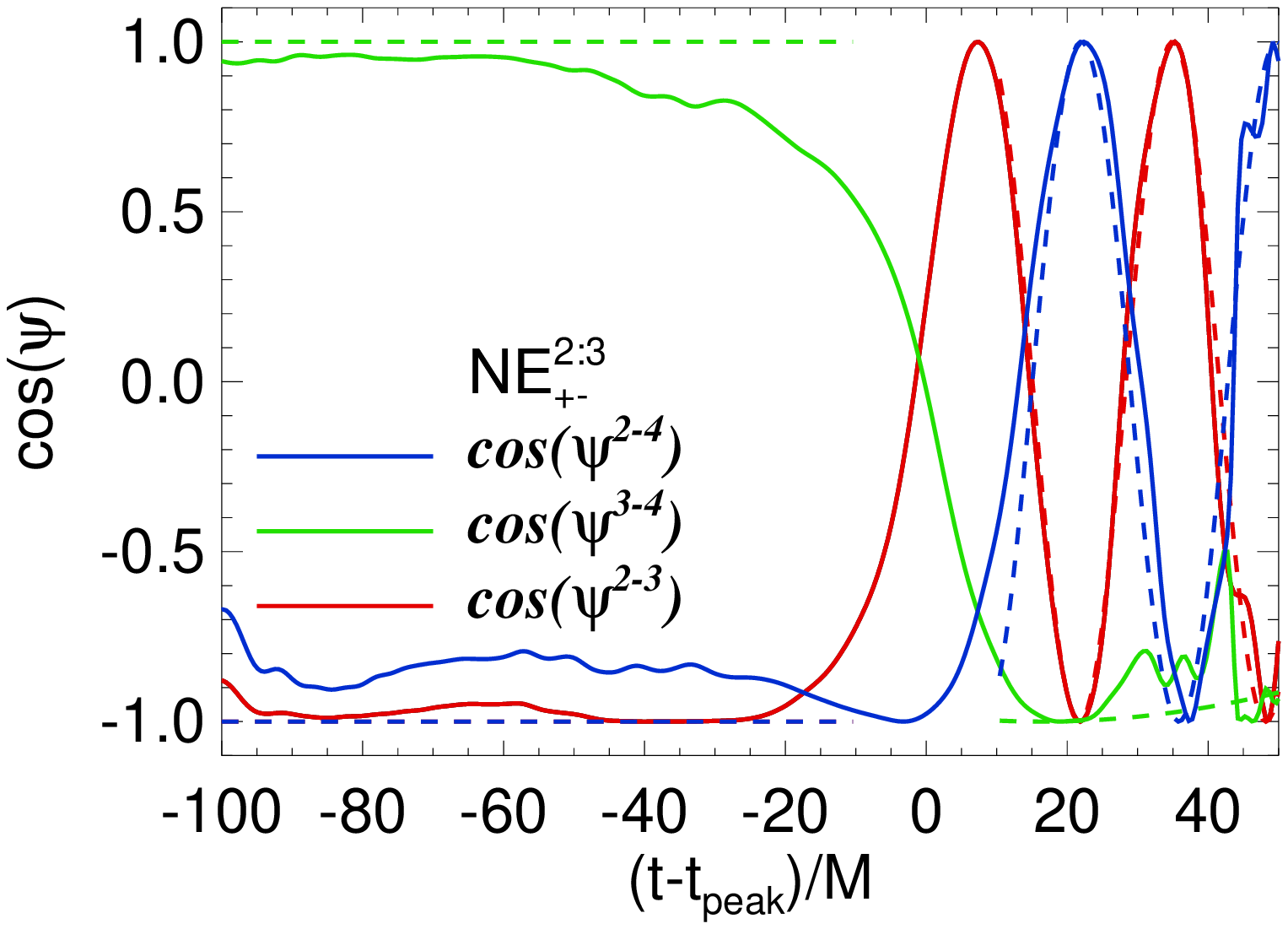}\\
\includegraphics[width=0.48\textwidth,clip=true]{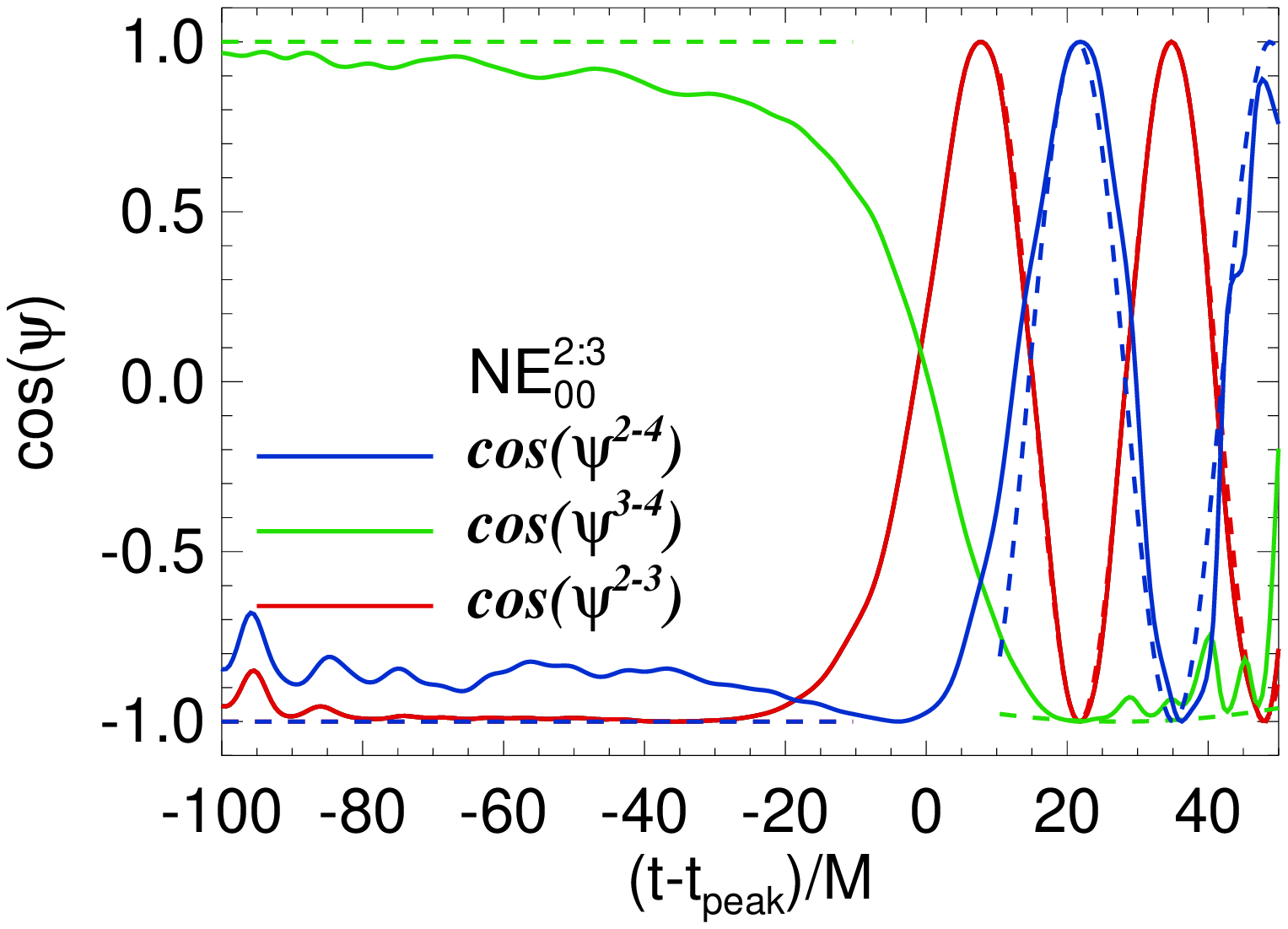}
\includegraphics[width=0.48\textwidth,clip=true]{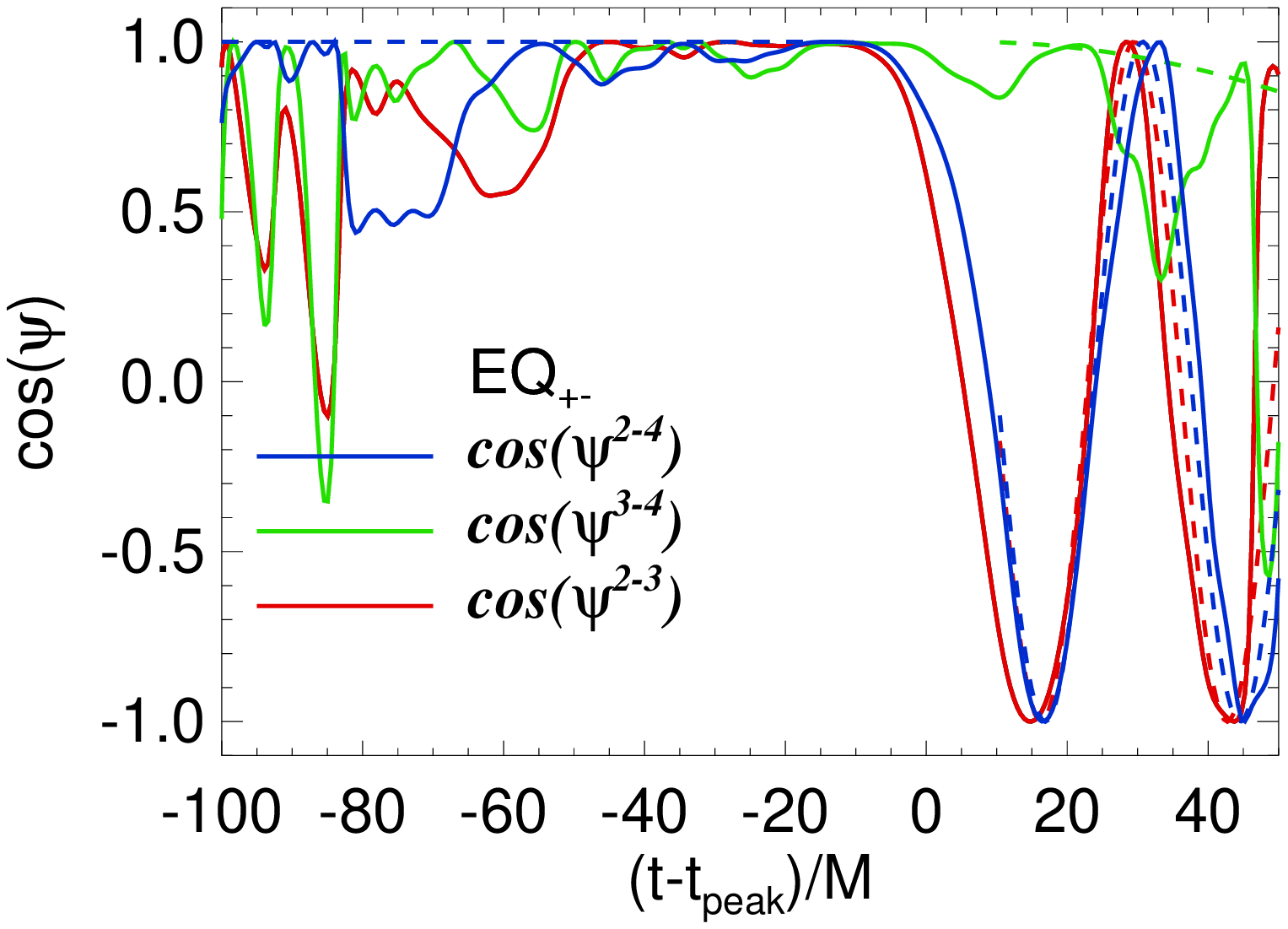}
\caption{\label{ne_modephase1} Phase differences between different
  mode-pair flux vectors, as defined by Eqs.~(\ref{phase_shifts1})--(\ref{phase_shifts3}). 
The data refer to the NE$_{+-}^{2:3}$ (upper left panel),
  NE$_{-+}^{2:3}$ (upper right panel), NE$_{00}^{2:3}$ (lower left
  panel), and EQ$_{+-}$ (lower right panel) runs. The
  dashed curves are the eN model predictions of
  Eqs.~(\ref{phase_shifts_insp}),(\ref{phase_shifts_RDe}).
We denote with $t_{\rm peak}$ the time at which $I^{22}$ reaches its maximum.}
\end{figure*} 

Even more intriguing, we find that for the unequal-mass simulations
described above, the phase relations during the inspiral and RD 
are almost identical, regardless of spin orientations. This can be
seen clearly in Fig.~\ref{ne_modephase1}, which plots $\cos\psi$ 
during inspiral, merger and RD for the different runs. 
The colinearity of the flux vectors is clear during the inspiral
phase, and the sinusoidal oscillations of the phases during RD agree
well with the analytic predictions (plotted in dashed
  curves in Fig.~\ref{ne_modephase1}). Since the analytic models are most reliable
during the inspiral and RD phases (but have more difficulty tracing
the merger portion), we omit in Fig.~\ref{ne_modephase1} the
transition region of $-10M\le (t-t_{\rm peak})\le 10 M$. The analytic phase
relations during inspiral are determined by
Eq.~(\ref{phase_shifts_insp}) and during ringdown by
Eqs.~(\ref{phase_shifts_RD1e})--(\ref{phase_shifts_RD3e}). Here we use a
$t_{\rm match}$ (and corresponding $\Phi_{\rm match}$) about $20M$ after $t_{\rm peak}$ to ensure that the
multipole moments are truly dominated by the fundamental QNMs, and thus
Eqs.~(\ref{phase_shifts_RD1e})--(\ref{phase_shifts_RD3e}) are valid. 
Note that the phase differences for EQ$_{+-}$ are particularly noisy
since the amplitude of the $I^{33}$ moment is zero to leading order, and
thus it is more difficult to extract a clear phase for that mode.

The feature that is most difficult to explain from an analytic model
alone (and is thus omitted from the eN curves in Fig.~\ref{ne_modephase1})
is the roughly 180-degree jump in phase between $F_{\rm insp}^{22,33}$ and 
$F_{\rm insp}^{33,44}$, beginning around $20M$ before the
peak. This appears to be a feature in all the
unequal-mass runs examined, but preliminary results suggest that is 
less significant (i.e., a smaller phase shift) for more extreme-mass
ratio systems, as we shall discuss in Appendix~\ref{app}.
We are not able to explain it with the additional RD
overtone modes described in Sec.~\ref{matching_RD}, but using
slightly different RD matching points for the different multipoles
may help explain the issue.

\subsection{The anti-kick}\label{antikick}

These flux amplitudes and phase relations can now be used to
understand the amplitude of the kick and anti-kick, by which we mean
the difference between the peak and the final recoil velocities (see
Fig.~\ref{intro_fig} for an example). Throughout the
inspiral phase, the amplitude and rotational frequency of the flux
vectors in Eq.~(\ref{newtonian_flux}) are monotonically increasing,
giving the familiar outward-spiraling trajectory for the velocity
vector. Then, in the RD phase, the dominant frequencies are nearly
constant while the amplitudes decay exponentially for each mode,
giving an inward-spiral that decays like a damped harmonic oscillator around
the final asymptotic recoil velocity. 

\begin{figure*}
\includegraphics[width=0.48\textwidth,clip=true]{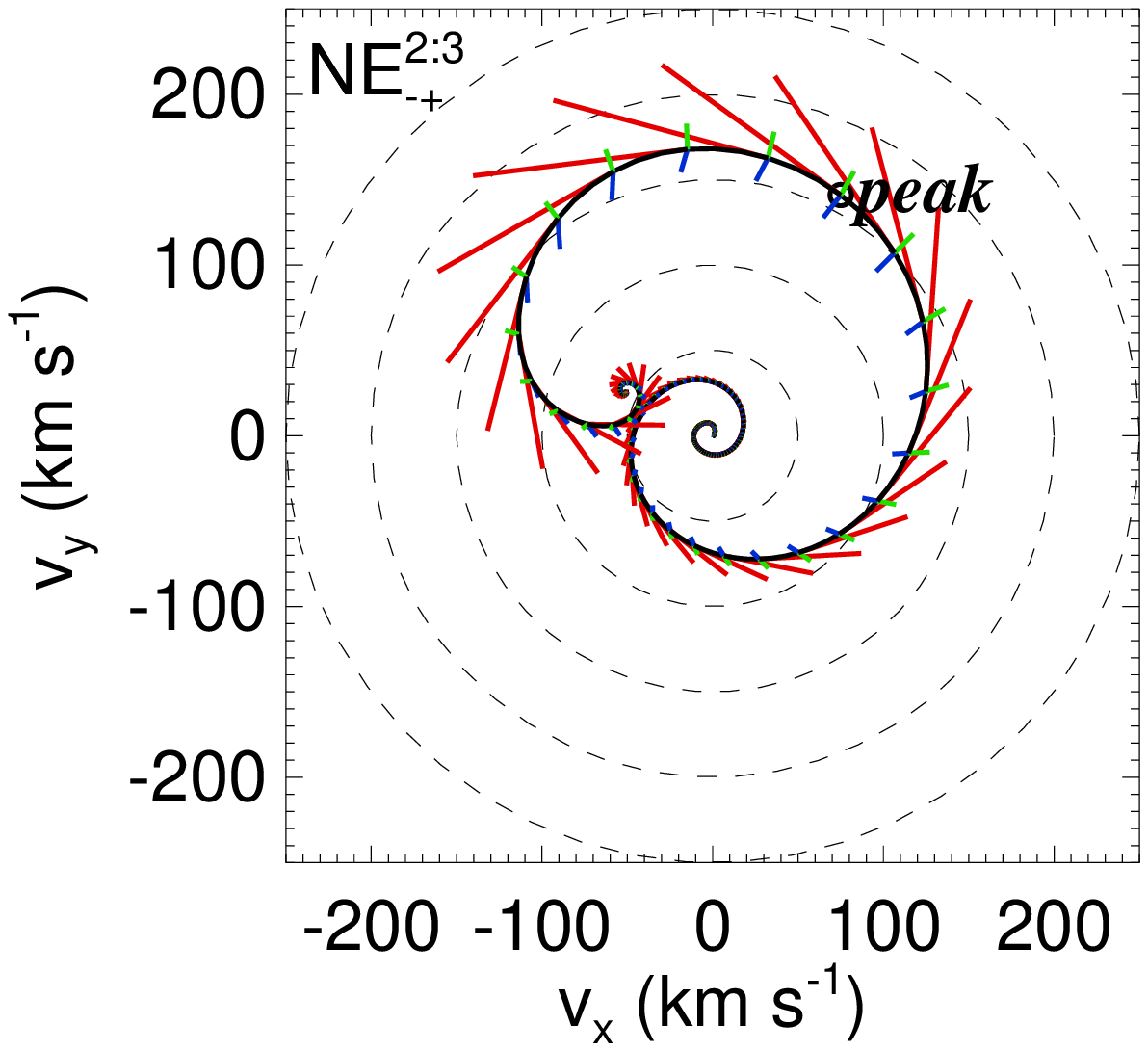}
\includegraphics[width=0.48\textwidth,clip=true]{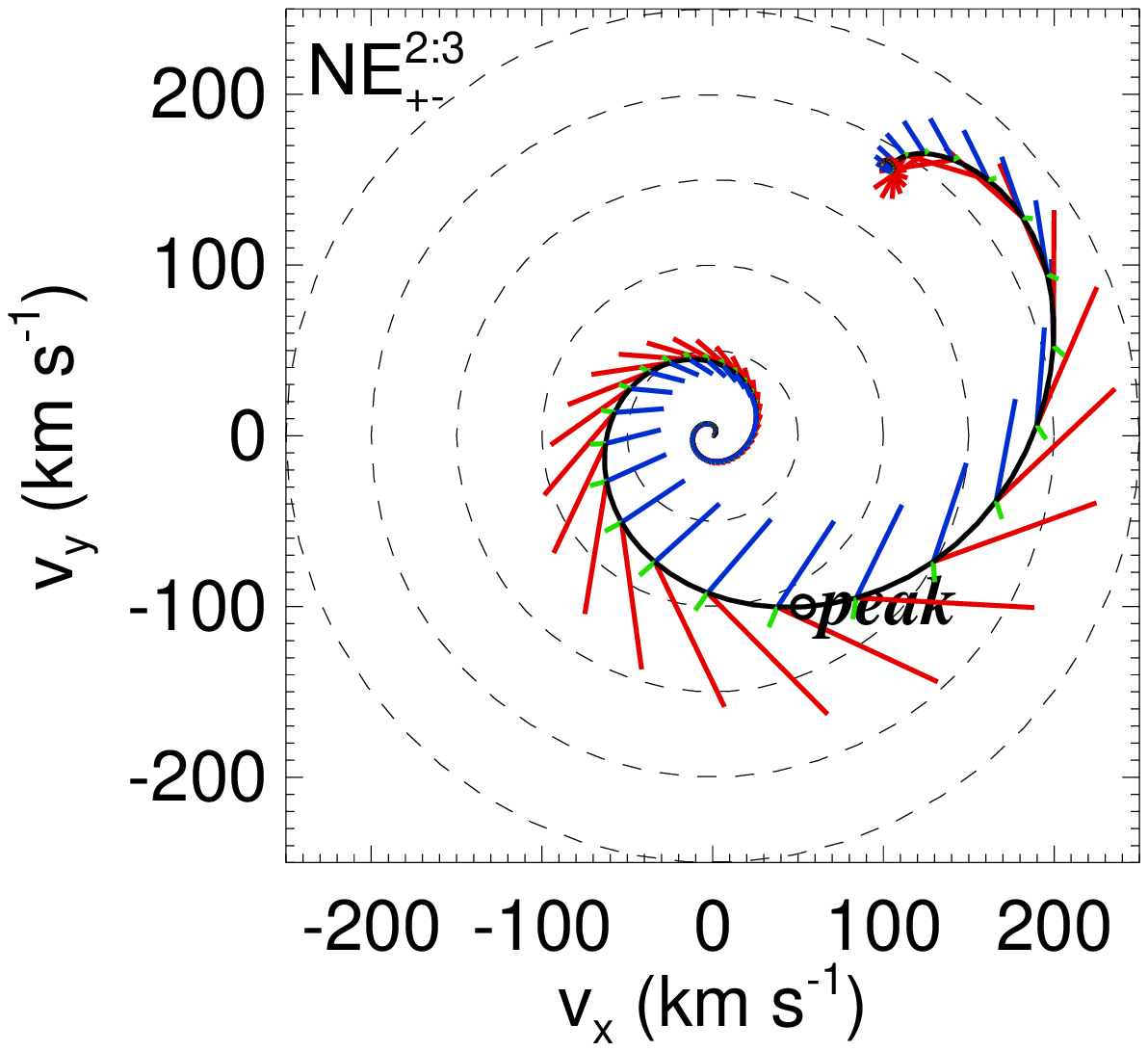}\\
\includegraphics[width=0.48\textwidth,clip=true]{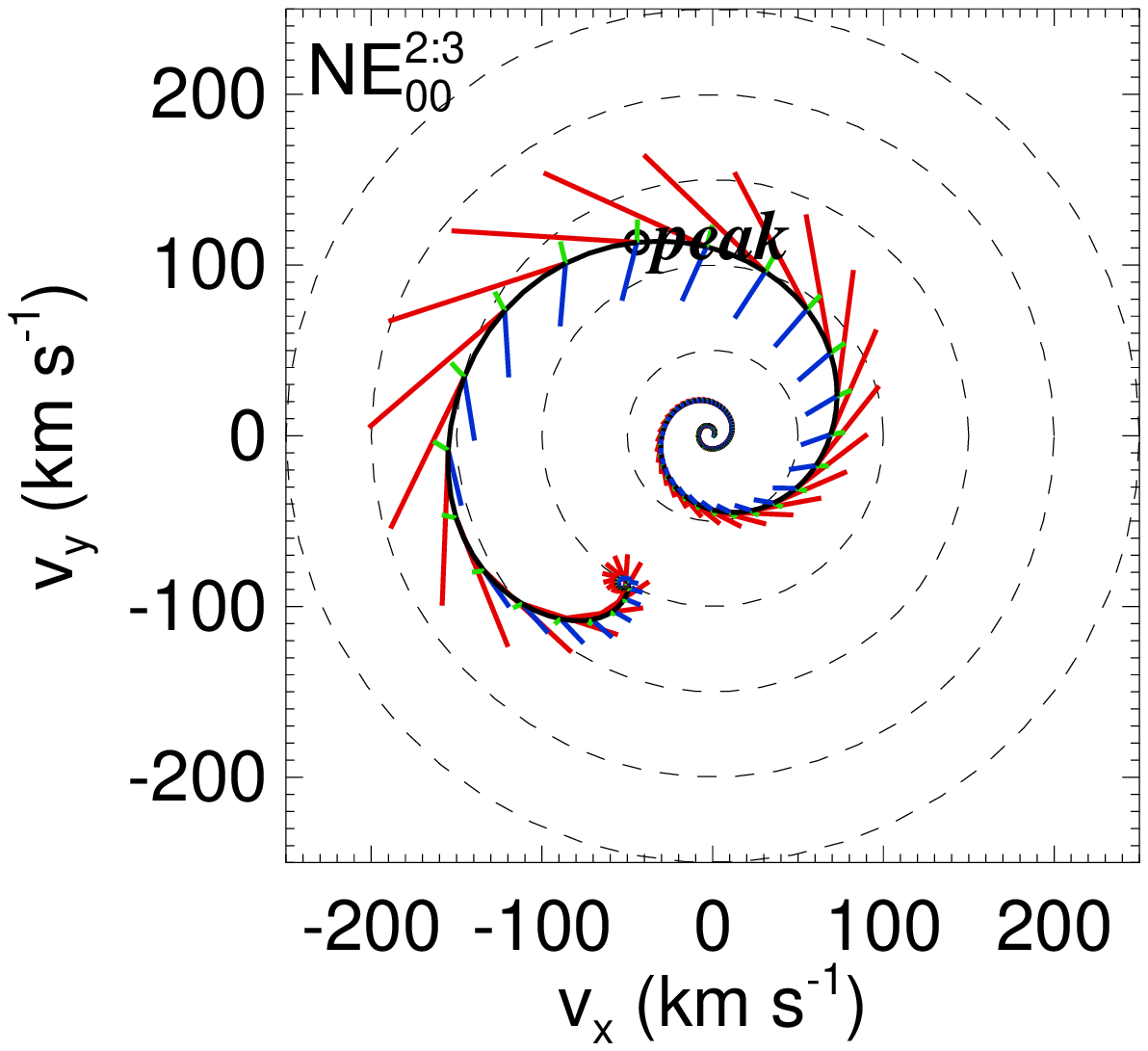}
\includegraphics[width=0.48\textwidth,clip=true]{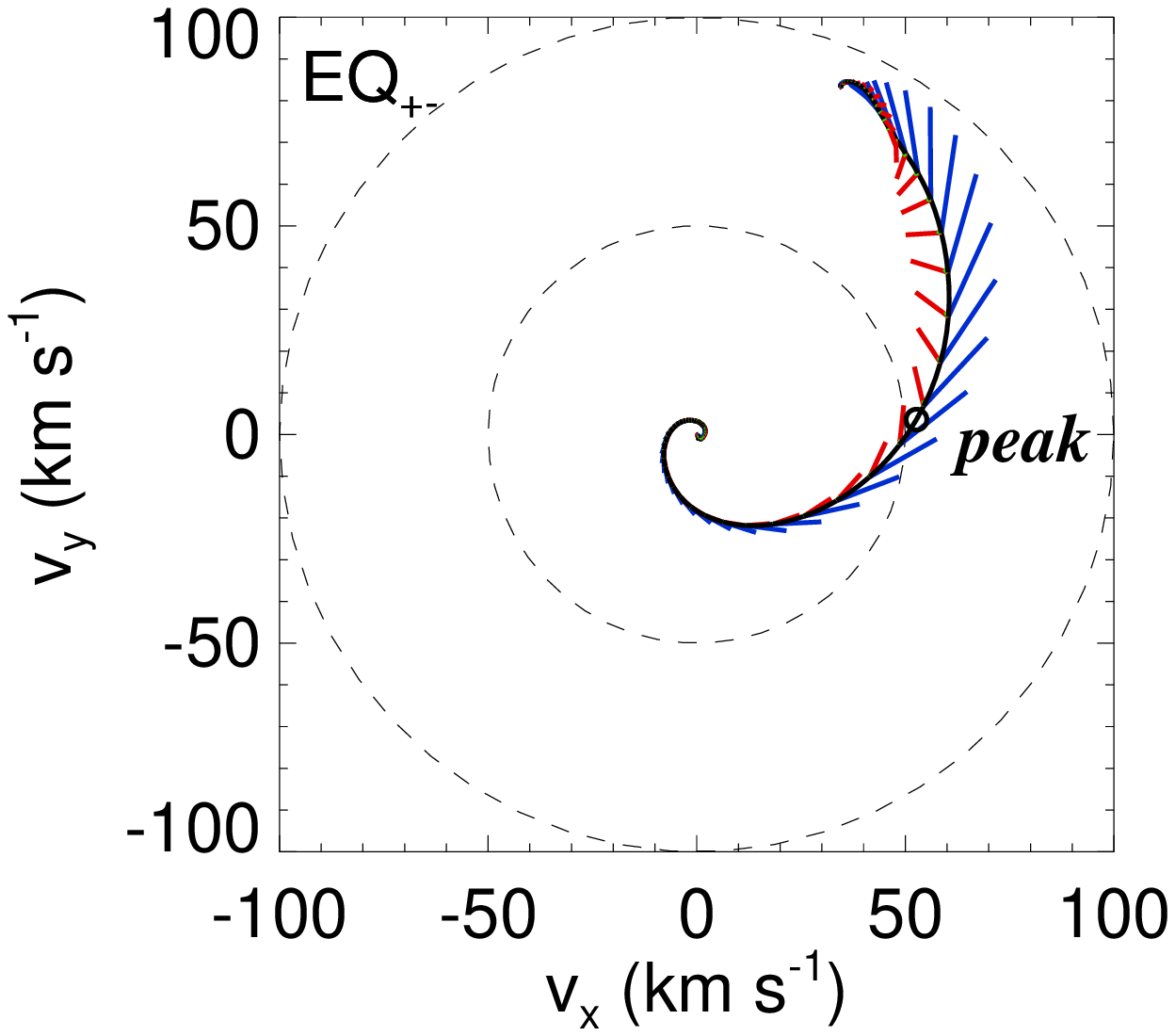}
\caption{\label{ne_spiralmodes1} The recoil velocity vector evolving in 
the $v_x$--$v_y$ plane ({\it black solid curve}), along with the flux 
vectors due to the three mode pairs at each time interval along the
velocity trajectory (same color
scheme as Fig.~\ref{ne_modeamps}). 
The data refer to the NE$_{+-}^{2:3}$ ({\it upper left panel}),
  NE$_{-+}^{2:3}$ ({\it upper right panel}), NE$_{00}^{2:3}$ ({\it lower left
  panel}), and EQ$_{+-}$ ({\it lower right panel}) runs.
  We denote with the label {\it peak} the time at which $I^{22}$ reaches
  its maximum.} 
\end{figure*}

These trajectories in velocity
space can be seen in Fig.~\ref{ne_spiralmodes1}, along with the
instantaneous flux vectors from the competing mode-pairs. Clearly,
even small changes in the mass ratios and spins orientations of the 
BHs can give a rather diverse selection of velocity
trajectories. Note in particular the difference between the
NE$_{-+}^{2:3}$ run, dominated by the $F^{22,33}$ flux and a large
anti-kick, and the EQ$_{+-}$ run, which in contrast is dominated by
the $F^{21,22}$ flux. We find that the EQ$_{+-}$ run has {\it no}
anti-kick, which can be explained by the slowly rotating flux vector
that does not spiral back inwards, but rather drifts off slowly
towards infinity during the ringdown. The difference between these
two runs can be explained entirely by examining the real part of
their fundamental QNM frequencies $\sigma_{\ell m 0}$, which in
turn determine the rotation rates of the flux vectors in
Eq.~(\ref{flux_RD}): EQ$_{+-}$ is dominated by
$\omega_{220}-\omega_{210}=0.08/M_{\rm f}$, a much 
slower frequency than $\omega_{330}-\omega_{220}=0.31/M_{\rm
  f}$, which causes the rapid inward-spiral of the NE$_{-+}^{2:3}$
run.

To calculate the recoil velocity, we must integrate the linear
momentum flux vectors in time. (For the initial velocity vector, we
integrate the post-Newtonian approximation for the momentum flux from
$t=-\infty$ to the beginning of the numerical simulation \cite{BQW}. This
effectively sets the centers of the spiral curves in
Fig.~\ref{ne_spiralmodes1} to correspond to the origin in velocity
space.) We can get a reasonable analytic
approximation by using Eqs.~(\ref{newtonian_flux}) and (\ref{flux_RD})
for the inspiral and RD phases, respectively. In the adiabatic
inspiral, the complex recoil velocity $v = v_x + iv_y$ can be written as
\begin{widetext}
\beq\label{v_insp}
v_{\rm insp}= \int_{-\infty}^{t_{\rm match}} F(t') \, dt' \simeq 
\frac{1}{i\omega_{\rm match}}\,F_{\rm match},  
\eeq
while for the RD portion we have

\bse\label{v_RD}
\bea
v_{\rm RD}(t) = \int_{t_{\rm match}}^{t} F(t')\,dt' &\simeq&
\sum_{\ell m,\ell'm'}\, 
\frac{i\, F^{\ell m,\ell'm'}_{\rm match}}{\sigma_{\ell
    m0}-\sigma_{\ell'm'0}^{*}}\, 
\left[e^{-i(\sigma_{\ell m 0}-\sigma_{\ell' m' 0}^*)(t-t_{\rm
      match})}-1\right], \\
v_{\rm f} \equiv v_{\rm RD}(t\to \infty) &\simeq &
\sum_{\ell m,\ell'm'}\, 
\frac{-i}{\sigma_{\ell m0}-\sigma_{\ell'm'0}^{*}}\,F^{\ell
  m,\ell'm'}_{\rm match},
\eea
\ese
summing the contributions from each pair of modes $(\ell m,\ell'm')$. Then the total velocity
in each of the dominant mode pairs is given by
\begin{subequations}\label{vtot_modes}
\bea 
\label{vtot_modes1}
\int F^{21,22}(t')\, dt' &=& \frac{16}{45}\, \frac{\mu^2}{M} \,R_{\rm match}^3\, \omega_{\rm match}^5
(2\delta m \, R_{\rm match}^2 \omega_{\rm match}+3\Delta^z) 
\left[1-\frac{i\omega_{\rm match}\, e^{i\phi^{21,22}_{\rm match}}}{\sigma_{210}-\sigma_{220}^{*}}
\right], \\
\label{vtot_modes2}
\int F^{22,33}(t')\, dt' &=& -\frac{36}{7}\, \frac{\mu^2}{M}\, R_{\rm
  match}^5\,\omega_{\rm match}^6 (\delta m + \omega_{\rm match} \Sigma_{33}^z)
\left[1-\frac{i\omega_{\rm match}\, e^{i\phi^{22,33}_{\rm match}}}{\sigma_{220}-\sigma_{330}^{*}}
\right], \\
\label{vtot_modes3}
\int F^{33,44}(t')\, dt' &=& -\frac{64}{7}\, \frac{\mu^2}{M}\, (1-3\eta)\,
R_{\rm match}^7\,\omega_{\rm match}^8\, (\delta m + \omega_{\rm match}
\Sigma_{33}^z) \left[1-\frac{i\omega_{\rm match}\,
e^{i\phi^{33,44}_{\rm match}}}
{\sigma_{330}-\sigma_{440}^{*}}\right]\,.
\eea
\end{subequations}
\end{widetext}
The phase $\phi^{21,22}_{\rm match}$ is defined as the angle made between the flux
vector $F^{21,22}$ and the velocity vector $\mathbf{v}$ at
the beginning of the ringdown (with other phases $\phi^{22,33}_{\rm match}$,
$\phi^{33,44}_{\rm match}$ defined analogously). Because of the anomalous phase
shifts and departure from adiabaticity at the transition from inspiral
to ringdown,
these angles are difficult to predict with an
independent analytic model, but can be calculated easily from plots
like Fig.~\ref{ne_spiralmodes1}. However, the accuracy of
Eq.~(\ref{vtot_modes}) is limited both by the adiabaticity condition of
Eq.~(\ref{v_insp}) as well as the accuracy of the spin-orbit
corrections to the eN model (see Fig.~\ref{r_eff_mp}). Therefore, in analyzing the anti-kick in
terms of RD modes, we find it more useful simply to integrate
Eq.~(\ref{v_insp}) directly from the numerical data during the
inspiral, and then attach the fundamental QNM terms from
Eq.~(\ref{v_RD}) at the matching point $t_{\rm match} = t_{\rm peak}$.

\begin{figure*}
\includegraphics[width=0.48\textwidth,clip=true]{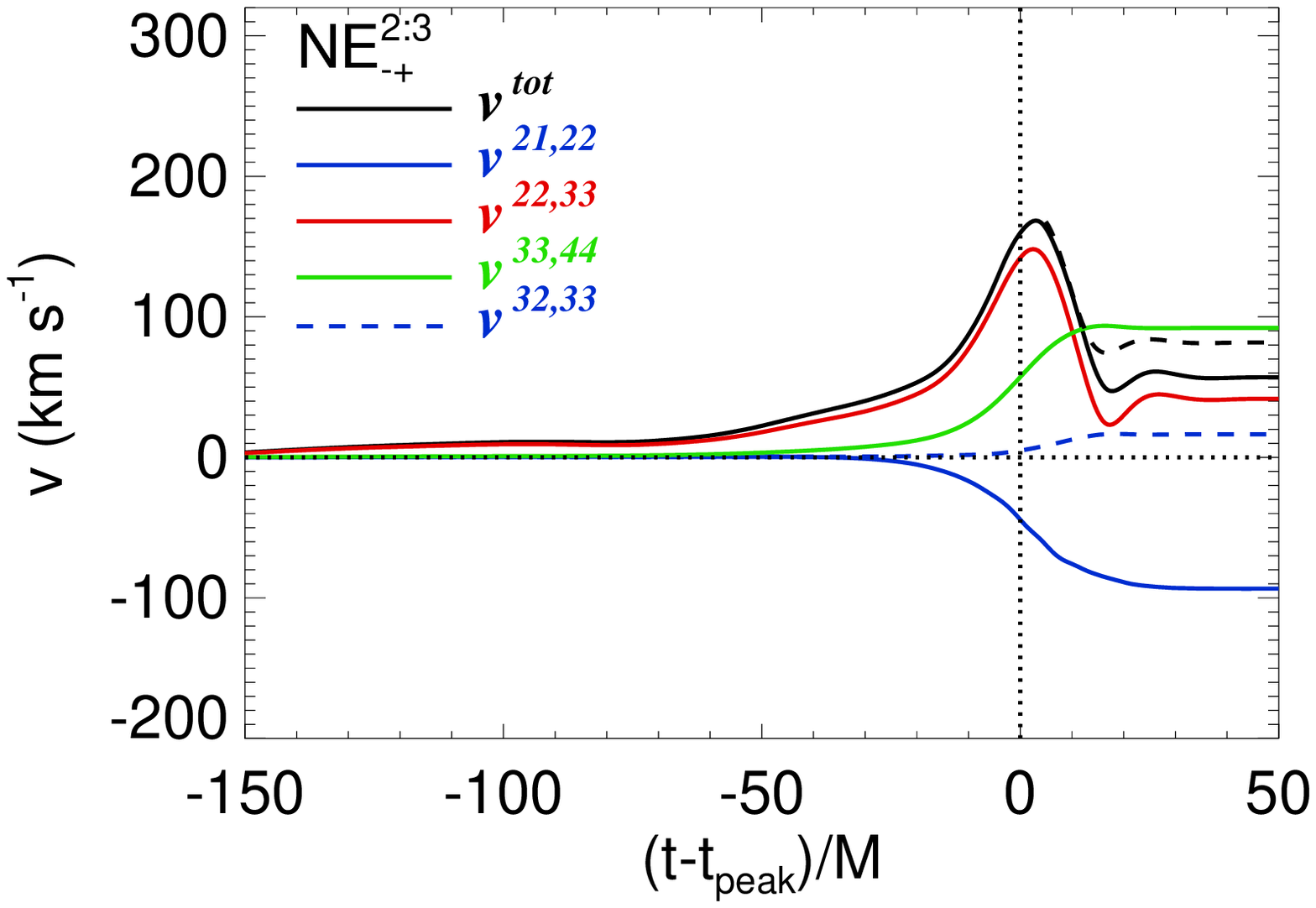}
\includegraphics[width=0.48\textwidth,clip=true]{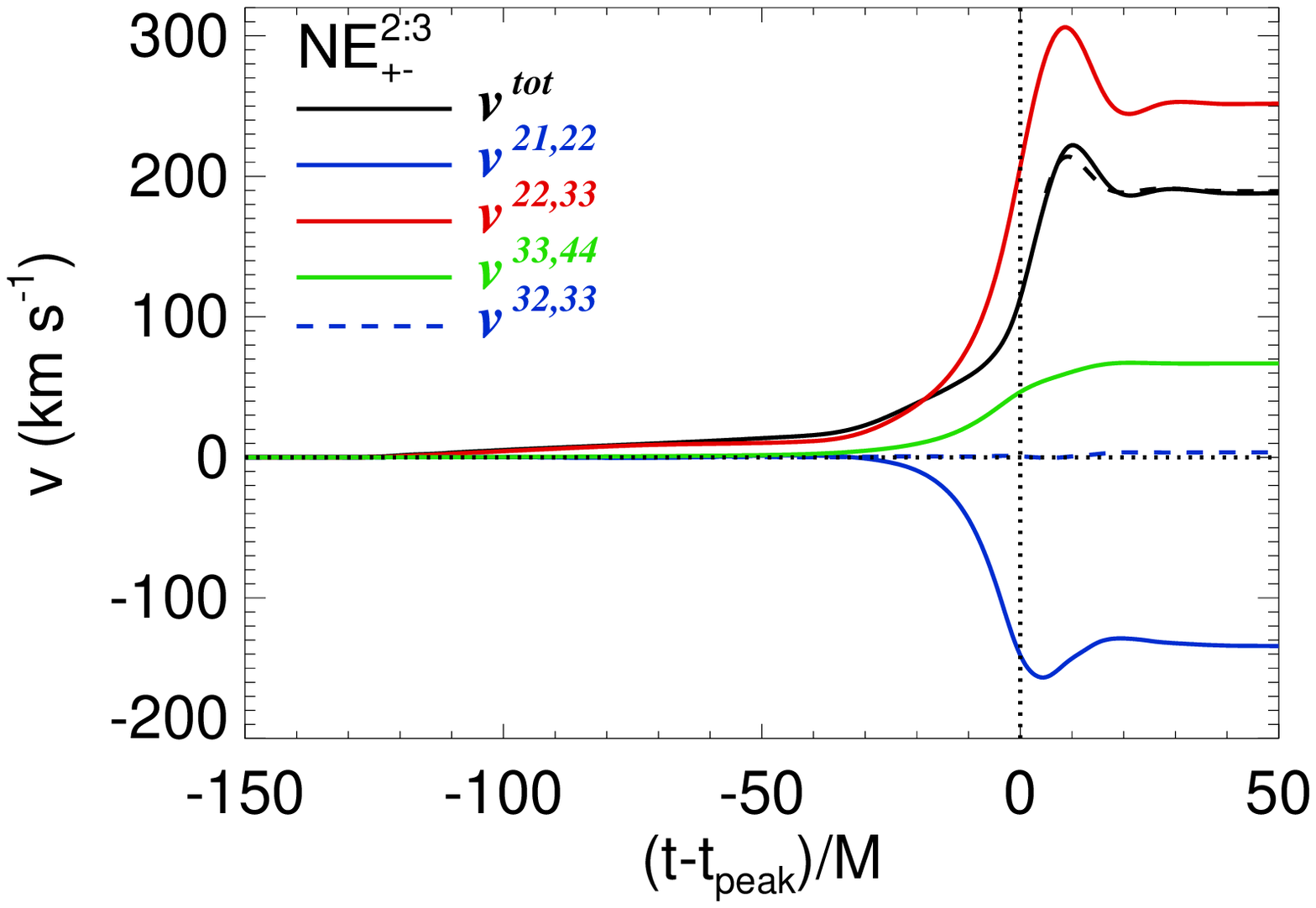} \\
\includegraphics[width=0.48\textwidth,clip=true]{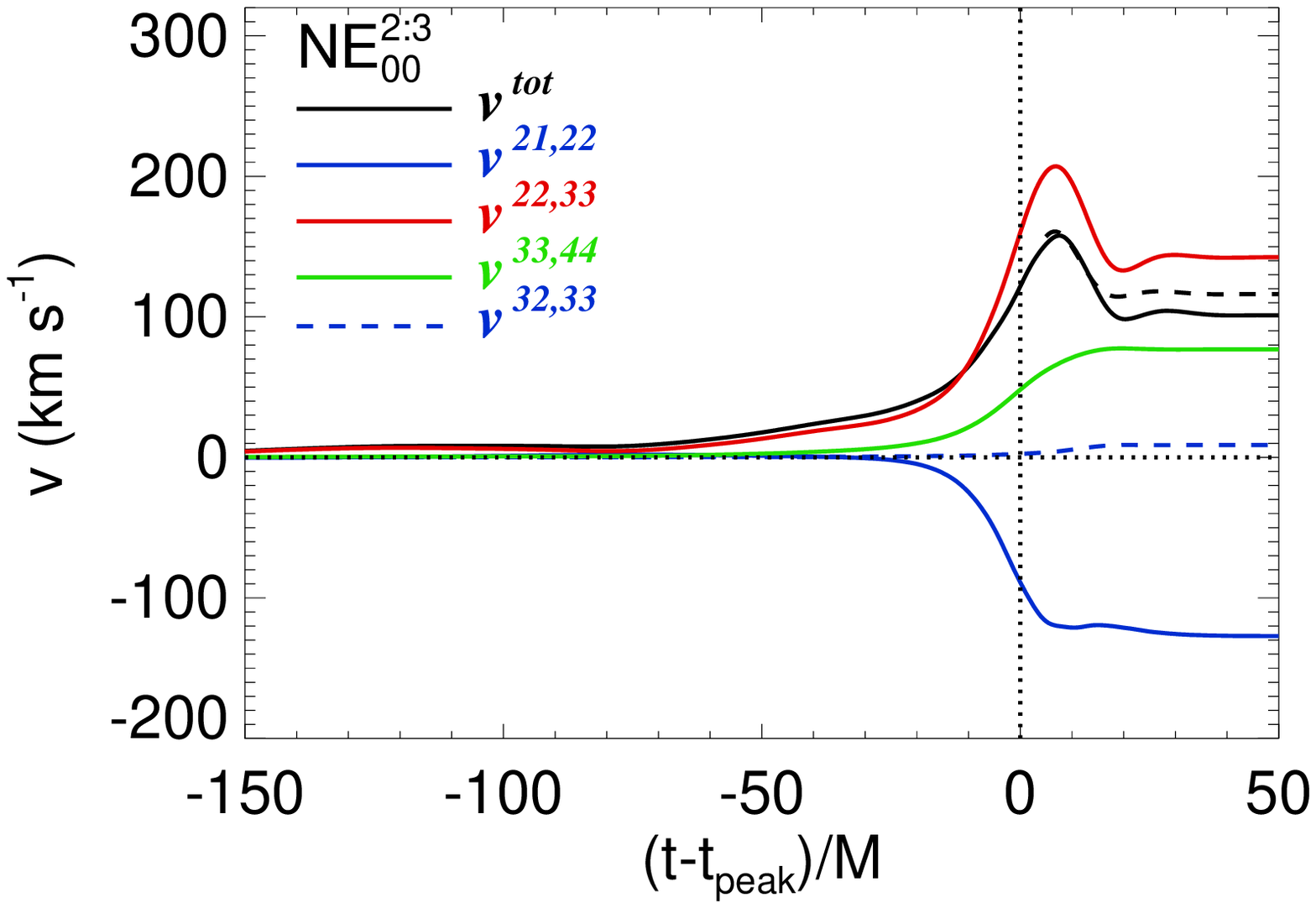}
\includegraphics[width=0.48\textwidth,clip=true]{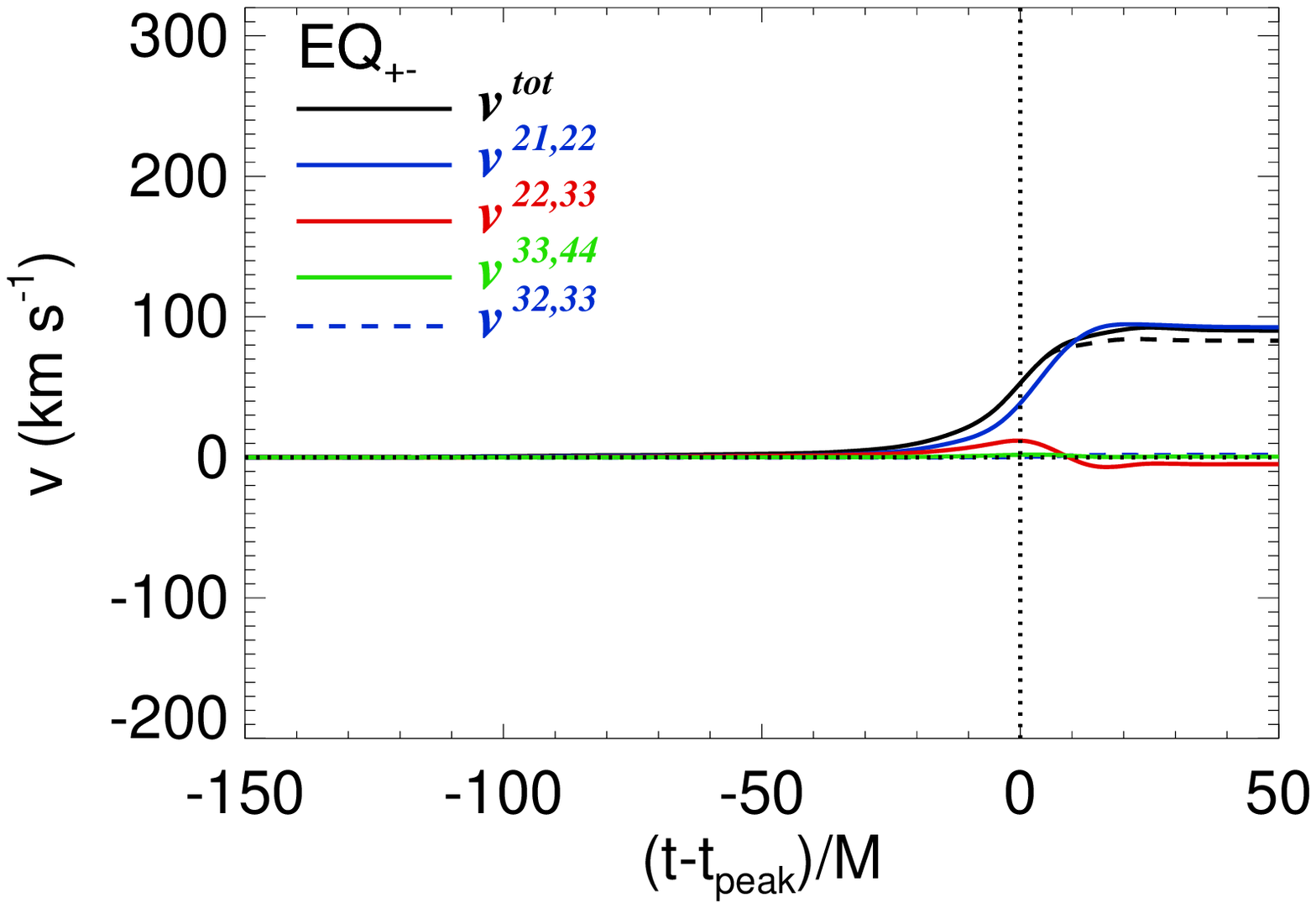}
\caption{\label{ne_kickmodes1} Relative contributions to the total
  recoil velocity from the different multipole
  mode-pairs. $I^{22}I^{33*}$ (red curve) is the dominant mode for
  unequal-mass 
  binary systems, while $S^{21}I^{22*}$ (blue curve) dominates for spinning,
  equal-mass binary systems. Also plotted is the
  contribution from the $S^{32}I^{33*}$ flux terms (blue dashed
  curve), demonstrating its very small contribution to the total
  recoil velocity. For $t>t_{\rm match}=t_{\rm peak}$, we include the
  quasi-analytic prediction for $v_{\rm RD}(t)$ (black dashed
  curves), based on the fundamental RD modes from Eq.~(\ref{v_RD}).
  The data refer to the NE$_{+-}^{2:3}$ (upper left panel),
  NE$_{-+}^{2:3}$ (upper right panel), NE$_{00}^{2:3}$ (lower left
  panel), and EQ$_{+-}$ (lower right panel) runs.
  We denote with $t_{\rm peak}$ the time at which $I^{22}$ reaches its maximum.}
\end{figure*}

Given $v_{\rm match}$ at the end of the inspiral, we can use this
quasi-analytic approach to predict the maximum and final recoil
velocities ($v_{\rm max}$ and $v_{\rm f}$, respectively). These
predictions are plotted as black dashed curves in
Fig.~\ref{ne_kickmodes1}, to be compared with the solid black curves
of the exact NR results. Within this
context, we define the anti-kick magnitude as 
\beq\label{f_ak}
f_{\rm ak}\equiv \frac{v_{\rm f}-v_{\rm max}}{v_{\rm max}}
\eeq
and the net ringdown contribution as 
\beq\label{f_RD}
f_{\rm RD}\equiv \frac{v_{\rm f}-v_{\rm match}}{v_{\rm match}}\, ,
\eeq
where $v_{\rm max}$ and $v_{\rm f}$ are the (real-valued) velocity
magnitudes calculated analytically from Eq.~(\ref{v_RD}).

In the case of the NE$_{-+}^{2:3}$ run, where the recoil is almost entirely 
dominated by the $F^{22,33}$ flux, we find a large anti-kick
with $f_{\rm ak}=-0.53$ and $f_{\rm RD}=-0.5$. On the other hand, for
the NE$_{+-}^{2:3}$ run, as can be seen in Fig.~\ref{ne_kickmodes1},
the net recoil velocity continues to increase after $t_{\rm
  match}=t_{\rm peak}$ before turning around for a small anti-kick of
$f_{\rm ak}=-0.11$. The total effect of the ringdown phase is actually
to increase the recoil with $f_{\rm RD}=0.68$. An intermediate effect
is seen for the NE$_{00}^{2:3}$ run, with $f_{\rm ak}=-0.28$ and
$f_{\rm RD}=-0.04$. However, as seen above in
Fig.~\ref{ne_spiralmodes1}, for the EQ$_{+-}$ run, we see no
anti-kick, with $f_{\rm ak}=-0.01$ and $f_{\rm RD}=0.58$.

In general, we find the magnitude
of the anti-kick is primarily dependent on the relative magnitudes of the
$S^{21}$ and $I^{33}$ moments. When $S^{21}$ dominates (e.g.,
  when $\delta m$ and $\Delta^z$ add constructively), the ringdown
rotation is slow and there is a small anti-kick, whereas a dominant
$I^{33}$ mode (e.g., large $\delta m$ or no spins) gives a rapidly rotating ringdown flux and thus a large
anti-kick. Furthermore, from Eq.~(\ref{newtonian_flux}), we see that for
non-spinning BHs, both the $S^{21}$ and $I^{33}$ modes share the same
mass and frequency scaling, so the relative size of the anti-kick should
be roughly independent of mass ratio (see Appendix \ref{app} for a caveat).

We would like a more quantitative picture of how these flux vectors add
constructively and destructively to give the total recoil velocity to
support the analytic estimates presented above.  
Using $\mathbf{v}=\int \mathbf{F} dt$, we can write
\begin{equation}
\label{dvhat_dt}
\frac{d}{dt}|\mathbf{v}| =
\frac{d}{dt}(\hat{\mathbf{v}}\cdot\mathbf{v}) = \hat{\mathbf{v}}\cdot
\mathbf{F}\,, 
\end{equation}
where $\hat{\mathbf{v}} \cdot \hat{\mathbf{v}} = 1$. 
Breaking $\mathbf{F}$ up into the contributions of the 
dominant modes as above, and then integrating in time gives
\bse
\begin{eqnarray}
\label{vdotf}
v^{21,22}&=&\int \hat{\mathbf{v}}\cdot\mathbf{F}^{21,22}\,dt \, ,\\
v^{22,33}&=&\int \hat{\mathbf{v}}\cdot\mathbf{F}^{22,33}\,dt \, ,\\
v^{33,44}&=&\int \hat{\mathbf{v}}\cdot\mathbf{F}^{33,44}\,dt \, ,
\end{eqnarray}
\ese
which add linearly to give to total recoil velocity:
\begin{equation}
|\mathbf{v}|=v^{21,22}+v^{22,33}+v^{33,44}.
\end{equation}
Note that with these definitions, the $v^{\ell
m,\ell' m'}$ are all real, but can be positive or negative. 
These different velocities are plotted in Fig.~\ref{ne_kickmodes1}, 
with the same color scheme as in Figs.~\ref{ne_modeamps} and \ref{ne_spiralmodes1}, 
along with the total recoil velocity in
  solid black curves. Also shown in Fig.~\ref{ne_kickmodes1} is the
  velocity $v^{32,33}$ (dashed blue curves), defined analogously to
  Eq.~(\ref{vdotf}) for the $S^{32}I^{33*}$ flux terms. The
  small contribution from this mode pair further justifies our focus
  on the more dominant pairs of Eq.~(\ref{flux_approx}) and
  Fig.~\ref{flux_lm}.

In the NE$_{-+}^{2:3}$ run, where the modal analysis shows the
$F^{21,22}$ and $F^{33,44}$ flux terms canceling out, we see that the
total recoil velocity (black curves in
Fig.~\ref{ne_kickmodes1}) is almost entirely dominated by the
$F^{22,33}$ flux (red curves). On the other hand, for the
NE$_{+-}^{2:3}$ run, the $F^{21,22}$ flux is much stronger, adding
destructively with the $F^{22,33}$ flux during the RD. This has the effect of both
increasing the peak velocity and also decreasing the relative strength
of the anti-kick, due to the slow rotation frequency of the $F^{21,22}$
flux during ringdown, as described above. As expected, the
NE$_{00}^{2:3}$ run displays behavior intermediate between these two
extremes. The EQ$_{+-}$ run, however, is entirely dominated by the
$F^{21,22}$ flux, and thus experiences {\it no} anti-kick, but
rather drifts off slowly in a nearly constant direction, as seen in the
bottom-right panel of Fig.~\ref{ne_spiralmodes1}.

\subsection{Application to non-planar kicks}\label{planarspins}

One of the most remarkable results from the recent renaissance in
numerical relativity was the prediction of extremely large kicks from
equal-mass BHs with spins pointing opposite to each other and normal
to the orbital angular momentum, producing a recoil out of the orbital
plane \cite{recoilRI,Bigrecoil,recoilFAU,recoilJena2}. While this configuration can
produce recoils of nearly $4000$ km/sec, the analogous non-precessing
configuation (EQ$_{+-}$ run in this paper) gives a kick of only $\sim
500$ km/sec in the case of maximal spin
\cite{recoilAEI,recoilPSU,recoilGoddard}. The 
multipole analysis tools developed above can be used for understanding
and explaining this remarkable difference.

First, we should note that leading-order PN
estimates of the linear momentum flux during inspiral suggest that the
discrepancy should be less than a factor of two. For example, Eq.~(3.31b)
of Kidder \cite{LK} gives the spin-orbit contribution to the momentum
flux for circular, Keplerian orbits as
\beq
\mathbf{F}_{\rm SO} = \frac{16}{15}\mu^2\, M\,\frac{\omega^2}{R^3}
[\hat{\mathbf{n}}\times \mathbf{\Delta}+
(\hat{\mathbf{n}}\times\hat{\mathbf{v}})
(\hat{\mathbf{v}}\cdot\mathbf{\Delta})],
\eeq
with $\hat{\mathbf{n}}$ and $\hat{\mathbf{v}}$ being the normalized
separation and velocity vectors, respectively. For spins parallel to
the orbital angular momentum, the term in square brackets has magnitude
$\Delta^z$, while for planar spins, it is
$2\Delta^p\sin\phi_{\Delta}$, where 
$\phi_{\Delta}$ is the angle between $\mathbf{\Delta}$ and
$\mathbf{n}$, and $\Delta^p$ is the
magnitude of $\mathbf{\Delta}$ in the orbital plane.

Not surprisingly, we get the exact same results from the multipole
analysis of Eqs.~(\ref{dPdt_radmoments}), (\ref{dPdtz_radmoments}),
(\ref{spinorbit_moments}), and one new multipole moment:
\beq\label{s22_SO}
S^{22}_{\rm SO} = 4i\sqrt{\frac{2\pi}{5}}\,\eta\, R\, \omega^3 \,e^{-i\phi}
(\Delta^x-i\Delta^y),
\eeq
while on the other hand, the $S^{21}$ mode is zero for the planar spin
configuration. Combining these equations, we get
\begin{widetext}
\beq\label{parallel_flux}
F_x+iF_y \approx \frac{1}{336\pi}(-14iS^{21}I^{22*})= 
\frac{16}{15}i \frac{\mu^2}{M}\, R^3\, \omega^6\, \Delta^z\, e^{i\phi}
\eeq
and using Eq.~(\ref{dPdtz_radmoments}) we obtain 
\bea\label{planar_flux}
F_z\approx \frac{1}{336\pi}[-28\Im(I^{22}S^{22*})] &=&
\frac{32}{15} \frac{\mu^2}{M}\, R^3\, \omega^6\, (\Delta^x\sin\phi
-\Delta^y\cos\phi) \nonumber\\
&=& \frac{32}{15} \frac{\mu^2}{M}\, R^3\, \omega^6\, \Delta^p\sin\phi_\Delta\, ,
\eea
\end{widetext}
where $\phi$ is the orbital phase of the binary.
So in both paradigms, we see that, when maximizing over
$\sin\phi_{\Delta}$, the planar-spin orientation should result in a
recoil twice as large as the parallel-spin case, leaving a factor
of roughly 4 difference unexplained. 

\begin{figure*}
\includegraphics[width=0.48\textwidth,clip=true]{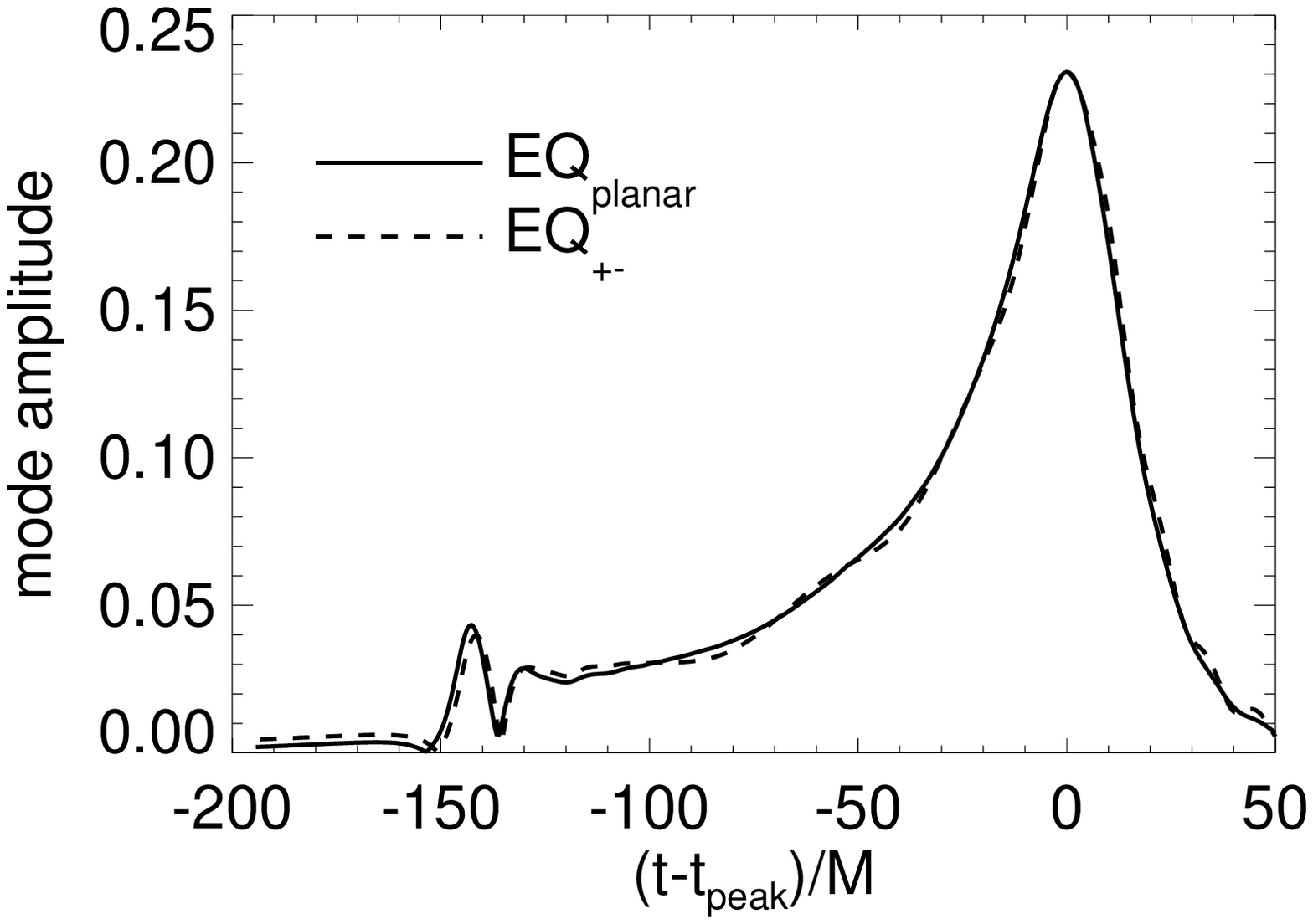}
\includegraphics[width=0.48\textwidth,clip=true]{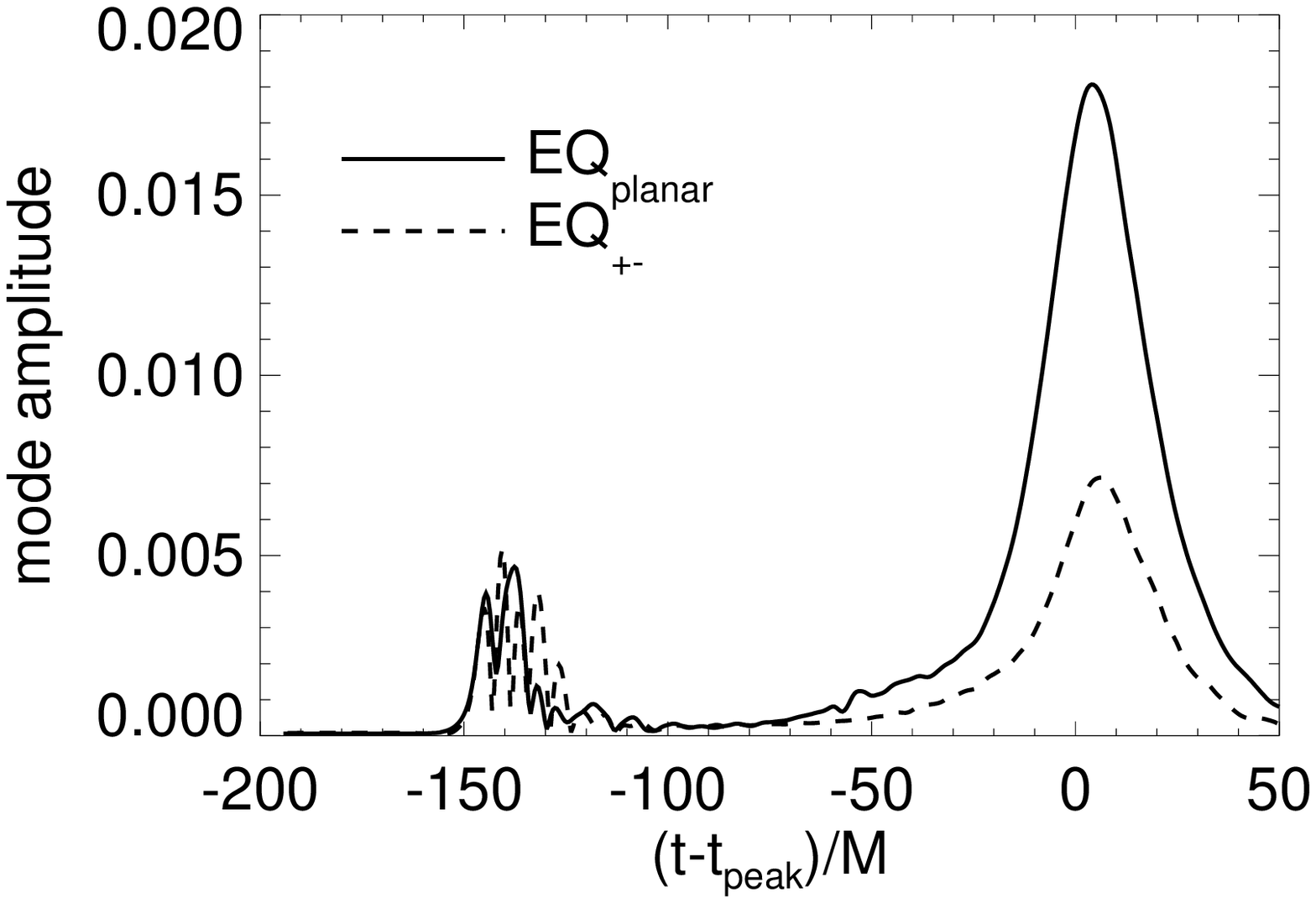}
\caption{\label{I22_planar} {\it left panel:} Comparison of the
  multipole amplitudes
  $I^{22}$ for the two different equal-mass simulations: EQ$_{\rm
    planar}$ (solid line) and EQ$_{+-}$ (dashed line). {\it right
    panel:} The $S^{22}$ amplitude from the planar-spins run (EQ$_{\rm
    planar}$, solid line) and the $S^{21}$ amplitude from the
parallel-spins run (EQ$_{+-}$, dashed line). We denote with $t_{\rm peak}$
  the time at which $I^{22}$ reaches its maximum.}
\end{figure*}

From Eqs.~(\ref{parallel_flux}),(\ref{planar_flux}) we see that the
only relevant modes involved should be $I^{22}$, $S^{21}$, and
$S^{22}$ (for these equal-mass systems the momentum flux is
dominated by a single mode pair, responsible for $\gtrsim 95$\% of the
final recoil value). In the left panel of Fig.~\ref{I22_planar} we plot the
amplitude of $I^{22}$
from the EQ$_{+-}$ simulation, along with that of a planar-spin
simulation EQ$_{\rm planar}$. All other binary parameters and the
initial conditions are the same. Remarkably, the mass-quadrupole
moments $I^{22}$ are nearly identical (and dominant) in both runs, and
this suggests that the energy and angular momentum fluxes are the same
[see Eqs.~(\ref{dEdt}),(\ref{dJzdt})]. 
This is in fact quite reasonable since the total spin of the system is zero in both
cases. However, we see in the right-hand panel of
Fig.~\ref{I22_planar} that the peak amplitude of the $S^{22}$ mode is a
factor of $\sim 2.5$ greater than that of the $S^{21}$ mode from the
EQ$_{\rm planar}$ and EQ$_{+-}$ runs, respectively. 

Yet Eqs.~(\ref{spinorbit_moments}),(\ref{s22_SO}) suggest that these
two modes should have exactly the same magnitudes, at least during the
inspiral phase, and presumably during the RD as well, since the RD
amplitudes are completely determined by the mode amplitudes at the
matching point. It appears from Fig.~\ref{I22_planar} that $S^{22}$
and $S^{21}$ do in fact have the same amplitude at early times, but
the relatively noisy data and short duration of the simulations make
it impossible to say for certain. If this is the case, one possible
explanation for the sudden remarkable increase in the amplitude of
$S^{22}$ might be mode-coupling with the dominant $I^{22}$ mode, as the
inspiral phase begins to transition to the RD phase. This coupling is 
analogous to that of $S^{32}$ and $I^{22}$ described above in
Sec.~\ref{RD_phase}, an effect that is apparently only
important between modes with the same $m$-number\cite{BCW,BCP}. We
hope to address this question in the future with longer
simulations to confirm the agreement at early times, as well as other
spin configurations that should enhance specific multipole modes and
may help identify other similar cases of mode amplification.

\begin{figure*}
\includegraphics[width=0.48\textwidth,clip=true]{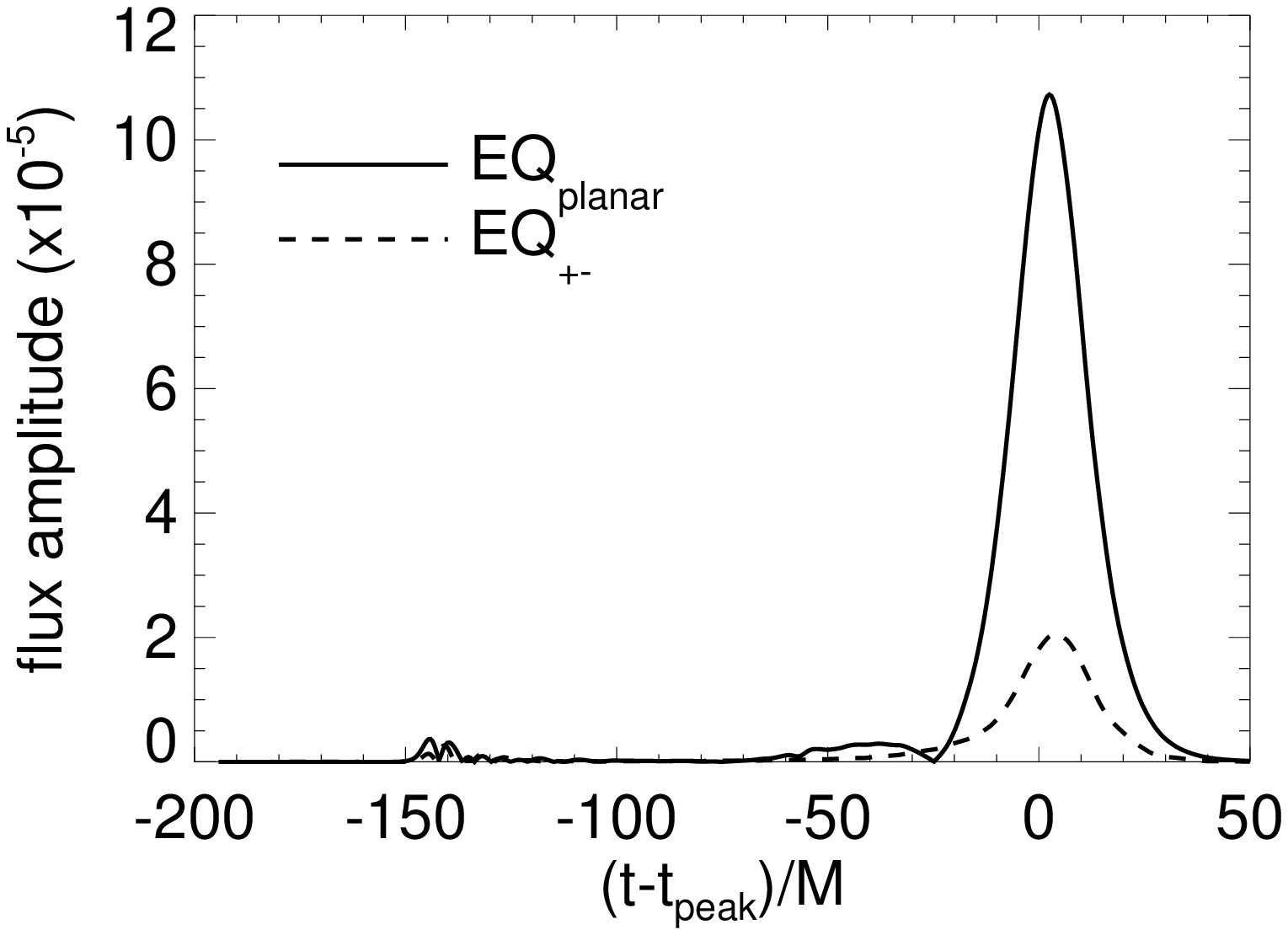}
\includegraphics[width=0.48\textwidth,clip=true]{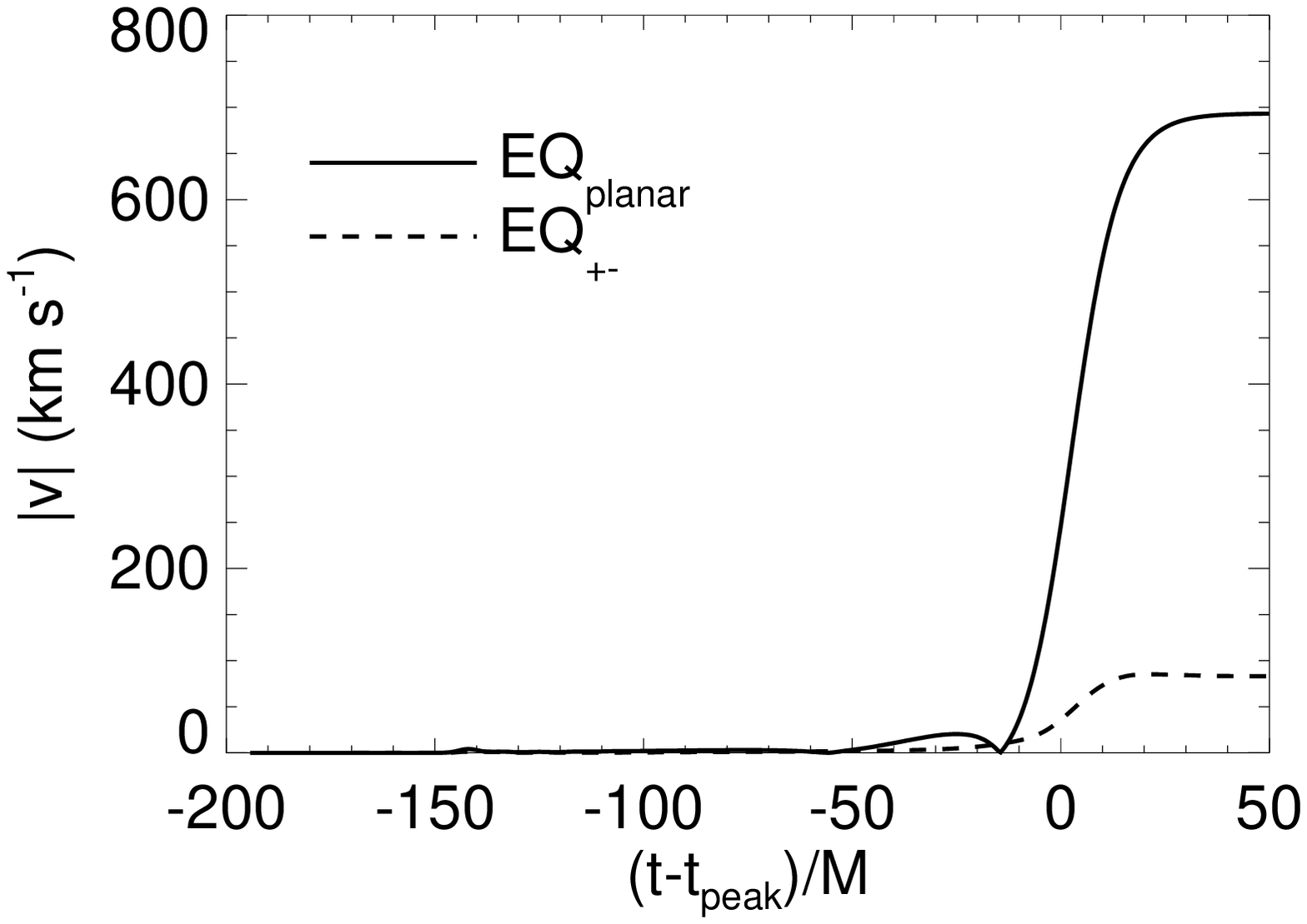}
\caption{\label{planar_recoil} {\it left panel:} Comparison of the
  linear momentum flux for the two different equal-mass simulations: EQ$_{\rm
    planar}$ (solid line) and EQ$_{+-}$ (dashed line). {\it right
    panel:} The total recoil velocity from the planar-spins run (EQ$_{\rm
    planar}$, solid line) and the parallel-spins run (EQ$_{+-}$,
  dashed line). We denote with $t_{\rm peak}$
  the time at which $I^{22}$ reaches its maximum.}
\end{figure*} 

Lastly, from the ringdown contribution to the velocity
[Eqs.~(\ref{v_RD}),(\ref{vtot_modes})], we can understand another
difference between the planar- and parallel-spin
orientations. 
Instead of having two different RD frequencies
$\sigma_{210}$ and $\sigma_{220}$ combine to give a slowly rotating
flux vector, for the planar-spin case, we have two {\it identical} RD
frequencies for $I^{22}$ and $S^{22}$ in Eq.~(\ref{planar_flux}), 
giving precisely zero rotation to the RD
flux. Furthermore, as the spin vector $\mathbf{\Delta}$ is precessing
faster and faster in a positive direction around the orbital angular
momentum vector, even during the inspiral the two modes $I^{22}$ and
$S^{22}$ become nearly locked in phase, producing a relatively
long-duration burst of linear momentum flux in a single direction
during the merger phase. Combined, these effects essentially
straighten out the spiral curve in the lower-right panel of
Fig.~\ref{ne_spiralmodes1}, providing another factor of $\sim 1.6$
of increased recoil velocity for planar spins. 

In Fig.~\ref{planar_recoil} we show the combination of the above
effects. In the left panel, we plot the linear momentum flux from
Eqs.~(\ref{parallel_flux}),(\ref{planar_flux}), showing the factor of
two increase predicted by the Kidder formula and our
Eqs.~(\ref{dPdt_radmoments}), (\ref{dPdtz_radmoments}), along with the factor of
$2.5$ increase in the amplitude of $S^{22}$ relative to $S^{21}$. In the
right panel, we plot the recoil velocity for both runs,
which includes the effect of flux rotation during the merger and
inspiral phases, accounting for another factor of $\sim 1.6$, giving a
total discrepancy of $v({\rm EQ}_{\rm planar})/v({\rm EQ}_{+-})\approx
2.5\times 2\times 1.6 = 8$. 

\section{Discussion}
\label{discussion}

In this paper we analysed several numerical simulations of binary BH
coalescence, focusing on the physics of the recoil. We developed tools,
based on the multipolar expansion~\cite{KT,BD,BDS,BS,JS}, that can be
used as a diagnostic of the numerical results, and  
understand how the recoil velocity evolves during the inspiral,
merger, and ringdown phases of the coalescence.

We wrote explicit expressions for the linear momentum flux expressed 
in terms of radiative multipole moments through $\ell=4$, valid for 
generic spinning, precessing BH binary systems. We found that 
these formulae are sufficient to obtain the total recoil velocity with
high accuracy. By comparing the amplitudes of the different 
multipole moments, we found that in the case of non-precessing
spins--and thus a recoil in the orbital plane--only three pairs
of modes contribute to most of the linear momentum flux, 
notably ${S^{21}I^{22*}}$, ${I^{22}I^{33*}}$ and ${I^{33}I^{44*}}$. 
Those modes account for the total recoil with an accuracy on the order
of $\sim 5-10\%$ throughout the simulations. (see Figs.~\ref{v_tot1},
\ref{v_tot2}).

The way in which the contribution from these three pairs of modes
builds up is not trivial, since not only the relative amplitudes, but 
especially, the relative phases are also quite important. We found that the 
relative phases between the three mode-pairs are nearly constant during 
the inspiral phase, but start diverging at the onset of the 
transition from inspiral to RD (see Fig.~\ref{ne_modephase1}).
The late-time evolution can be described reasonably well with
analytic formula obtained expressing the mode-pairs in terms 
of fundamental QNMs of a Kerr BH. We showed that it is the relative 
magnitude of the current-quadrupole mode $S^{21}$ and the mass-octupole
mode $I^{33}$, together with the differences of the 
QNM fundamental frequencies for each of the dominant modes,
that determine the difference between the recoil at the 
peak of the linear momentum flux, and the final recoil velocity, i.e., 
the magnitude of the anti-kick.

With the final goal of improving analytic PN models, we also explored 
whether simple modifications of the Newtonian formula for the 
linear momentum flux allow us to match the numerical 
results all along the binary evolution. We found that, 
if we treat the binary radial separation in the Newtonian multipole 
modes (\ref{S21})--(\ref{I33}) with an effective radius, which is 
computed from the numerical simulations assuming that each multipole
mode is described by a dominant frequency (see
Fig.~\ref{omega_modes}), the leading Newtonian modes reproduce quite
well the numerical ones (see Figs.~\ref{compare_ring}, \ref{kick_match}) 
up to the end of the inspiral phase. We also found, confirming the results
in Ref.~\cite{BCP}, that a superposition of three 
QNMs can fit the numerical waveforms very well from the peak 
of the radiation through the RD phase.

The tools developed in this paper will be employed to 
improve current analytic predictions for the recoil 
velocity~\cite{BQW,DG} using PN analytic models
~\cite{LB} and the EOB approach~\cite{BD1,DJS,DIS98,BD2,BCD,
EOB4PN}. An accurate, fully analytic description of the recoil velocity 
can be adopted in fast Monte Carlo 
simulations to predict recoil distributions from 
BH mergers with uncertainties smaller than in Ref.~\cite{SB}. 
Those recoil distributions can in turn be included in simulations 
of hierarchical merger models of supermassive BHs providing 
more robust predictions for LISA.

\acknowledgments 
We thank Emanuele
Berti for providing us with tabulated data for the Kerr QNM
frequencies. We would like to thank the anonymous referee for their
careful and constructive comments.

A.B.\ and J.D.S.\ acknowledge support from NSF grant PHYS-0603762, and
A.B.\ was also supported by the Alfred P. Sloan Foundation. The work at
Goddard was supported in part by NASA grant 05-BEFS-05-0044 and
06-BEFS06-19. The simulations were carried out using
Project Columbia at the NASA Advanced Supercomputing Division (Ames
Research Center) and at
the NASA Center for Computational Sciences (Goddard
Space Flight Center).  B.J.K.\ was supported by the NASA Postdoctoral Program at
the Oak Ridge Associated Universities. S.T.M.\ was supported in part
by the Leon A.\ Herreid Graduate Fellowship.

\appendix

\section{Results from 1:4 mass ratio}
\label{app}

In addition to the simulations presented in the main body of this
paper, we have also recently analyzed a non-spinning
system with mass ratio 1:4 ($\eta = 0.16$). The
results of this analysis are
presented briefly in this appendix, as well as in
Tables~\ref{table:idparams}--\ref{table:QNM_freqs} (labeled
appropriately as NE$_{00}^{1:4}$). More details can be 
found in Ref.~\cite{EOB4PN}.

\begin{figure}
\includegraphics[width=0.48\textwidth,clip=true]{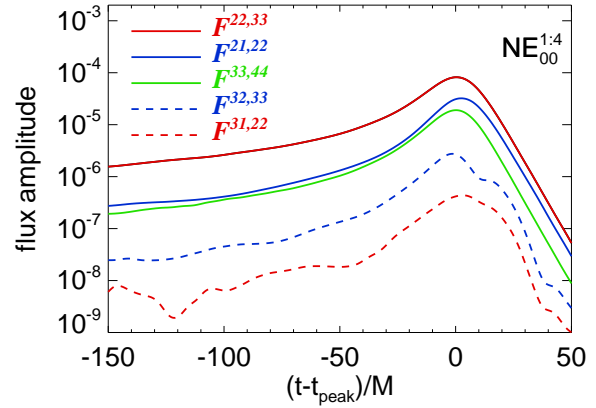}
\caption{\label{app_fig1} Flux amplitudes from the NE$_{00}^{1:4}$
  run, as in Fig.~\ref{flux_lm}.}
\end{figure} 

\begin{figure}
\includegraphics[width=0.48\textwidth,clip=true]{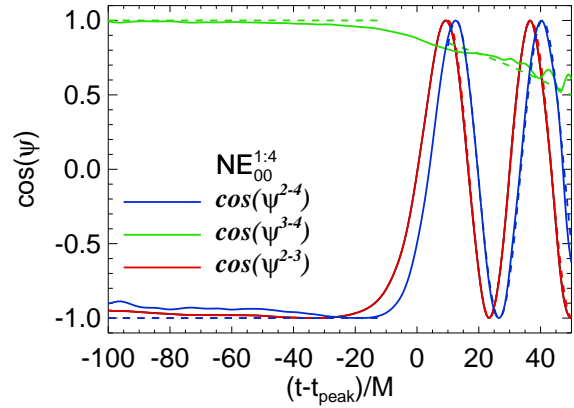}
\caption{\label{app_fig2} Phase differences from the NE$_{00}^{1:4}$
  run, as in Fig.~\ref{ne_modephase1}.}
\end{figure} 

\begin{figure}
\includegraphics[width=0.48\textwidth,clip=true]{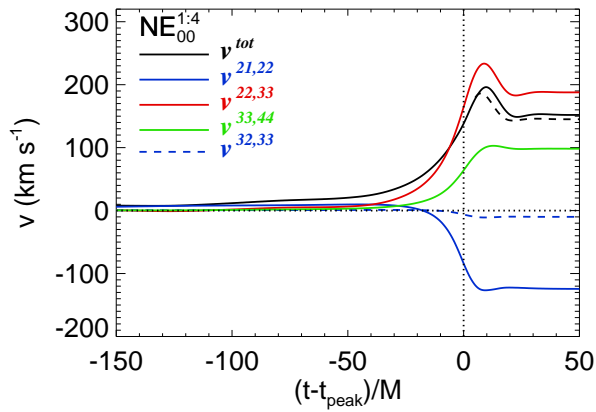}
\caption{\label{app_fig3} Relative contributions to the total
  recoil velocity from the different multipole mode-pairs for the NE$_{00}^{1:4}$
  run, as in Fig.~\ref{ne_kickmodes1}.}
\end{figure} 

In Fig.~\ref{app_fig1} we show the flux amplitudes from the different
modes, as in Fig.~\ref{flux_lm} above. We find the relative amplitudes
almost identical to those of the NE$_{00}^{1:2}$ run, with a slightly
stronger contribution from the $I^{44}$ mode, as expected from
Eq.~(\ref{I44}), which predicts a maximum in the $I^{44}$ amplitude for
$\eta=0.167$.

In Fig.~\ref{app_fig2} we plot the phase relations between the
different flux vectors, defined in
Eqs.~(\ref{phase_shifts1})-(\ref{phase_shifts3}). As anticipated in
Sec.~\ref{transition} above, we find a smaller phase shift in
$\psi^{3-4}$ at
the transition from inspiral to ringdown for this more extreme
mass-ratio system. The other phases appear to behave as expected.

Lastly, in Fig.~\ref{app_fig3}, we show the total recoil velocity
along with the relative contributions from the dominant modes for the
NE$_{00}^{1:4}$ run. Again, the qualitative behavior is quite similar
to the NE$_{00}^{2:3}$ and NE$_{00}^{1:2}$ runs, but we can now
identify a clear trend of a smaller anti-kick for smaller values of
$\eta$.  As mentioned above in Section \ref{antikick}, the
amplitude of the anti-kick is most strongly dependent on the relative
amplitudes of the $S^{21}$ and $I^{33}$ modes, but for non-spinning BH
binaries, these modes both scale the same with mass ratio. However,
the amplitude of the $I^{22}$ mode decreases with decreasing
$\eta$, while the amplitude of $I^{44}$ increases with
decreasing $\eta$, at least over the range considered here. Thus the
amplitude of the $F^{33,44}$ flux increases relative to the
$F^{22,33}$ flux for more extreme mass ratios. From
Figs.~\ref{ne_kickmodes1} and \ref{app_fig3}, we see that the
$F^{22,33}$ flux dominates the anti-kick, while the 
$F^{33,44}$ flux contributes almost nothing to it, so by
increasing the relative amplitude of $F^{33,44}$, we have effectively
decreased the size of the anti-kick.

\renewcommand{\prd}{\emph{Phys. Rev. D}}

\end{document}